\documentclass[a4paper,12pt]{article}
\usepackage[colorlinks,pdfpagelabels,pdfstartview=FitH,bookmarks=false,linkcolor=blue,plainpages=false,hypertexnames=false,citecolor=blue, urlcolor=blue]{hyperref}
\usepackage[utf8]{inputenc} 
\usepackage{amsfonts}
\usepackage{adjustbox}
\usepackage{amsmath}
\usepackage{graphicx}
\usepackage{amssymb}
\usepackage{longtable}
\usepackage{authblk}
\usepackage{pdflscape}
\usepackage{soul}
\usepackage{rotating}
\usepackage{siunitx}
\usepackage{float}
\usepackage[round]{natbib}
\usepackage{booktabs}
\usepackage{threeparttable}
\usepackage{booktabs}
\usepackage{threeparttable}
\usepackage{xcolor}
\usepackage{colortbl}
\usepackage[hmargin=2.5cm,vmargin=3.5cm]{geometry}
\usepackage{setspace}
\usepackage{chngcntr}
\usepackage{verbatim}
\usepackage{multirow}
\usepackage{subcaption}
\usepackage{chngcntr}
\usepackage[title]{appendix}
\usepackage[gen]{eurosym}
\usepackage{comment}
\usepackage[bottom]{footmisc}
\usepackage{gensymb}
\usepackage{textcomp,mathcomp}
\usepackage{wrapfig}
\usepackage{pdflscape}
\usepackage{rotating}
\usepackage{caption}
\usepackage{subcaption}
\usepackage{longtable,tabularx,ltxtable,ragged2e}
\emergencystretch=\maxdimen
\renewcommand{\floatpagefraction}{0.85}
\begin{document}

\pagenumbering{arabic}
\author[1] {Leonardo N. Ferreira}
\affil[1]{Central Bank of Brazil }
\author[2] {Haroon Mumtaz}
\affil[2]{Queen Mary University of London}
\author[3] {Ana Skoblar}
\affil[3]{European Central Bank}

\title{Stochastic Volatility-in-mean VARs with Time-Varying Skewness\footnote{Corresponding author: Haroon Mumtaz (\href{mailto:h.mumtaz@qmul.ac.uk} {h.mumtaz@qmul.ac.uk}). The views expressed in this paper are those of the authors and do not represent those of the European Central Bank, the Central Bank of Brazil, or any of their Committees.
}}

\date{This version: \today.}
\maketitle


\begin{abstract}
\noindent This paper introduces a Bayesian vector autoregression (BVAR) with stochastic volatility-in-mean and time-varying skewness. Unlike previous approaches, the proposed model allows both volatility and skewness to directly affect macroeconomic variables. We provide a Gibbs sampling algorithm for posterior inference and apply the model to quarterly data for the US and the UK. Empirical results show that skewness shocks have economically significant effects on output, inflation and spreads, often exceeding the impact of volatility shocks. In a pseudo-real-time forecasting exercise, the proposed model outperforms existing alternatives in many cases. Moreover, the model produces sharper measures of tail risk, revealing that standard stochastic volatility models tend to overstate uncertainty. These findings highlight the importance of incorporating time-varying skewness for capturing macro-financial risks and improving forecast performance.

\medskip
 
\vspace{1ex}

\end{abstract}



\clearpage
\pagenumbering{arabic} 
\setcounter{page}{2}

\section{\label{intro}Introduction}
Bayesian vector autoregressions (BVARs) with stochastic volatility have become a workhorse model both for forecasting and estimating the impact of economic shocks. \cite{clark2011} shows that stochastic volatility is a crucial ingredient for BVARs designed for density forecasting. This finding has been confirmed by a large subsequent literature. As shown in \cite{Koopman_uspensky}, \cite{Mumtaz_Surico2018} and \cite{carriero2016} amongst others, these models can be extended to allow stochastic volatility of the disturbances to affect the level of endogenous variables. These BVARs with stochastic volatility in mean have been used to estimate the effects of uncertainty shocks.
\newline In contrast to these models with time-varying second moments, considerably less attention has been devoted to BVARs where the distribution of the error term also features skewness. \cite{KARLSSON2023104580} introduce skewed and heavy-tailed distributions in BVARs and show that allowing for skewness improves forecasting performance. \cite{Galdon_Ortega} model structural disturbances in a VAR allowing for time-varying skewness, and find that this feature is pervasive in Euro-Area data. A similar VAR model is also considered by \cite{Renzetti} who report evidence for the presence of time-varying skewness in recursively identified shocks using US data.\footnote{\cite{ISERINGHAUSEN2020275} introduces a univariate stochastic volatility model with time-varying skewness to model exchange rate returns. \cite{ISERINGHAUSEN2024229} extends this model to a panel setting. }
\newline One feature that is common in these papers is that the time-variation in skewness is treated as an exogenous process. However, in a recent contribution \cite{kostas_skew} show that a measure of the skewness of macroeconomic variables has a dynamic relationship with real activity and inflation, contradicting this assumption.
\newline In this paper, we propose a BVAR that features time-varying volatility \textit{and} skewness. Unlike existing contributions, lagged shock volatility and skewness can affect the endogenous variables.
\newline We provide a Gibbs sampling algorithm to approximate the posterior distribution of the parameters. The model is applied to US and UK data where we consider the density forecasting performance of the proposed model relative to a BVAR that only features stochastic volatility but no time-varying skewness and to a BVAR with no dynamic relationship between the endogenous variables and skewness.
\newline Results show that allowing skewness in mean enhances the forecasting performance of BVARs. In the US, gains are most evident for the Gross National Product price deflator (GNP deflator) and,
during periods of economic stress, for the Real Gross National Product (GNP). For the UK, the benefits are even more
pronounced, with consistent improvements across variables and horizons. Finally, tail
risk measures show that the proposed framework provides a sharper characterisation of
downside and upside risks.
\newline The paper is organised as follows. The model is described in Section \ref{model} with the estimation algorithm summarised in Section \ref{Gibbs}. Section \ref{data} describes the data. Section 
\ref{Results} presents the empirical results while Section \ref{conc}
 concludes.
\section{Empirical Model} \label{model}
We model skewness in the disturbances of the VAR model by using the multivariate skew-normal distribution proposed by \cite{Arellano-Valle}. \cite{Arellano-Valle} show that if a vector $E$ has a skew-normal distribution with location vector $\mu$, scale matrix $\Sigma$ and a (diagonal) skewness matrix $\Delta$, it can be written as:
\begin{equation}
    E\overset{d}{=}\Delta|X_{0}|+X_{1}
\end{equation} 
where $X_{0}$ and $X_{1}$ are independent and $X_{0} \sim N(0,1)$ and $X_{1} \sim N(\mu,\Sigma)$. Thus, if skewness (i.e. diagonal elements of $\Delta$) is equal to zero, the distribution collapses to the multivariate normal.\newline With this definition in hand, we consider the following state-space model:
\begin{eqnarray}
\beta_{t} &=&\alpha +\theta \beta_{t-1}+%
\sum_{j=1}^{Q}d_{j}Y_{t-j}+\eta _{t}  \label{eq1} \\
Y_{t} &=&c+\sum_{j=1}^{P}B _{j}Y_{t-j}+\sum_{l=1}^{L}b_{l}\tilde{h}_{t-l}+\sum_{l=1}^{L}a_{L}\tilde{d}_{t-l}+A^{-1}E_{t} \label{eq2} 
\end{eqnarray}
where $Y_{t}$ denotes the vector of $N$ endogenous variables included in the VAR. The error term of the observation equation (\ref{eq2}) follows a skewed normal distribution and is defined as $E_{t}=\tilde{d}_{t}\odot \tau_{T}+e_{t} $ with $\tau_{t}=|\Theta_{t}|, \Theta_{t} \sim N(0,1)$ and $e_{t}$ is a multivariate normal as described below. Note that $A$ is a lower triangular matrix with ones on the main diagonal.
\newline The state variables are denoted by $\underbrace{\beta_{t}}_{K\times 1}=\begin{pmatrix}
 \tilde{h}_{t} \\
 \tilde{d_{t}}
\end{pmatrix}$ and $\underbrace{\tau_{t}}_{N\times 1}$ .
The stochastic volatilities are denoted by 
$\tilde{h}_{t}=[h_{1t},h_{2t},\dots,h_{N,t}]$ and $H_{t}=diag\left( \exp \left(\tilde{h}_{t}\right) \right) $. Time-varying skewness is denoted by $\tilde{d_{t}}=[d_{1t},d_{2t},\dots,d_{N,t}]$. These state variables evolve according to equation \ref{eq1}. As evident, $\tilde{h}_{t}$ and $\tilde{d_{t}}$ are allowed to have a lagged impact on the endogenous variables in equation \ref{eq2}. 
\newline The disturbances $\varepsilon _{t}=\left( 
\begin{array}{c}
\underbrace{\eta _{t}}_{K\times 1} \\ 
\underbrace{e_{t}}_{N\times 1}%
\end{array}%
\right) $ are distributed normally $N\left(\begin{pmatrix} 0 \\
0 \\
\end{pmatrix},
\begin{pmatrix}
    Q & 0 \\
    0 & H_{t}
\end{pmatrix}
\right)$

\subsection{Gibbs Sampling Algorithm} \label{Gibbs}
We use a Gibbs sampling algorithm to approximate the posterior distributions of the model parameters. The algorithm, which is based on \cite{Mumtaz_Surico2018} is described in detail in the technical appendix. Here, we present a summary of the conditional posterior distributions.
\paragraph{Coefficients of the VAR in the transition equation (\ref{eq1}) and observation equation (\ref{eq2}):} We exploit the fact that conditional on the state vector, the transition equation is a VAR and the conditional posterior is normal. 
\newline Conditional on the states, we rewrite the VAR in (\ref{eq2}) as:
\begin{equation} Y_{t}-\left(\tilde{d}_{t}\odot \tau_{T} \right)A^{-1\prime} =c+\sum_{j=1}^{P}B _{j}Y_{t-j}+\sum_{l=1}^{L}b_{l}\tilde{h}_{t-l}+\sum_{l=1}^{L}a_{L}\tilde{d}_{t-l}+\tilde{e}_{t} \end{equation}
The residuals of this VAR are Gaussian but with a known form of heteroscedasticity with $var(\tilde{e}_{t})=\Sigma_{t}=A^{-1}H_{t}A^{-1\prime}$. We use the Kalman filter to calculate the mean and variance of the conditionally Gaussian posterior.

\paragraph{Elements of $A$ and $Q$:} Conditional on the states and VAR parameters, the residuals $V_{t}$ can be calculated where $AV_{t}=E_{t}$. As shown in \cite{cogley-sargent-05}, the elements of the $A$ matrix are coefficients in regressions among these residuals and conditional on the skewness parameters $\tilde{d}_t$ and $\tau_{t}$ the conditional posterior is standard. The conditional posterior of $Q$ is inverse Wishart and is easily sampled.
 
\paragraph{ States $\tilde{h}_{t} ,\tilde{d_{t}}, \tau_{t}$    :}
Conditional on the coefficients and fixed variances, the model has the non-linear state-space representation depicted in equations (\ref{eq1}) and (\ref{eq2}). As in \cite{mumtaz2018}, we draw from the conditional posterior of the state vector using the particle Gibbs of \cite{RePEc:bla:jorssb:v:72:y:2010:i:3:p:269-342} and \cite{JMLR:v15:lindsten14a}.

\section{Data}
\label{data}
We estimate the model using quarterly data on real activity growth (GNP), inflation (GNP deflator) and the spread between the corporate bond yield and a long-term government bond yield, included as a measure of financial stress. We use data for the US and the UK.
\newline For the US, the sample runs from 1875Q1 to 2023Q4. The series are constructed by splicing pre-1947 data from the NBER's \hyperlink{https://www.nber.org/research/data/tables-american-business-cycle}{tables from the "The American Business Cycle."} with more recent data from \hyperlink{https://fred.stlouisfed.org/}{FRED} and \hyperlink{https://globalfinancialdata.com/}{Global Financial Database (GFD)}. The measure of real activity for the US is real GNP growth, while inflation is constructed using the GNP deflator. We use the variables RGNP72 and GNPD72 from the NBER pre-1947, splicing them with GNPC1 and GNPDEF from FRED. Similarly, the corporate bond yield is a spliced series using CORPYIELD from the NBER pre-1947 and BAA from FRED post 1947. We obtain the 10-year government bond yield from GFD (IGUSA10d).
\newline The full sample for the UK is 1920Q2 to 2023Q4. Data on GDP growth and CPI inflation are obtained from the Bank of England's database \hyperlink{https://www.bankofengland.co.uk/-/media/boe/files/statistics/research-datasets/a-millennium-of-macroeconomic-data-for-the-uk.xlsx}{A Millennium of Macroeconomic data} and the Office of National Statistics (ONS). Real GDP is available at a quarterly frequency from 1920-1938, but only annually over the period 1939-1954 from the Bank of England database. We linearly interpolate the GDP data over the Second World War (WWII) period and splice it with the quarterly series. The post-1955 real GDP data are taken from the ONS with code ABMI. CPI data from 1920-1916 are obtained from the Bank of England database and combined with the series with code D7BT from the ONS for the subsequent period. UK spreads are corporate bond yields from the Bank of England database spliced with data from GFD, minus 10 year yield from GFD.

\section{Empirical Results} \label{Results}
\subsection{Estimated State Variables}
We begin by presenting the estimated state variables from the model estimated on the full sample of data for each country.
\subsubsection{Stochastic volatility}

\begin{figure}[ht]
    \centering
    \includegraphics[width=\textwidth]{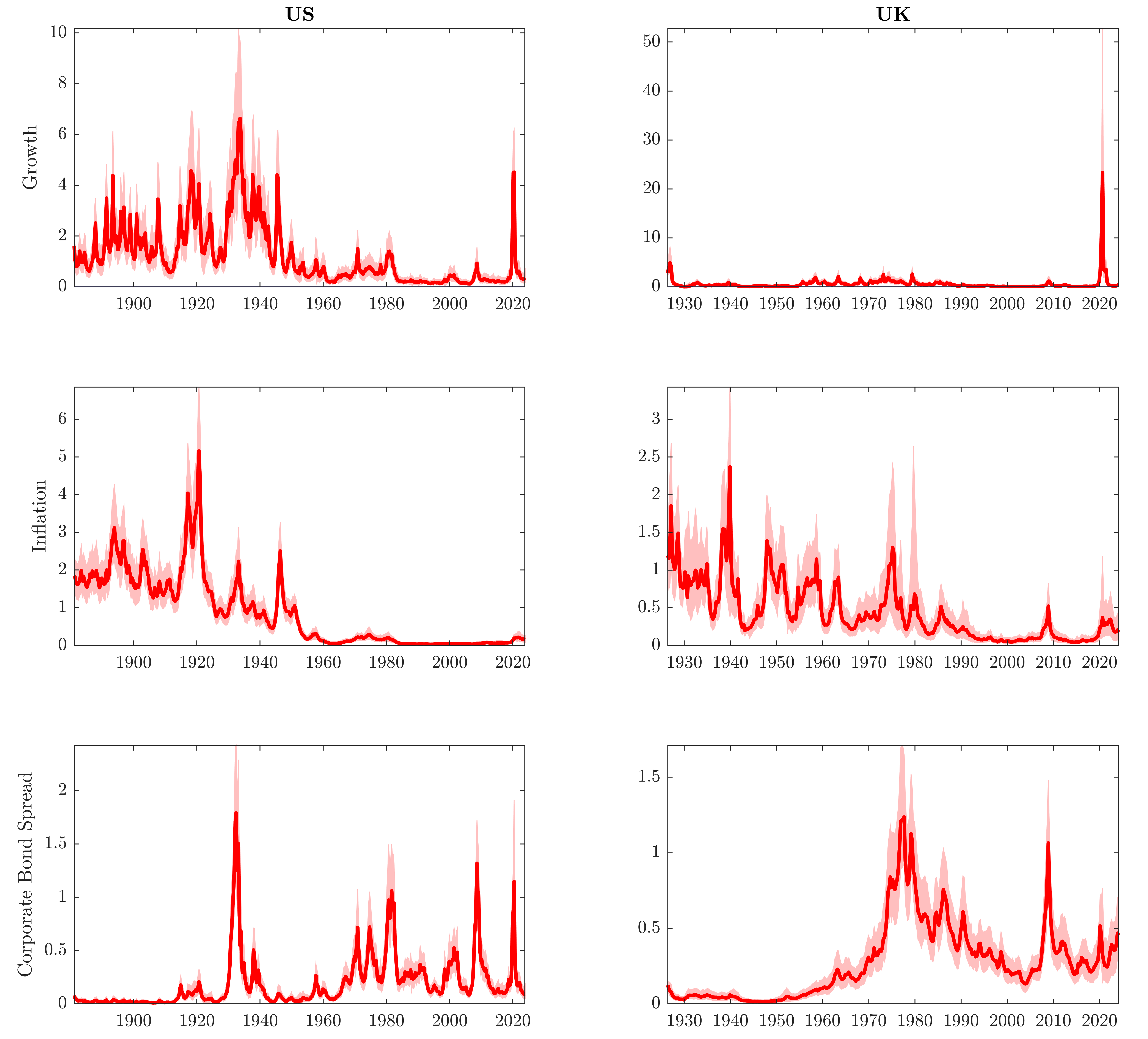}
    \caption{Estimate standard deviation of the shocks.}
    \label{figure1}
    \scriptsize 
    \textit{Note: The solid lines are the medians while the shaded area represents the 68\% error bands.} 
\end{figure}

Figure \ref{figure1} displays the (square root of) the estimated stochastic volatility. The largest peaks in US GNP volatility occurred at the end of the First World War (WWI), during the Great Depression and then the WWII and most recently during the Covid-19 crisis. Volatility of the US inflation shock was highest during the pre-1980 period, with the largest peak during the WWI. Volatility of the US spread shock peaked during the Great Depression and then during the Great Financial Crisis (GFC) and the Covid-19 period. Note that, unlike the macro variables, the spread disturbance appears to be more volatile in the post-1980 period. 
For the UK, the volatility of the GDP shock during the Covid period dwarfs the variance estimated in the previous years. The volatility of the UK inflation shock follows a pattern similar to that of the US and was at its highest in the pre-WWII period and during the 1970s and early 1980s. The volatility of the spread was higher, on average, after 1970 with peaks in the mid-1970s, mid-1980s, early 1990s and during the GFC.

\subsubsection{Time-varying Skewness}
\begin{figure}[ht]
    \centering
    \includegraphics[width=\textwidth]{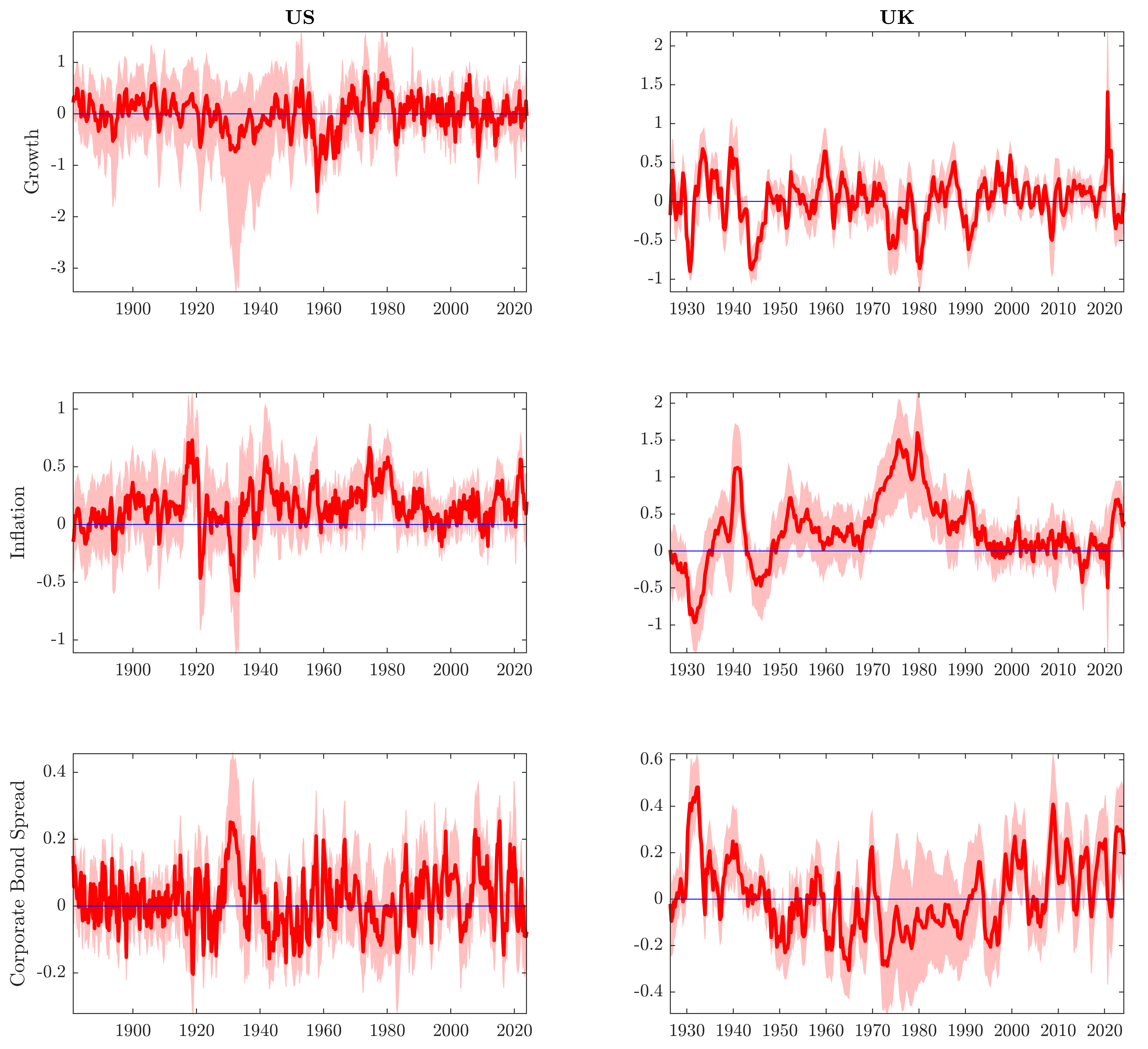}
    \caption{Estimate skewness of the shocks.}
    \label{figure2}
    \scriptsize 
    \textit{Note: The solid lines are the medians while the shaded area represents the 68\% error bands.} 
\end{figure}

The top left panel of Figure \ref{figure2} shows that estimated skewness of output shocks in the US largely fluctuates around zero with (imprecisely estimated) negative movements during the Great Depression, the mid-1950s, the early 1980s and the GFC. Inflation shock skewness was positive and statistically different from zero after WWI, the early 1940s and the mid 1950s. Inflation shock skewness remained above zero during the mid-1970s and the early 1980s and showed an increase during the post-Covid period. The skewness of the spread shock is estimated to be positive during the Great Depression and the GFC.

It is interesting to note that the skewness of the UK output shock shows more variation than the associated volatility over the full sample. The shock is negatively skewed during the large recession of the mid-1940s, the early 1970s and 1980s, in 1990s and then during the GFC. The large positive spike in 2020Q3, picks up the rebound in GDP growth after the initial decline associated with the Covid-19 crisis that began in 2020Q1. The skewness of the inflation shock in the UK shows a persistent increase during the 1970s, the early 1980s and 1990s and then after the Covid-19 period in 2022. Spread shocks were positively skewed over most of the pre-WWII period, and the same feature is evident after 1998.

\subsection{Impulse Response Functions}
\begin{figure}[h!]
    \centering
    \includegraphics[width=6.3in,height=7.6in,scale = 0.3, clip=true,trim = 0cm 0cm 0cm 0cm]{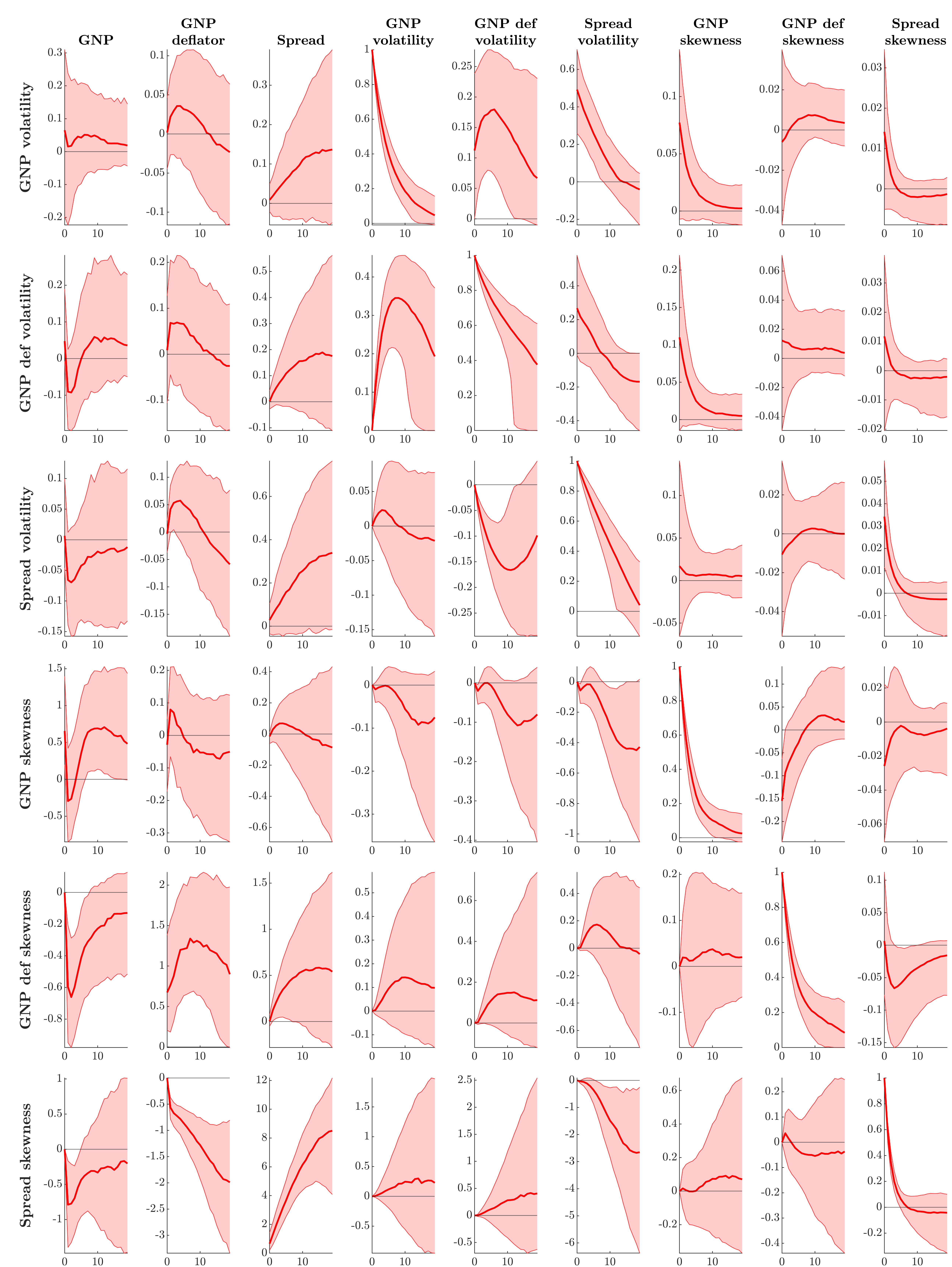}
     \caption{Impulse responses of shocks to the transition equation for the US.}
    \scriptsize 
    \textit{Note: The solid lines are the medians while the shaded area represents the 68\% error bands. Each row corresponds to a shock, and each column corresponds to a variable receiving the shock.} 
\label{irf1}
   \end{figure}

\begin{figure}[h!]
    \centering
    \includegraphics[width=6.3in,height=7.6in,scale = 0.3, clip=true,trim = 0cm 0cm 0cm 0cm]{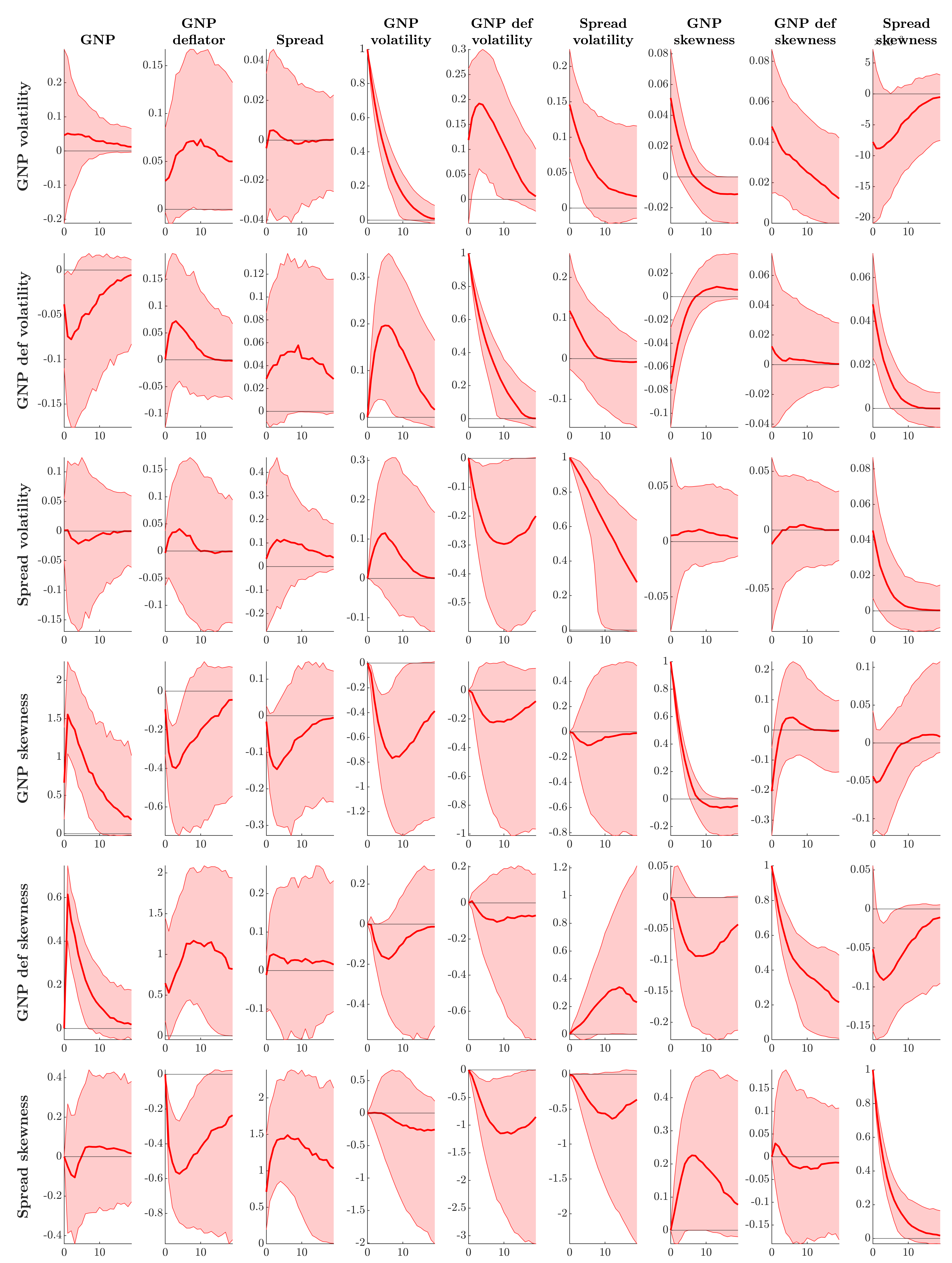}
    
     \caption{Impulse responses of shocks to the transition equation for the UK.}
    \scriptsize 
    \textit{Note: The solid lines are the medians while the shaded area represents the 68\% error bands. Each row corresponds to a shock, and each column corresponds to a variable receiving the shock.} 
\label{irf2}
   \end{figure}

Before moving on to the forecasting exercise, we present impulse response functions to a skewness shock. This structural analysis contributes to a deeper understanding of how skewness influences the economy.
To identify the shocks, we adopt a simple recursive scheme in which we use a Cholesky decomposition of $Q$ to compute the impulse responses to shocks to the transition equation. 
Impulse responses are calculated using Monte Carlo integration as described in \cite{koop-pesaran-potter-96} and are conditioned on the last data point in the sample. Figure \ref{irf1} shows the responses to shocks to volatilities and skewnesses for the US, and Figure \ref{irf2} shows these responses for the UK. 

In the US, stochastic volatility shocks primarily influence other stochastic volatilities, with the exception of spread volatility, which affects spread skewness and the GNP deflator. In contrast, skewness shocks exhibit a broader impact, affecting both volatilities and the main variables. For example, an increase in the GNP skewness increases the growth of GNP. GNP deflator skewness shocks, on the other hand, lead to a decline in GNP growth, while driving up both the GNP deflator and the spread. Finally, spread skewness causes a reduction in both GNP growth and its deflator and an increase in spread. It is worth noting that, among these, the shocks to the GNP deflator and spread skewness exhibit the most pronounced effects.

\vspace{1ex}

In the UK, stochastic volatilities are also important drivers of skewness. However, as in the US, skewness shocks have a more pervasive influence. Specifically, GNP skewness increases GNP growth while decreasing the GNP deflator. The skewness of the GNP deflator increases both the growth of GNPs and its deflator, whereas the spread skewness reduces the GNP deflator and increases the spread level.
Taken together, the evidence from both countries indicates that skewness has a substantial impact on core economic indicators. This underscores the importance of time-varying skewness in mean to help explain the dynamics of real economic activity, inflation, and corporate bond spreads. Consequently, accounting for this feature may also enhance the forecasting performance of BVARs -- a hypothesis we explore further in the next subsection.

\subsection{Forecast evaluation}
Our main application is a pseudo-real-time density forecasting experiment. For each country, we use a training sample that runs from the start of the available sample to 1974Q4. We then estimate the forecasting models recursively with the final estimation sample ending in 2021Q4. At each iteration we construct the predictive density up to eight quarters ahead and evaluate the forecasting performance. For output and inflation, which are modeled in first log-differences, we look at cumulative growth rates.

We first consider the accuracy of point forecasts (defined as posterior means), using root mean squared errors (RMSEs). We then evaluate density forecasts based on log scores and continuous ranked probability scores (CRPS), as in \cite{clark2015} and \cite{carriero2024addressing}, among many others. We 
focus on weighted log scores and weighted CRPS. Weighted scoring rules extend the standard log scores and CRPS by applying a weighting function to emphasise specific regions of the predictive distribution (\cite{Amisano-Giacomini-07, Gneiting2011}). In particular, we target the evaluation of forecast accuracy at the tails of the distribution, by assigning higher weights to these areas.\footnote{More details as well as standard log scores and CRPS, and log scores and CRPS focusing on other areas are presented in the appendix.} Statistical significance is assessed based on \cite{giacomini2006tests}'s tests. Nonetheless, p-values are only indicative as the models are estimated recursively, rather than using a fixed rolling window.

\subsubsection{Competing Models}
We compare the forecasting performance of the proposed model (which we denote as $VAR^{\sigma,\kappa}$) with two alternatives, which are restricted versions of the proposed model.\footnote{$\sigma$ denotes stochastic volatility, and $\kappa$ denotes time-varying skewness.} First, we restrict the coefficients $d_{j},b_{l}$ and $a_{l}$ in equations \ref{eq1} and \ref{eq2} to be equal to zero, thus assuming no dynamic relationship between the endogenous variables and the states. Thus, this model relates to \cite{BOTELHO2024104849} who do not allow an in-mean effect.\footnote{Note, however, that \cite{BOTELHO2024104849} do allow risk factors to influence the time-variation in skewness.} We denote this model as $VAR^{\sigma,\kappa}_{restricted}$. Second, we consider a VAR model where the disturbances are assumed to be conditional Gaussian with time-varying volatility only with skewness assumed to be zero (denoted as $VAR^{\sigma}$). Like the proposed model, the stochastic volatility is allowed to have a dynamic relationship with the endogenous variables $Y$.

\subsubsection{US results}
Table \ref{tab:combined_forecast_comparison} presents the average forecast performance relative to the competing models over the entire evaluation period. In terms of point forecast accuracy, $VAR^{\sigma,\kappa}$ tends to be marginally more accurate than $VAR^{\sigma}$ for GNP and the GNP deflator. When compared to $VAR^{\sigma,\kappa}_{restricted}$, improvements in RMSE are modest and primarily concentrated in GNP forecasts at horizons 2--7, with reductions of 1--4\%. Against $VAR^{\sigma}$, the proposed model shows more substantial gains, achieving RMSE reductions of 3--4\% for GNP at most horizons and 2--4\% for the GNP deflator across all forecast horizons.

Allowing stochastic volatilities and skewness to affect the endogenous variables plays a more significant role for metrics based on the full distributions. For weighted log scores emphasising both tails, the results reveal heterogeneous performance across variables and horizons. Relative to $VAR^{\sigma,\kappa}_{restricted}$, the proposed model delivers improvements for GNP deflator forecasts, with gains ranging from 0.1\% to 0.8\% across horizons, becoming statistically significant at longer horizons (H6--H8). Against $VAR^{\sigma}$, the GNP deflator shows consistent and statistically significant improvements, with gains of 0.3--1.3\%, particularly strong at shorter horizons where significance reaches the 1\% level.

The weighted CRPS results reinforce these findings. Compared to $VAR^{\sigma,\kappa}_{restricted}$, the proposed model achieves CRPS reductions of 6--14\% for the GNP deflator at longer horizons (H6--H8), with statistical significance. Against $VAR^{\sigma}$, improvements are even more pronounced, with CRPS reductions of 8--18\% for the GNP deflator and 6--16\% for the spread variable, many statistically significant at the 1\% level. Overall, scoring rules reveal that the proposed model consistently outperforms the alternatives for GNP deflator forecasts.

The analysis over time in Table \ref{tab:combined_period_comparison} shows several periods when the proposed model outperforms the alternatives for GNP and the spread forecasts as well. During the Great Inflation period, $VAR^{\sigma,\kappa}$ shows a strong performance across all variables, with cumulative log score gains of 9.8 for GNP, 69.0 for the GNP deflator, and 108.0 for the spread against $VAR^{\sigma,\kappa}_{restricted}$. The Covid-19 period shows particularly notable improvements for GNP, with cumulative log score gains of 1108.2 relative to $VAR^{\sigma,\kappa}_{restricted}$, averaging 221.6 per quarter. The most recent period (since 2021Q3) shows consistent improvements across all variables relative to $VAR^{\sigma,\kappa}_{restricted}$, with statistically significant gains.

A key takeaway is that skewness is a valuable asset for density forecasting across variables and horizons. The results demonstrate that incorporating time-varying skewness with dynamic linkages to macroeconomic fundamentals enhances forecast accuracy, particularly for density forecasts and during periods of economic stress.

\begin{table}[H]
\centering
\caption{US Forecast Performance Comparison Across Multiple Metrics}
\label{tab:combined_forecast_comparison}
\begin{threeparttable}
\scriptsize
\begin{tabular}{l*{8}{c}}
\toprule
\textbf{Variable} & \multicolumn{8}{c}{\textbf{Forecast Horizon}} \\
\cmidrule(lr){2-9}
 & \textbf{H1} & \textbf{H2} & \textbf{H3} & \textbf{H4} & \textbf{H5} & \textbf{H6} & \textbf{H7} & \textbf{H8} \\
\midrule
\multicolumn{9}{c}{\cellcolor{gray!15}\textbf{Panel A: RMSE}} \\
\midrule
\multicolumn{9}{c}{\textbf{vs. $VAR^{\sigma,\kappa}_{restricted}$}} \\
\midrule
\textit{GNP} & 1.009 & \textbf{\textcolor{blue}{0.996}} & \textbf{\textcolor{blue}{0.982}} & \textbf{\textcolor{blue}{0.983}} & \textbf{\textcolor{blue}{0.987}} & \textbf{\textcolor{blue}{0.990}} & \textbf{\textcolor{blue}{0.997}} & 1.008 \\
\textit{GNP def} & 1.033 & 1.045 & 1.057 & 1.064 & 1.072 & 1.071 & 1.072 & 1.075 \\
\textit{SPREAD} & 1.004 & 1.009 & 1.011 & 1.024 & 1.029 & 1.037 & 1.052 & 1.073 \\
\midrule
\multicolumn{9}{c}{\textbf{vs. $VAR^{\sigma}$}} \\
\midrule
\textit{GNP} & 1.011 & \textbf{\textcolor{blue}{0.984}} & \textbf{\textcolor{blue}{0.970}} & \textbf{\textcolor{blue}{0.968}} & \textbf{\textcolor{blue}{0.963}} & \textbf{\textcolor{blue}{0.961}} & \textbf{\textcolor{blue}{0.960}} & \textbf{\textcolor{blue}{0.964}} \\
\textit{GNP def} & \textbf{\textcolor{blue}{0.962}} & \textbf{\textcolor{blue}{0.966}} & \textbf{\textcolor{blue}{0.964}} & \textbf{\textcolor{blue}{0.968}} & \textbf{\textcolor{blue}{0.975}} & \textbf{\textcolor{blue}{0.975}} & \textbf{\textcolor{blue}{0.976}} & \textbf{\textcolor{blue}{0.980}} \\
\textit{SPREAD} & \textbf{\textcolor{blue}{1.000}} & 1.010 & 1.011 & 1.010 & 1.025$^{**}$ & 1.026$^{***}$ & 1.031$^{**}$ & 1.035$^{**}$ \\
\midrule
\multicolumn{9}{c}{\cellcolor{gray!15}\textbf{Panel B: Weighted Both Tails Log Score}} \\
\midrule
\multicolumn{9}{c}{\textbf{vs. $VAR^{\sigma,\kappa}_{restricted}$}} \\
\midrule
\textit{GNP} & -29.9\% & \textbf{\textcolor{blue}{66.7\%}} & \textbf{\textcolor{blue}{14.4\%}} & \textbf{\textcolor{blue}{0.7\%}} & -4.1\% & -0.3\% & -1.1\% & -0.5\% \\
\textit{GNP def} & \textbf{\textcolor{blue}{0.1\%}} & \textbf{\textcolor{blue}{0.2\%}} & \textbf{\textcolor{blue}{0.3\%}} & \textbf{\textcolor{blue}{0.3\%}} & \textbf{\textcolor{blue}{0.3\%}} & \textbf{\textcolor{blue}{0.5\%$^{*}$}} & \textbf{\textcolor{blue}{0.6\%$^{*}$}} & \textbf{\textcolor{blue}{0.8\%$^{*}$}} \\
\textit{SPREAD} & -127.0\% & -1.2\% & -0.9\% & -0.1\% & -3.7\% & -7.1\% & \textbf{\textcolor{blue}{26.8\%}} & -11.1\% \\
\midrule
\multicolumn{9}{c}{\textbf{vs. $VAR^{\sigma}$}} \\
\midrule
\textit{GNP} & \textbf{\textcolor{blue}{30.8\%}} & -161.8\% & -55.5\% & -0.5\% & -3.9\% & -0.5\% & -1.4\% & -1.0\% \\
\textit{GNP def} & \textbf{\textcolor{blue}{1.3\%$^{***}$}} & \textbf{\textcolor{blue}{1.1\%$^{***}$}} & \textbf{\textcolor{blue}{0.9\%$^{***}$}} & \textbf{\textcolor{blue}{0.7\%}} & \textbf{\textcolor{blue}{0.5\%}} & \textbf{\textcolor{blue}{0.4\%}} & \textbf{\textcolor{blue}{0.3\%}} & \textbf{\textcolor{blue}{0.4\%}} \\
\textit{SPREAD} & -136.8\% & \textbf{\textcolor{blue}{6.9\%$^{**}$}} & \textbf{\textcolor{blue}{6.1\%$^{*}$}} & \textbf{\textcolor{blue}{5.0\%}} & -1.4\% & -9.1\% & \textbf{\textcolor{blue}{1.4\%}} & -7.8\% \\
\midrule
\multicolumn{9}{c}{\cellcolor{gray!15}\textbf{Panel C: Weighted Both Tails CRPS}} \\
\midrule
\multicolumn{9}{c}{\textbf{vs. $VAR^{\sigma,\kappa}_{restricted}$}} \\
\midrule
\textit{GNP} & 1.026 & 1.006 & 1.005 & \textbf{\textcolor{blue}{0.996}} & 1.025 & \textbf{\textcolor{blue}{0.983}} & 1.026 & 1.027 \\
\textit{GNP def} & \textbf{\textcolor{blue}{0.966}} & \textbf{\textcolor{blue}{0.942}} & \textbf{\textcolor{blue}{0.944}} & \textbf{\textcolor{blue}{0.941}} & \textbf{\textcolor{blue}{0.945}} & \textbf{\textcolor{blue}{0.899$^{*}$}} & \textbf{\textcolor{blue}{0.889$^{*}$}} & \textbf{\textcolor{blue}{0.865$^{*}$}} \\
\textit{SPREAD} & 1.011 & 1.036 & 1.034 & 1.043 & 1.040 & 1.035 & 1.041 & 1.054 \\
\midrule
\multicolumn{9}{c}{\textbf{vs. $VAR^{\sigma}$}} \\
\midrule
\textit{GNP} & 1.031 & \textbf{\textcolor{blue}{0.999}} & 1.042 & 1.053 & 1.071 & 1.046 & 1.082 & 1.090 \\
\textit{GNP def} & \textbf{\textcolor{blue}{0.819$^{***}$}} & \textbf{\textcolor{blue}{0.826$^{***}$}} & \textbf{\textcolor{blue}{0.852$^{*}$}} & \textbf{\textcolor{blue}{0.880}} & \textbf{\textcolor{blue}{0.914}} & \textbf{\textcolor{blue}{0.894}} & \textbf{\textcolor{blue}{0.907}} & \textbf{\textcolor{blue}{0.892}} \\
\textit{SPREAD} & \textbf{\textcolor{blue}{0.843$^{***}$}} & \textbf{\textcolor{blue}{0.860$^{***}$}} & \textbf{\textcolor{blue}{0.867$^{**}$}} & \textbf{\textcolor{blue}{0.888$^{*}$}} & \textbf{\textcolor{blue}{0.935}} & \textbf{\textcolor{blue}{0.947}} & \textbf{\textcolor{blue}{0.955}} & \textbf{\textcolor{blue}{0.950}} \\
\bottomrule
\end{tabular}
\begin{tablenotes}
\scriptsize
\item \textbf{Notes:} For RMSE and CRPS, values shown are ratios (Proposed model/Competitor); values less than 1 (\textcolor{blue}{\textbf{bold blue}}) indicate the $VAR^{\sigma,\kappa}$ model outperforms the competitor. For Log Score, values shown are percentage differences ($VAR^{\sigma,\kappa}$ - Competitor); positive values (\textcolor{blue}{\textbf{bold blue}}) indicate superior performance. Significance levels from Giacomini-White unconditional test: *** p$<$0.01, ** p$<$0.05, * p$<$0.1. For the conditional version, the reader is referred to the appendix.
\end{tablenotes}
\end{threeparttable}
\end{table}

\begin{table}[H]
\centering
\caption{Forecast Performance by Period: Log Score and CRPS Comparison}
\label{tab:combined_period_comparison}
\begin{threeparttable}
\scriptsize
\begin{tabular}{@{}l*{6}{c}@{}}
\toprule
 & \multicolumn{6}{c}{\textbf{Variables}} \\
\cmidrule(lr){2-7}
\textbf{Period} & \multicolumn{2}{c}{\textbf{GNP}} & \multicolumn{2}{c}{\textbf{GNP Deflator}} & \multicolumn{2}{c}{\textbf{SPREAD}} \\
\cmidrule(lr){2-3} \cmidrule(lr){4-5} \cmidrule(lr){6-7}
 & \textbf{$VAR^{\sigma,\kappa}_{rest}$} & \textbf{$VAR^{\sigma}$} & \textbf{$VAR^{\sigma,\kappa}_{rest}$} & \textbf{$VAR^{\sigma}$} & \textbf{$VAR^{\sigma,\kappa}_{rest}$} & \textbf{$VAR^{\sigma}$} \\
\midrule
\multicolumn{7}{c}{\cellcolor{gray!15}\textbf{Panel A: Cumulative Weighted Both Tails Log Score Differences}} \\
\midrule
Full Sample & \textbf{\textcolor{blue}{1085.0}} & -4577.3 & \textbf{\textcolor{blue}{77.9}}$^{*}$ & \textbf{\textcolor{blue}{130.6}}$^{***}$ & -2936.1 & -3236.5$^{**}$ \\
1975Q2--2023Q1 & (\textbf{\textcolor{blue}{5.7}}) & (-24.2) & (\textbf{\textcolor{blue}{0.4}}) & (\textbf{\textcolor{blue}{0.7}}) & (-15.5) & (-17.2) \\
\addlinespace
Great Inflation & \textbf{\textcolor{blue}{9.8}}$^{*}$ & -28.3$^{**}$ & \textbf{\textcolor{blue}{69.0}}$^{**}$ & \textbf{\textcolor{blue}{108.1}}$^{***}$ & \textbf{\textcolor{blue}{108.0}} & \textbf{\textcolor{blue}{440.2}}$^{**}$ \\
1975Q2--1984Q4 & (\textbf{\textcolor{blue}{0.3}}) & (-0.8) & (\textbf{\textcolor{blue}{2.0}}) & (\textbf{\textcolor{blue}{3.0}}) & (\textbf{\textcolor{blue}{3.0}}) & (\textbf{\textcolor{blue}{12.3}}) \\
\addlinespace
Volcker Disinflation & \textbf{\textcolor{blue}{4.9}} & -16.2$^{***}$ & \textbf{\textcolor{blue}{1.9}}$^{***}$ & \textbf{\textcolor{blue}{10.9}}$^{**}$ & -24.4$^{***}$ & -51.6$^{**}$ \\
1981Q3--1983Q2 & (\textbf{\textcolor{blue}{0.6}}) & (-2.0) & (\textbf{\textcolor{blue}{0.2}}) & (\textbf{\textcolor{blue}{1.4}}) & (-3.1) & (-6.5) \\
\addlinespace
Great Moderation & -6.6$^{***}$ & \textbf{\textcolor{blue}{19.7}}$^{***}$ & -4.0$^{***}$ & \textbf{\textcolor{blue}{45.0}}$^{***}$ & -16.4 & \textbf{\textcolor{blue}{1114.3}}$^{***}$ \\
1985Q1--2007Q4 & (-0.1) & (\textbf{\textcolor{blue}{0.2}}) & (-0.0) & (\textbf{\textcolor{blue}{0.5}}) & (-0.2) & (\textbf{\textcolor{blue}{12.1}}) \\
\addlinespace
Housing Boom & \textbf{\textcolor{blue}{1.7}}$^{***}$ & \textbf{\textcolor{blue}{6.6}}$^{***}$ & -0.4$^{***}$ & \textbf{\textcolor{blue}{11.8}}$^{***}$ & \textbf{\textcolor{blue}{15.9}} & \textbf{\textcolor{blue}{261.7}}$^{***}$ \\
2004Q1--2007Q4 & (\textbf{\textcolor{blue}{0.1}}) & (\textbf{\textcolor{blue}{0.4}}) & (-0.0) & (\textbf{\textcolor{blue}{0.7}}) & (\textbf{\textcolor{blue}{1.0}}) & (\textbf{\textcolor{blue}{16.4}}) \\
\addlinespace
GFC & -29.8 & -29.8 & -0.2$^{*}$ & \textbf{\textcolor{blue}{20.4}}$^{***}$ & \textbf{\textcolor{blue}{51.3}}$^{*}$ & -1277.2 \\
2008Q1--2014Q4 & (-1.1) & (-1.1) & (-0.0) & (\textbf{\textcolor{blue}{0.7}}) & (\textbf{\textcolor{blue}{1.8}}) & (-45.6) \\
\addlinespace
COVID-19 & \textbf{\textcolor{blue}{1108.2}}$^{*}$ & -4530.3 & \textbf{\textcolor{blue}{1.5}} & \textbf{\textcolor{blue}{6.1}}$^{***}$ & -8.2$^{***}$ & \textbf{\textcolor{blue}{4.8}}$^{***}$ \\
2020Q1--2021Q2 & (\textbf{\textcolor{blue}{221.6}}) & (-906.1) & (\textbf{\textcolor{blue}{0.3}}) & (\textbf{\textcolor{blue}{1.2}}) & (-1.6) & (\textbf{\textcolor{blue}{1.0}}) \\
\addlinespace
Since 2021Q3 & \textbf{\textcolor{blue}{7.3}}$^{*}$ & \textbf{\textcolor{blue}{1.8}}$^{*}$ & \textbf{\textcolor{blue}{10.3}}$^{***}$ & -60.7$^{**}$ & \textbf{\textcolor{blue}{1.2}}$^{**}$ & \textbf{\textcolor{blue}{58.4}}$^{***}$ \\
2021Q3--2024Q1 & (\textbf{\textcolor{blue}{1.1}}) & (\textbf{\textcolor{blue}{0.1}}) & (\textbf{\textcolor{blue}{0.4}}) & (-8.4) & (\textbf{\textcolor{blue}{0.1}}) & (\textbf{\textcolor{blue}{8.8}}) \\
\midrule
\multicolumn{7}{c}{\cellcolor{gray!15}\textbf{Panel B: Weighted CRPS Both Tails Ratios}} \\
\midrule
Full Sample & 1.012 & 1.052 & \textbf{\textcolor{blue}{0.924}}$^{*}$ & \textbf{\textcolor{blue}{0.873}}$^{***}$ & 1.037 & \textbf{\textcolor{blue}{0.909}}$^{***}$ \\
1975Q2--2023Q1 &  & &  & &  & \\
\addlinespace
Great Inflation & \textbf{\textcolor{blue}{0.939}}$^{*}$ & 1.090$^{**}$ & \textbf{\textcolor{blue}{0.892}}$^{**}$ & \textbf{\textcolor{blue}{0.812}}$^{***}$ & \textbf{\textcolor{blue}{0.946}} & \textbf{\textcolor{blue}{0.825}}$^{***}$ \\
1975Q2--1984Q4 &  & &  & &  & \\
\addlinespace
Volcker Disinflation & \textbf{\textcolor{blue}{0.962}}$^{**}$ & 1.188$^{***}$ & 1.031$^{**}$ & \textbf{\textcolor{blue}{0.913}}$^{**}$ & 1.096$^{***}$ & 1.053 \\
1981Q3--1983Q2 &  & &  & &  & \\
\addlinespace
Great Moderation & 1.045$^{**}$ & \textbf{\textcolor{blue}{0.914}}$^{***}$ & 1.053$^{*}$ & \textbf{\textcolor{blue}{0.583}}$^{***}$ & 1.039 & \textbf{\textcolor{blue}{0.631}}$^{***}$ \\
1985Q1--2007Q4 &  & &  & &  & \\
\addlinespace
Housing Boom & \textbf{\textcolor{blue}{0.936}}$^{**}$ & \textbf{\textcolor{blue}{0.809}}$^{***}$ & \textbf{\textcolor{blue}{0.992}}$^{***}$ & \textbf{\textcolor{blue}{0.431}}$^{***}$ & \textbf{\textcolor{blue}{0.879}}$^{*}$ & \textbf{\textcolor{blue}{0.398}}$^{***}$ \\
2004Q1--2007Q4 &  & &  & &  & \\
\addlinespace
GFC & 1.060 & 1.066 & 1.055 & \textbf{\textcolor{blue}{0.654}}$^{***}$ & 1.069$^{*}$ & 1.155 \\
2008Q1--2014Q4 &  & &  & &  & \\
\addlinespace
COVID-19 & 1.044$^{**}$ & 1.070$^{**}$ & \textbf{\textcolor{blue}{0.991}} & \textbf{\textcolor{blue}{0.917}}$^{***}$ & 1.102$^{**}$ & \textbf{\textcolor{blue}{0.934}}$^{***}$ \\
2020Q1--2021Q2 &  & &  & &  & \\
\addlinespace
Since 2021Q3 & \textbf{\textcolor{blue}{0.851}}$^{*}$ & \textbf{\textcolor{blue}{0.986}} & \textbf{\textcolor{blue}{0.996}}$^{**}$ & 1.257$^{**}$ & \textbf{\textcolor{blue}{0.942}}$^{**}$ & \textbf{\textcolor{blue}{0.451}}$^{***}$ \\
2021Q3--2024Q1 &  & &  & &  & \\
\bottomrule
\end{tabular}
\begin{tablenotes}
\scriptsize
\item \textbf{Notes:} Panel A shows cumulative weighted both tails log score differences ($VAR^{\sigma,\kappa}$ - Competitor) over specified periods. Values in parentheses show average score differences per quarter within each period. Positive values (\textcolor{blue}{\textbf{bold blue}}) indicate superior performance. Panel B shows weighted CRPS both tails ratios ($VAR^{\sigma,\kappa}$ WCRPS-LR / Competitor WCRPS-LR). Values less than 1 (\textcolor{blue}{\textbf{bold blue}}) indicate superior performance. All values are averaged across forecast horizons 1-8. Significance levels from Giacomini-White unconditional test: *** p$<$0.01, ** p$<$0.05, * p$<$0.1.
\end{tablenotes}
\end{threeparttable}
\end{table}

\subsubsection{UK results}

Table \ref{tab:combined_forecast_comparison_uk} presents the forecast performance for the UK throughout the evaluation period. In contrast to the US results, the UK shows more substantial and consistent improvements in point forecast accuracy. Compared to $VAR^{\sigma,\kappa}_{restricted}$, the proposed model achieves RMSE reductions of 4--13\% for GNP across all horizons, 5--12\% for the GNP deflator (with larger improvements at longer horizons), and modest improvements of 1--2\% for the spread variable. Against $VAR^{\sigma}$, the gains are even more pronounced, with RMSE reductions of 8--15\% for GNP (becoming statistically significant at horizons 6--8), 2--7\% for the GNP deflator, and 1--3\% for the spread across all forecast horizons.

The density forecast metrics reveal that incorporating time-varying skewness with dynamic linkages yields substantial benefits for the UK economy. For weighted log scores emphasising both tails, the GNP deflator exhibits consistent and statistically significant improvements relative to $VAR^{\sigma,\kappa}_{restricted}$, with gains of 1.6--2.5\% across horizons 2--8, significant at the 5--10\% level. Against $VAR^{\sigma}$, the improvements are similar, ranging from 1.2--3.2\%, with statistical significance at multiple horizons. The GNP variable shows more heterogeneous performance, with notable improvements at horizons 3, 5, 6, and particularly at horizon 7 (23.3\% gain relative to $VAR^{\sigma,\kappa}_{restricted}$).

The weighted CRPS results strongly favour the proposed model for the GNP deflator, demonstrating improvements of 12--28\% relative to $VAR^{\sigma,\kappa}_{restricted}$ across horizons, with many results statistically significant. The most substantial gains appear at longer horizons (H6--H8), where CRPS reductions reach 23--28\%. Against $VAR^{\sigma}$, the GNP deflator improvements range from 7--16\%, with statistical significance at horizon 2. For GNP, the proposed model achieves CRPS reductions of 2--11\% relative to $VAR^{\sigma,\kappa}_{restricted}$ at most horizons, and 2--12\% against $VAR^{\sigma}$ at shorter to medium horizons.

The period-by-period analysis in Table \ref{tab:combined_period_comparison_uk} reveals important patterns across different economic episodes. During the stagflation period (1975Q2--1979Q4), the proposed model shows strong performance for the GNP deflator, with cumulative log score gains of 29.9 and 21.1 against the two competitors, both statistically significant. The ERM crisis period demonstrates substantial improvements across variables, with cumulative log score gains of 13.3 for GNP, 8.3 for the GNP deflator, and strong CRPS performance against both competitors.

The Global Financial Crisis period reveals particularly striking results, with cumulative log score gains of 171.2 for GNP relative to $VAR^{\sigma,\kappa}_{restricted}$, averaging 7.1 per quarter. The GNP deflator also shows consistent improvements during this period, with gains of 5.8 and 25.4 against the two competitors. The Covid-19 period mirrors this pattern, with dramatic improvements for GNP (cumulative log score of 287.0, averaging 43.8 per quarter) and consistent gains for the GNP deflator. The most recent period (since 2021Q3) shows statistically significant improvements for both the GNP deflator (cumulative log score of 54.4) and the spread variable (72.3), highlighting the continued relevance of time-varying skewness in the current economic environment.

\begin{table}[ht]
\centering
\caption{UK Forecast Performance Comparison Across Multiple Metrics}
\label{tab:combined_forecast_comparison_uk}
\begin{threeparttable}
\scriptsize
\begin{tabular}{l*{8}{c}}
\toprule
\textbf{Variable} & \multicolumn{8}{c}{\textbf{Forecast Horizon}} \\
\cmidrule(lr){2-9}
 & \textbf{H1} & \textbf{H2} & \textbf{H3} & \textbf{H4} & \textbf{H5} & \textbf{H6} & \textbf{H7} & \textbf{H8} \\
\midrule
\multicolumn{9}{c}{\cellcolor{gray!15}\textbf{Panel A: RMSE}} \\
\midrule
\multicolumn{9}{c}{\textbf{vs. $VAR^{\sigma,\kappa}_{restricted}$}} \\
\midrule
\textit{GNP} & \textbf{\textcolor{blue}{0.936}} & \textbf{\textcolor{blue}{0.927}} & \textbf{\textcolor{blue}{0.951}} & \textbf{\textcolor{blue}{0.944}} & \textbf{\textcolor{blue}{0.951}} & \textbf{\textcolor{blue}{0.949}} & \textbf{\textcolor{blue}{0.960}} & \textbf{\textcolor{blue}{0.969}} \\
\textit{GNP def} & \textbf{\textcolor{blue}{0.954}} & \textbf{\textcolor{blue}{0.944}} & \textbf{\textcolor{blue}{0.947}} & \textbf{\textcolor{blue}{0.925}} & \textbf{\textcolor{blue}{0.917}} & \textbf{\textcolor{blue}{0.903}} & \textbf{\textcolor{blue}{0.885}} & \textbf{\textcolor{blue}{0.876}} \\
\textit{SPREAD} & \textbf{\textcolor{blue}{0.984}} & \textbf{\textcolor{blue}{0.989}} & \textbf{\textcolor{blue}{0.987}} & \textbf{\textcolor{blue}{0.999}} & 1.002 & \textbf{\textcolor{blue}{0.999}} & \textbf{\textcolor{blue}{0.999}} & \textbf{\textcolor{blue}{0.991}} \\
\midrule
\multicolumn{9}{c}{\textbf{vs. $VAR^{\sigma}$}} \\
\midrule
\textit{GNP} & \textbf{\textcolor{blue}{0.919}} & \textbf{\textcolor{blue}{0.885}} & \textbf{\textcolor{blue}{0.877}} & \textbf{\textcolor{blue}{0.864}} & \textbf{\textcolor{blue}{0.865}} & \textbf{\textcolor{blue}{0.854$^{*}$}} & \textbf{\textcolor{blue}{0.857$^{*}$}} & \textbf{\textcolor{blue}{0.857$^{*}$}} \\
\textit{GNP def} & \textbf{\textcolor{blue}{0.981}} & \textbf{\textcolor{blue}{0.964}} & \textbf{\textcolor{blue}{0.975}} & \textbf{\textcolor{blue}{0.961}} & \textbf{\textcolor{blue}{0.954}} & \textbf{\textcolor{blue}{0.939}} & \textbf{\textcolor{blue}{0.929}} & \textbf{\textcolor{blue}{0.933}} \\
\textit{SPREAD} & \textbf{\textcolor{blue}{0.969}} & \textbf{\textcolor{blue}{0.983}} & \textbf{\textcolor{blue}{0.977}} & \textbf{\textcolor{blue}{0.972}} & \textbf{\textcolor{blue}{0.991}} & \textbf{\textcolor{blue}{0.995}} & \textbf{\textcolor{blue}{0.991}} & \textbf{\textcolor{blue}{0.997}} \\
\midrule
\multicolumn{9}{c}{\cellcolor{gray!15}\textbf{Panel B: Weighted Both Tails Log Score}} \\
\midrule
\multicolumn{9}{c}{\textbf{vs. $VAR^{\sigma,\kappa}_{restricted}$}} \\
\midrule
\textit{GNP} & -0.3\% & -6.3\% & \textbf{\textcolor{blue}{1.3\%}} & -1.0\% & \textbf{\textcolor{blue}{3.7\%}} & \textbf{\textcolor{blue}{3.6\%}} & \textbf{\textcolor{blue}{23.3\%}} & -3.7\% \\
\textit{GNP def} & -7.6\% & \textbf{\textcolor{blue}{1.6\%$^{*}$}} & \textbf{\textcolor{blue}{1.9\%$^{*}$}} & \textbf{\textcolor{blue}{2.3\%$^{**}$}} & \textbf{\textcolor{blue}{1.9\%$^{**}$}} & \textbf{\textcolor{blue}{2.0\%$^{**}$}} & \textbf{\textcolor{blue}{2.3\%$^{*}$}} & \textbf{\textcolor{blue}{2.5\%$^{*}$}} \\
\textit{SPREAD} & -3.6\%$^{**}$ & -14.3\% & -0.9\% & -30.2\% & \textbf{\textcolor{blue}{11.9\%}} & -10.5\% & -44.8\% & \textbf{\textcolor{blue}{5.8\%}} \\
\midrule
\multicolumn{9}{c}{\textbf{vs. $VAR^{\sigma}$}} \\
\midrule
\textit{GNP} & -1.3\% & -7.8\% & -2.1\% & -177.0\% & -90.3\% & -35.9\% & -37.9\% & -15.4\% \\
\textit{GNP def} & -3.2\% & \textbf{\textcolor{blue}{3.2\%$^{***}$}} & \textbf{\textcolor{blue}{2.5\%$^{**}$}} & \textbf{\textcolor{blue}{2.3\%$^{**}$}} & \textbf{\textcolor{blue}{2.0\%$^{*}$}} & \textbf{\textcolor{blue}{1.6\%}} & \textbf{\textcolor{blue}{1.5\%$^{**}$}} & \textbf{\textcolor{blue}{1.2\%$^{***}$}} \\
\textit{SPREAD} & -4.8\% & -18.4\% & -6.4\% & -36.5\% & -15.9\% & -106.5\% & -79.5\% & -9.4\% \\
\midrule
\multicolumn{9}{c}{\cellcolor{gray!15}\textbf{Panel C: Weighted CRPS}} \\
\midrule
\multicolumn{9}{c}{\textbf{vs. $VAR^{\sigma,\kappa}_{restricted}$}} \\
\midrule
\textit{GNP} & \textbf{\textcolor{blue}{0.904}} & 1.025 & \textbf{\textcolor{blue}{0.983}} & \textbf{\textcolor{blue}{0.895}} & \textbf{\textcolor{blue}{0.950}} & \textbf{\textcolor{blue}{0.959}} & \textbf{\textcolor{blue}{0.998}} & 1.002 \\
\textit{GNP def} & \textbf{\textcolor{blue}{0.934}} & \textbf{\textcolor{blue}{0.875$^{*}$}} & \textbf{\textcolor{blue}{0.884}} & \textbf{\textcolor{blue}{0.799$^{**}$}} & \textbf{\textcolor{blue}{0.806$^{**}$}} & \textbf{\textcolor{blue}{0.771$^{*}$}} & \textbf{\textcolor{blue}{0.725$^{*}$}} & \textbf{\textcolor{blue}{0.717}} \\
\textit{SPREAD} & 1.008 & 1.011 & 1.002 & 1.020 & 1.004 & \textbf{\textcolor{blue}{0.999}} & 1.004 & \textbf{\textcolor{blue}{0.992}} \\
\midrule
\multicolumn{9}{c}{\textbf{vs. $VAR^{\sigma}$}} \\
\midrule
\textit{GNP} & \textbf{\textcolor{blue}{0.934}} & \textbf{\textcolor{blue}{0.982}} & \textbf{\textcolor{blue}{0.932}} & \textbf{\textcolor{blue}{0.942}} & \textbf{\textcolor{blue}{0.966}} & \textbf{\textcolor{blue}{0.991}} & 1.056 & 1.034 \\
\textit{GNP def} & \textbf{\textcolor{blue}{0.924}} & \textbf{\textcolor{blue}{0.875$^{**}$}} & \textbf{\textcolor{blue}{0.909}} & \textbf{\textcolor{blue}{0.848}} & \textbf{\textcolor{blue}{0.851}} & \textbf{\textcolor{blue}{0.848}} & \textbf{\textcolor{blue}{0.867}} & \textbf{\textcolor{blue}{0.926}} \\
\textit{SPREAD} & 1.012 & 1.042 & 1.046 & 1.047 & 1.058 & 1.048 & 1.047 & 1.041 \\
\bottomrule
\end{tabular}
\begin{tablenotes}
\scriptsize
\item \textbf{Notes:} For RMSE and CRPS, values shown are ratios (Proposed model/Competitor); values less than 1 (\textcolor{blue}{\textbf{bold blue}}) indicate the $VAR^{\sigma,\kappa}$ model outperforms the competitor. For Log Score, values shown are percentage differences ($VAR^{\sigma,\kappa}$ - Competitor); positive values (\textcolor{blue}{\textbf{bold blue}}) indicate superior performance. Significance levels from Giacomini-White unconditional test: *** p$<$0.01, ** p$<$0.05, * p$<$0.1. For the conditional version, the reader is referred to the appendix.
\end{tablenotes}
\end{threeparttable}
\end{table}

\begin{table}[H]
\centering
\caption{Forecast Performance by Period: Log Score and CRPS Comparison}
\label{tab:combined_period_comparison_uk}
\begin{threeparttable}
\scriptsize
\begin{tabular}{@{}l*{6}{c}@{}}
\toprule
 & \multicolumn{6}{c}{\textbf{Variables}} \\
\cmidrule(lr){2-7}
\textbf{Period} & \multicolumn{2}{c}{\textbf{GNP}} & \multicolumn{2}{c}{\textbf{GNP Deflator}} & \multicolumn{2}{c}{\textbf{SPREAD}} \\
\cmidrule(lr){2-3} \cmidrule(lr){4-5} \cmidrule(lr){6-7}
 & \textbf{$VAR^{\sigma,\kappa}_{rest}$} & \textbf{$VAR^{\sigma}$} & \textbf{$VAR^{\sigma,\kappa}_{rest}$} & \textbf{$VAR^{\sigma}$} & \textbf{$VAR^{\sigma,\kappa}_{rest}$} & \textbf{$VAR^{\sigma}$} \\
\midrule
\multicolumn{7}{c}{\cellcolor{gray!15}\textbf{Panel A: Cumulative Weighted Both Tails Log Score Differences}} \\
\midrule
Full Sample & \textbf{\textcolor{blue}{490.0}}$^{*}$ & -4781.9 & \textbf{\textcolor{blue}{161.9}}$^{**}$ & \textbf{\textcolor{blue}{262.4}}$^{***}$ & -2092.0$^{**}$ & -6453.8 \\
1975Q2--2023Q1 & (\textbf{\textcolor{blue}{2.6}}) & (-25.5) & (\textbf{\textcolor{blue}{0.9}}) & (\textbf{\textcolor{blue}{1.4}}) & (-11.1) & (-34.5) \\
\addlinespace
Stagflation & -2.4$^{*}$ & \textbf{\textcolor{blue}{7.5}}$^{**}$ & \textbf{\textcolor{blue}{29.9}}$^{**}$ & \textbf{\textcolor{blue}{21.1}}$^{**}$ & -51.5$^{*}$ & -169.4$^{**}$ \\
1975Q2--1979Q4 & (-0.0) & (\textbf{\textcolor{blue}{0.5}}) & (\textbf{\textcolor{blue}{4.9}}) & (\textbf{\textcolor{blue}{2.7}}) & (-3.3) & (-10.7) \\
\addlinespace
Thatcher & \textbf{\textcolor{blue}{2.7}} & \textbf{\textcolor{blue}{22.4}} & -101.0 & -144.6 & -51.9$^{*}$ & -23.9$^{**}$ \\
1979Q1--1989Q4 & (\textbf{\textcolor{blue}{0.1}}) & (\textbf{\textcolor{blue}{0.5}}) & (-2.3) & (-3.3) & (-1.2) & (-0.5) \\
\addlinespace
ERM crisis & \textbf{\textcolor{blue}{13.3}}$^{***}$ & \textbf{\textcolor{blue}{25.2}}$^{**}$ & \textbf{\textcolor{blue}{8.3}}$^{**}$ & \textbf{\textcolor{blue}{14.3}}$^{***}$ & -7.0$^{***}$ & \textbf{\textcolor{blue}{99.5}}$^{***}$ \\
1990Q1--1992Q4 & (\textbf{\textcolor{blue}{1.1}}) & (\textbf{\textcolor{blue}{2.1}}) & (\textbf{\textcolor{blue}{0.7}}) & (\textbf{\textcolor{blue}{1.2}}) & (-0.6) & (\textbf{\textcolor{blue}{8.3}}) \\
\addlinespace
Great Moderation & -2.9$^{***}$ & \textbf{\textcolor{blue}{19.8}}$^{***}$ & -5.0$^{***}$ & \textbf{\textcolor{blue}{110.5}}$^{***}$ & -253.3$^{**}$ & \textbf{\textcolor{blue}{35.4}} \\
1993Q1--2007Q4 & (-0.0) & (\textbf{\textcolor{blue}{0.3}}) & (-0.1) & (\textbf{\textcolor{blue}{1.8}}) & (-4.2) & (\textbf{\textcolor{blue}{0.6}}) \\
\addlinespace
GFC & \textbf{\textcolor{blue}{171.2}}$^{*}$ & -216.8 & \textbf{\textcolor{blue}{5.8}}$^{***}$ & \textbf{\textcolor{blue}{25.4}}$^{***}$ & -1765.2 & -6287.8$^{*}$ \\
2007Q1--2012Q4 & (\textbf{\textcolor{blue}{7.1}}) & (-9.0) & (\textbf{\textcolor{blue}{0.2}}) & (\textbf{\textcolor{blue}{1.1}}) & (-73.5) & (-262.0) \\
\addlinespace
Brexit & -6.1$^{***}$ & \textbf{\textcolor{blue}{11.6}}$^{***}$ & -3.5 & \textbf{\textcolor{blue}{75.1}}$^{***}$ & \textbf{\textcolor{blue}{25.3}}$^{***}$ & \textbf{\textcolor{blue}{249.3}}$^{***}$ \\
2013Q1--2019Q4 & (-0.2) & (\textbf{\textcolor{blue}{0.4}}) & (-0.1) & (\textbf{\textcolor{blue}{2.7}}) & (\textbf{\textcolor{blue}{0.9}}) & (\textbf{\textcolor{blue}{8.9}}) \\
\addlinespace
COVID-19 & \textbf{\textcolor{blue}{287.0}}$^{***}$ & -4710.2$^{**}$ & \textbf{\textcolor{blue}{5.1}}$^{***}$ & \textbf{\textcolor{blue}{19.6}}$^{***}$ & -69.0$^{*}$ & -208.9 \\
2020Q1--2021Q2 & (\textbf{\textcolor{blue}{43.8}}) & (-966.0) & (\textbf{\textcolor{blue}{1.0}}) & (\textbf{\textcolor{blue}{3.9}}) & (-13.8) & (-41.8) \\
\addlinespace
Since 2021Q3 & \textbf{\textcolor{blue}{7.3}} & \textbf{\textcolor{blue}{3.1}}$^{*}$ & \textbf{\textcolor{blue}{54.4}}$^{***}$ & -9.8$^{*}$ & \textbf{\textcolor{blue}{72.3}}$^{***}$ & -322.8$^{**}$ \\
2021Q3--2024Q1 & (\textbf{\textcolor{blue}{0.5}}) & (\textbf{\textcolor{blue}{0.3}}) & (\textbf{\textcolor{blue}{7.1}}) & (-1.2) & (\textbf{\textcolor{blue}{8.9}}) & (-41.8) \\
\midrule
\multicolumn{7}{c}{\cellcolor{gray!15}\textbf{Panel B: Weighted CRPS Both Tails Ratios}} \\
\midrule
Full Sample & \textbf{\textcolor{blue}{0.964}} & \textbf{\textcolor{blue}{0.979}} & \textbf{\textcolor{blue}{0.814}}$^{**}$ & \textbf{\textcolor{blue}{0.880}}$^{**}$ & 1.007 & 1.039 \\
1975Q2--2023Q1 &  & &  & &  & \\
\addlinespace
Stagflation & 1.017 & \textbf{\textcolor{blue}{0.897}}$^{**}$ & \textbf{\textcolor{blue}{0.752}}$^{**}$ & \textbf{\textcolor{blue}{0.823}}$^{*}$ & 1.011$^{**}$ & 1.045$^{***}$ \\
1975Q2--1979Q4 &  & &  & &  & \\
\addlinespace
Thatcher & \textbf{\textcolor{blue}{0.992}} & \textbf{\textcolor{blue}{0.959}} & \textbf{\textcolor{blue}{0.863}} & \textbf{\textcolor{blue}{0.976}} & \textbf{\textcolor{blue}{1.000}} & \textbf{\textcolor{blue}{0.971}}$^{**}$ \\
1979Q1--1989Q4 &  & &  & &  & \\
\addlinespace
ERM crisis & \textbf{\textcolor{blue}{0.931}}$^{***}$ & \textbf{\textcolor{blue}{0.781}}$^{***}$ & \textbf{\textcolor{blue}{0.920}}$^{**}$ & \textbf{\textcolor{blue}{0.922}}$^{***}$ & 1.021$^{***}$ & \textbf{\textcolor{blue}{0.821}}$^{***}$ \\
1990Q1--1992Q4 &  & &  & &  & \\
\addlinespace
Great Moderation & 1.001 & \textbf{\textcolor{blue}{0.902}}$^{***}$ & 1.044$^{**}$ & \textbf{\textcolor{blue}{0.699}}$^{***}$ & 1.075$^{***}$ & \textbf{\textcolor{blue}{0.991}} \\
1993Q1--2007Q4 &  & &  & &  & \\
\addlinespace
GFC & 1.066$^{**}$ & 1.091$^{*}$ & \textbf{\textcolor{blue}{0.902}}$^{***}$ & \textbf{\textcolor{blue}{0.783}}$^{***}$ & \textbf{\textcolor{blue}{0.997}} & 1.168$^{*}$ \\
2007Q1--2012Q4 &  & &  & &  & \\
\addlinespace
Brexit & 1.129$^{***}$ & \textbf{\textcolor{blue}{0.856}}$^{***}$ & 1.048 & \textbf{\textcolor{blue}{0.732}}$^{***}$ & \textbf{\textcolor{blue}{0.971}}$^{**}$ & \textbf{\textcolor{blue}{0.819}}$^{***}$ \\
2013Q1--2019Q4 &  & &  & &  & \\
\addlinespace
COVID-19 & \textbf{\textcolor{blue}{0.962}}$^{***}$ & 1.042$^{***}$ & \textbf{\textcolor{blue}{0.968}}$^{***}$ & \textbf{\textcolor{blue}{0.716}}$^{***}$ & \textbf{\textcolor{blue}{0.988}}$^{**}$ & 1.174 \\
2020Q1--2021Q2 &  & &  & &  & \\
\addlinespace
Since 2021Q3 & 1.007 & \textbf{\textcolor{blue}{0.968}} & \textbf{\textcolor{blue}{0.744}}$^{***}$ & 1.081$^{**}$ & \textbf{\textcolor{blue}{0.958}}$^{***}$ & 1.209$^{***}$ \\
2021Q3--2024Q1 &  & &  & &  & \\
\bottomrule
\end{tabular}
\begin{tablenotes}
\scriptsize
\item \textbf{Notes:} Panel A shows cumulative weighted both tails log score differences ($VAR^{\sigma,\kappa}$ - Competitor) over specified periods. Values in parentheses show average score differences per quarter within each period. Positive values (\textcolor{blue}{\textbf{bold blue}}) indicate superior performance. Panel B shows weighted CRPS both tails ratios ($VAR^{\sigma,\kappa}$ WCRPS-LR / Competitor WCRPS-LR). Values less than 1 (\textcolor{blue}{\textbf{bold blue}}) indicate superior performance. All values are averaged across forecast horizons 1-8. Significance levels from Giacomini-White unconditional test: *** p$<$0.01, ** p$<$0.05, * p$<$0.1.
\end{tablenotes}
\end{threeparttable}
\end{table}

Overall, UK results provide strong evidence that incorporating time-varying skewness with dynamic linkages to macroeconomic fundamentals substantially improves forecast accuracy. The improvements are more consistent and pronounced than in the US case, particularly for point forecasts and density forecasts of the GNP deflator. Once more, the model performs especially well during periods of economic stress and structural change, such as the ERM crisis, GFC, and Covid-19 pandemic, underscoring the value of capturing asymmetries in the predictive distribution during turbulent times.

\subsection{Measures of Risk}
\begin{figure}[H]
    \centering
    \includegraphics[width=6.5in,height=5.5in,scale = 0.4, clip=true,trim = 0cm 0cm 0cm 0cm]{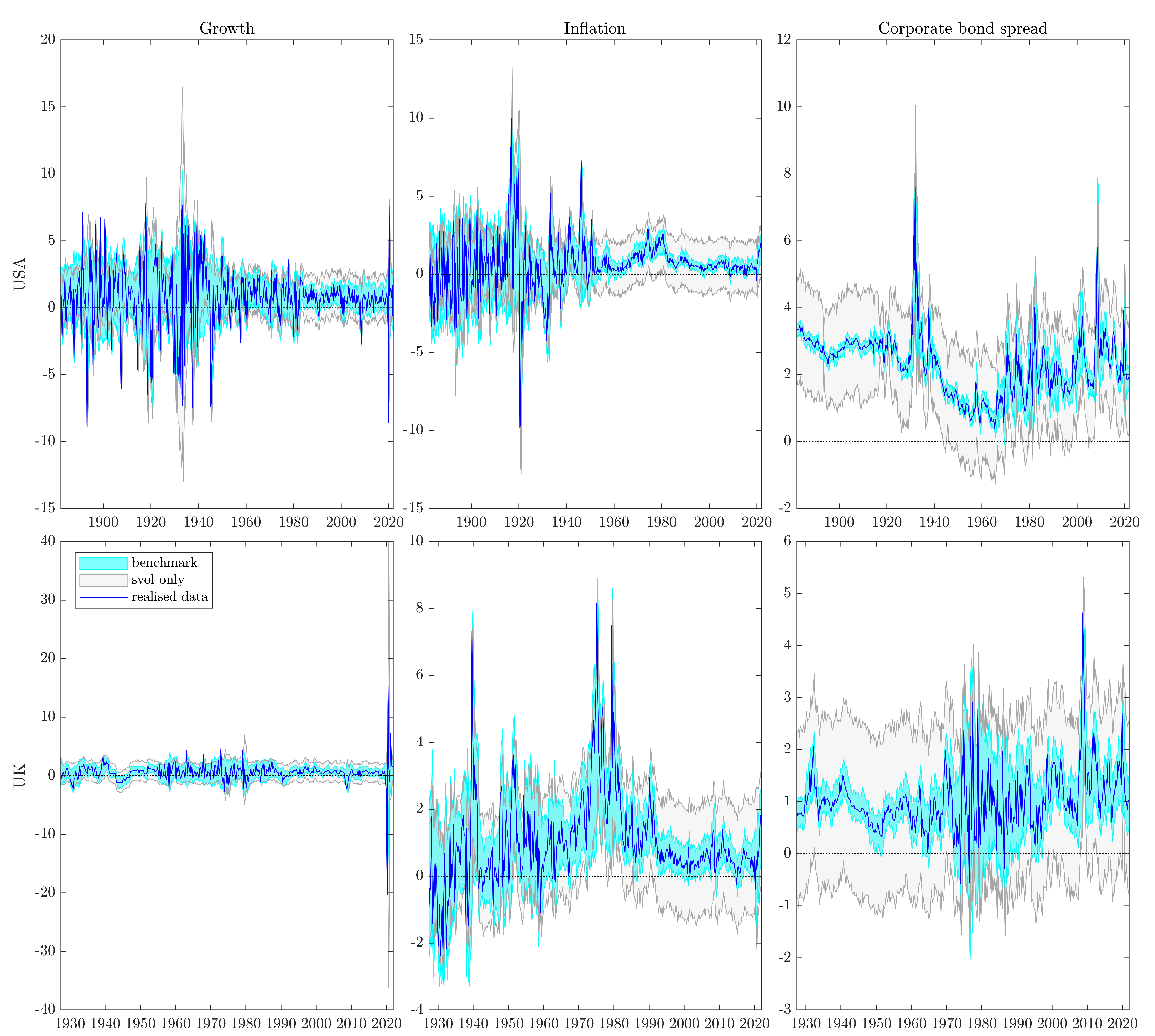}
    
     \caption{Percentiles of the one step ahead forecast distribution} 
     \justify
     {\scriptsize Note: The shaded area represents the region between the $5^{th}$ and $95^{th}$ percentile of forecast distribution.}
\label{figure_risk}
\end{figure}

In this section, we use the proposed model to construct measures of tail risk for the endogenous variables. Following \cite{CaldaraMumtazZhong2024}, we construct the forecast distribution for the endogenous variables at each point in time using the full-sample estimates. The $5^{th}$ and $95^{th}$ percentiles of the forecast distribution capture the variation in the left and the right tail over time. Figure \ref{figure_risk} displays these percentiles estimated using the proposed model and compares them to those obtained from the stochastic volatility in mean model without skewness. 

A number of key differences in the estimated risk measures are apparent. The forecast distribution from the stochastic volatility in mean model is, in general, more dispersed than the estimate from the proposed model. As a consequence, the stochastic volatility model points to higher uncertainty rather than highlighting tail risk. This is especially apparent when considering the forecast distribution in the post-second world war period for both countries. The proposed model consistently points to higher right tail risk during the 1970s, the early 1980s/1990s and the post-Covid period. In contrast, the stochastic volatility model also assigns substantially more mass to downside risk during these periods. For example, in March 2021, the proposed model assigns a probability of 0.7 to annualized inflation in the US exceeding 5\% next quarter. In contrast, this probability is estimated to be 0.5 using the stochastic volatility model.

\section{Conclusion} \label{conc}
This paper proposes a BVAR that incorporates both stochastic volatility-in-mean and time-varying skewness, allowing these features to directly affect macroeconomic variables. Our results, based on US and UK data, demonstrate that time-varying skewness plays a crucial role in shaping the dynamics of output, inflation, and financial spreads. Impulse responses show that skewness shocks significantly impact output, inflation and spreads in both the US and the UK, often with larger effects than volatility shocks.

The forecasting evaluation further underscores the value of incorporating time-varying skewness. In the US, gains are most evident for the GNP deflator and, during periods of economic stress, for the GNP. For the UK, the benefits are even more pronounced, with consistent improvements across variables and horizons. Finally, tail risk measures show that the proposed framework provides a sharper characterisation of downside and upside risks, offering a useful tool for policy analysis. Overall, this paper contributes to the growing literature emphasising the importance of non-Gaussian features in macroeconomic models and shows that time-varying skewness is a key dimension for understanding and forecasting macro-financial dynamics.

\newpage

\bibliographystyle{ecta}
\bibliography{references/references, references/references1,references/references1zz,references/LR_biblio,references/LR_biblio1,references/LR_biblio2,references/LR_biblio25,references/LR_biblio3,references/LR_biblio4,references/LR_bibliox,references/referenceszz,references/biblio}

\clearpage
\appendix
\counterwithin{figure}{section}
\counterwithin{table}{section}
\counterwithin{equation}{section}
\section{Tables and Figures}
\begin{figure}[ht]
    \centering
    \includegraphics[width=5.5in,height=5.5in,scale = 0.38, clip=true,trim = 0cm 0cm 0cm 0cm]{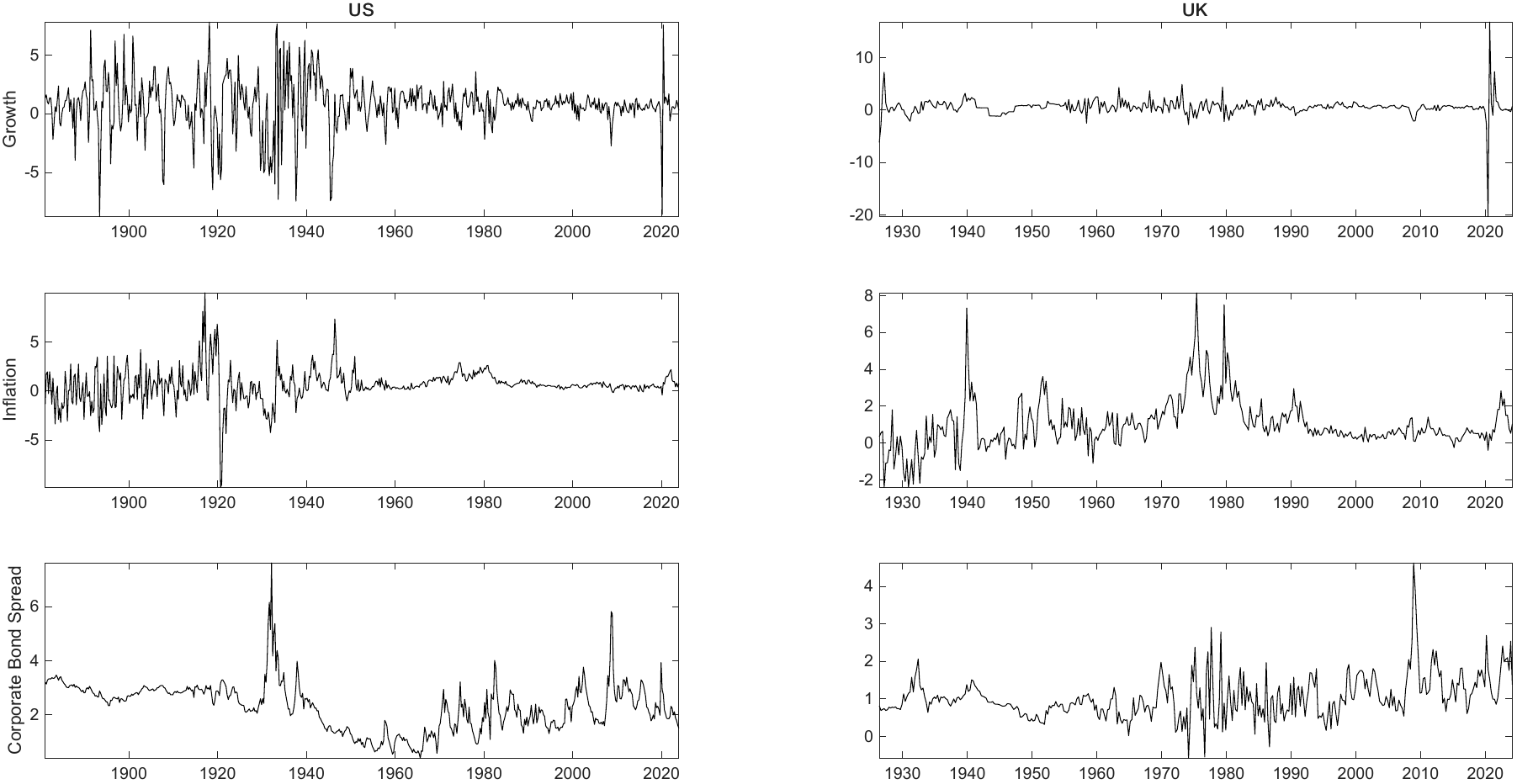}
    
     \caption{Data series used for each country}
\label{figure_data}
   \end{figure}

\newpage
The data sources for the US are summarised in Table \ref{Table_1}.

\begin{table}[ht]
\centering
\caption{Data sources}
\label{Table_1}
\begin{tabular}{c c c c}
\hline
Variables & NBER & FRED & GFD \\
 & 1875-1946 & 1947-2019 & 1875-2019 \\
\hline
Real GNP & RGNP72 & GNPC1 &   \\
GNP deflator & GNPD72 & GNPDEF &   \\
CB rate & CORPYIELD & BAA &  \\
10 year rate &  &  & IGUSA10D \\
\hline
\end{tabular}
\end{table}

\clearpage

\section{\label{appendixA} Appendix: Model Estimation} 
\subsection{Description of the Model}
We consider the following state-space model:

\begin{eqnarray}
\beta_{t} &=&\alpha +\theta \beta_{t-1}+%
\sum_{j=1}^{Q}d_{j}Y_{t-j}+\eta _{t}  \label{eq1-app} \\
\tau_{t}&=&|\Theta_{t}|, \Theta_{t} \sim N(0,1) \label{eq3_app} \\
Y_{t} &=&c+\sum_{j=1}^{P}\beta _{j}Y_{t-j}+\sum_{l=1}^{L}b_{l}\tilde{h}_{t-l}+\sum_{l=1}^{L}a_{L}\tilde{d}_{t-l}+V_{t}  \label{eq2-app} \\
V_{t}&=&A^{-1}E_{t} \\
E_{T}&=&\tilde{d}_{t}\odot \tau_{T}+e_{t} \\
\varepsilon _{t}&=&\left( 
\begin{array}{c}
\eta _{t} \\ 
e_{t}%
\end{array}%
\right) \sim N \left(0,  \begin{pmatrix}
Q & 0 \\
0 & H_{t}
\end{pmatrix}  \right) 
\end{eqnarray}

where $Y_{t}$ denotes the $N$ endogenous variables. The composite orthogonalised error term in the observation equation \ref{eq2-app} $E_{t}=\tilde{d}_{t}\odot \tau_{T}+e_{t}$ follows a skewed Normal distribution as described in Proposition 1 of \cite{Arellano-Valle}. \newline The state variables are denoted by $\underbrace{\beta_{t}}_{K\times 1}=\begin{pmatrix}
 \tilde{h}_{t} \\
 \tilde{d_{t}}
\end{pmatrix}$ and $\underbrace{\tau_{t}}_{N\times 1}$ .
\newline The stochastic volatilities are denoted by 
$\tilde{h}_{t}=[h_{1t},h_{2t},\dots,h_{N,t}]$ and $H_{t}=diag\left( \exp \left( 
\tilde{h}_{t}\right) \right) $. Time-varying skewness is denoted by $\tilde{d_{t}}=[d_{1t},d_{2t},\dots,d_{N,t}]$. These state-variables evolve
according to equation \ref{eq1-app}. We assume that the autoregressive coefficient matrix $\theta$ is block diagonal with dynamic relationships allowed amongst $N$ volatilities and skewness parameters, respectively. The shocks to these equation have a full variance-covariance matrix 
$Q$. As evident, $\tilde{h}_{t}$ and $\tilde{d_{t}}$ are allowed to
have an impact on the endogenous variables in equation \ref{eq3_app}. 

\subsection{Priors}

\paragraph{VAR coefficients}
Let $\Gamma =vec\left( [\beta _{j};b_{l};a_{l};c]\right) $. Following \cite{Banbura-Giannone-Reichlin-10Paper}, we employ a Normal prior. The priors are
implemented by the dummy observations $y_{D}$ and $x_{D}$ that are defined
as:

\begin{equation}
y_{D}=\left[ 
\begin{array}{c}
\frac{diag\left( \gamma _{1}s_{1}...\gamma _{n}s_{n}\right) }{\tau } \\ 
0_{N\times \left( P-1\right) \times N} \\ 
.............. \\ 
0_{EX\times N}%
\end{array}%
\right] ,\ \ \ \ \ \ x_{D}=\left[ 
\begin{array}{c}
\frac{J_{P}\otimes diag\left( s_{1}...s_{n}\right) }{\tau }\text{ }%
0_{NP\times EX} \\ 
\text{ }0_{N\times (NP)+EX}\ \ \ \ \ \ \  \\ 
.............. \\ 
0_{EX\times NP}\text{ \ \ \ \ \ \ \ \ }I_{EX}\times 1/c%
\end{array}%
\right]
\end{equation}%
where $\gamma _{1}$ to $\gamma _{n}$ denote the prior mean for the
parameters on the first lag obtained by estimating individual AR(1)
regressions, $\tau $ measures the tightness of the prior on the VAR
coefficients, and $c$ is the tightness of the prior on the exogenous and
pre-determined regressors. $EX$ denotes the number of exogenous and
pre-determined regressors in each equation. N denotes the total number of
endogenous variables and P is the lag length. We set $\tau =0.1$. We use a
different value of $c$ for the coefficients on the lagged volatilities and
the remaining pre-determined regressors. For the coefficients on the lagged
volatilities $c$ is set equal to $0.1.$ A flat prior is used for the
intercept terms and the corresponding tightness is set equal to $c=1000$.
Note that these dummies do not directly implement a prior belief on the VAR
error covariance matrix which is time-varying in our setting.

The priors for the coefficients are thus: $N\left( \Gamma _{0},P_{0}\right) $
where $\Gamma _{0}=\left( x_{D}^{\prime }x_{D}\right) ^{-1}\left(
x_{D}^{\prime }y_{D}\right) $ and $P_{0}=S\otimes \left( x_{D}^{\prime
}x_{D}\right) ^{-1}$ where $S$ is a diagonal matrix with an estimate of the
variance of $Y_{t}$ (obtained using a training sample ) on
the main diagonal.

\paragraph{Elements of $\beta_{t},\tau_{t}$}
The state vector used in the state-space representation below is given by $\mathcal{B}_{t}=\begin{pmatrix}
 \Theta_{t} \\
 \beta_{t} \\
 \beta_{t-1}\\
 \vdots \\
 \beta_{t-K}
\end{pmatrix}$. The initial state is defined as $\mathcal{B}_{0} \sim N(\mathcal{B}_{0|0},\mathcal{P}_{0|0})$. The prior mean $\mathcal{B}_{0|0}$ is obtained by first estimating a VAR model with stochastic volatility and time-varying skewness. This simpler model is defined as:
\begin{equation}
    Y_{t} =c+\sum_{j=1}^{P}\beta _{j}Y_{t-j}+A^{-1}(\tilde{d}_{t}\odot \tau_{T}+u_{t})
\end{equation}
where $\tau_{t}=|\Theta_{t}|, \Theta_{t} \sim N(0,1)$, $u_{t}$ is Normal with $VAR(u_t)=H_{t}$ where $A$ is lower triangular and $H$ is diagonal with the stochastic volatility on the main diagonal. A Particle Gibbs sampler is used to draw from the posterior of the VAR coefficients, elements of $A$ and the states $\tilde{d}_{t},\tau_{t}, H_{t}$. The estimate states at time $1$ from these initial models are used to set the prior mean. The prior variance $\mathcal{P}_{0|0}$ is set to an identity matrix.

\paragraph{Elements of $A $}

Using the lags of these initial estimates of the state-variables, we estimate the VAR model in equation \ref{eq2-app} by OLS and obtain an
initial estimate of the residuals of the VAR, denoted $v_t$. The prior for the
off-diagonal elements $A$ is \ $A_{0}\sim N\left( \hat{a}^{ols},V\left( 
\hat{a}^{ols}\right) \right) $ where $\hat{a}^{ols}$ are the off-diagonal
elements of the inverse of the Cholesky decomposition of $var\left(
v_{t}\right) $ where each row of the decomposition is divided by the
corresponding element on the diagonal. $V\left( \hat{a}^{ols}\right) $ is
assumed to be diagonal with the elements set equal to 1.

\paragraph{Parameters of the transition equation \protect\ref{eq1-app}}
The prior for VAR coefficients $\tilde{\Gamma}=vec\left( [\alpha ;\theta
;d_{j}]\right) $ is set as above for the VAR coefficients $\Gamma $. That is the prior is normal $N\left( \tilde{\Gamma} _{0},p_{0}\right) $. We use the prior to impose the block-diagonality of $\theta$, by setting the relevant elements of $p_{0}$ to be very small and the corresponding elements of $\tilde{\Gamma} _{0}$ to $0$. We assume an inverse Wishart
prior for $Q:IW(v_{0},T_{0})$, where $T_{0}=K+1$ and $v_{0}$ is set to the diagonal of the covariance of the shocks to the state transition equation obtained using the initial estimation described above.

\subsection{Simulating the posterior distributions}

\paragraph{Coefficients of the transition equation}
Conditional on $\beta,S$ and $Q$, the model can be written as a VAR
\begin{eqnarray}
\beta_{t} &=&\alpha +\theta \beta_{t-1}+\sum_{j=1}^{Q}d_{j}Y_{t-j}+\eta _{t} \label{VARtrans-app} \\
var\left( \eta _{t}\right) &=&Q  \notag
\end{eqnarray}%

Given a normal prior for the parameters $\tilde{\Gamma}=vec\left( [\theta
;d_{j};\alpha]\right) $, the posterior is normal with mean and variance given, respectively,  by:
\begin{equation*}
M=\left( p_{0}^{-1}+Q^{ -1} \otimes x^{\prime }x\right) ^{-1}\left(
p_{0}^{-1}\tilde{\Gamma}_{0}+Q^{-1} \otimes x^{\prime }x\hat{b}\right)
\end{equation*}

\begin{equation*}
V=\left( p_{0}^{-1}+Q^{ -1} \otimes x^{\prime }x\right) ^{-1}
\end{equation*}
where $x$ collects all the regressors and $\hat{b}$ denotes the vectorised OLS estimates of the coefficients of the VAR in equation \ref{VARtrans-app}. 

\paragraph{Elements of $Q$}
The conditional posterior of $Q$ is inverse Wishart with scale matrix $\eta'\eta+v_{0}$ and degrees of freedom $T+T_{0}$

\paragraph{Element of $A$}

Given a draw for $\Gamma ,$ $\tilde{d}_{t}$ and $\tilde{h}_{t}$ the
VAR model can be written as $A\left( V_{t}\right) =E_{t}$. This is a system of linear equations with skew normal residuals. The conditional distributions for a linear
regression apply to each equation of this system after a simple
transformation to make the errors Gaussian and homoscedastic. The $kth$ equation of this
system is given as $V_{t}^{k}=-V_{t}^{-k}\alpha +E_{t}^{k}$ where the
superscript $k$ denotes the $kth$ column of the residual matrix while $-k$
denotes columns $1$ to $k-1$. Note that $E_t^{k}=\tilde{d}_{t}^{k} \odot \tau_{t}^{k}+e_{t}^{k}$ where $e_{t}\sim N(0,h_{t}^{k})$. Using this, the model can be re-written as $V_{t}^{k}-\tilde{d}_{t}^{k} \odot \tau_{t}^{k}=-V_{t}^{-k}\alpha +e_{t}^{k}$. A GLS
transformation involves dividing both sides of the equation by $\sqrt{\exp
\left( h_{t}^{k}\right) }$ to produce $%
V_{t}^{k\ast }=-V_{t}^{-k\ast }\alpha +\tilde{e}_{t}^{\ast }$ where * denotes the
transformed variables and $var\left( \tilde{e}_{t}^{\ast }\right) =1.$ The
conditional posterior for $\alpha $ is normal with mean and variance given
by $M^{\ast }$ and $V^{\ast }:$

\begin{eqnarray*}
M^{\ast } &=&\left( V\left( \hat{a}^{ols}\right) ^{-1}+V_{t}^{-k\ast \prime
}V_{t}^{-k\ast }\right) ^{-1}\left( V\left( \hat{a}^{ols}\right) ^{-1}\hat{a%
}^{ols}+V_{t}^{-k\ast \prime }V_{t}^{k\ast }\right) \\
V^{\ast } &=&\left( V\left( \hat{a}^{ols}\right) ^{-1}+V_{t}^{-k\ast \prime
}V_{t}^{-k\ast }\right) ^{-1}
\end{eqnarray*}

\paragraph{Coefficients of the VAR in equation \ref{eq2-app} }
We write equation \ref{eq2-app} conditional on the skewness parameters:
\begin{eqnarray}
Y_{t}^\ast &=&c+\sum_{j=1}^{P}\beta
_{j}Y_{t-j}+\sum_{k=1}^{K}b_{k}\beta_{t-k}+\tilde{e}_{t}  \notag \\
var\left( \tilde{e}_{t}\right)  &=&\Sigma _{t}=A^{-1}H_{t}A^{-1 \prime}  \notag
\end{eqnarray}
where: $Y_{t}^\ast=Y_{t}-A^{-1}(\tilde{d}_{t}\odot \tau_{T})$.

This is a VAR model with a known form of heteroscedasticity and conditional posterior is normal once the heteroscedasticity is accounted for. We use the Kalman filter to calculate the mean and variance of the conditional posterior and then draw the VAR coefficients from the normal distribution.

\paragraph{Elements of $\beta_{t}$}

Conditional on the parameters, the model has a multivariate non-linear state-space
representation. It is convenient to express the state-space as:%
\begin{eqnarray}
\digamma _{t} &=&C+\Psi \digamma _{t-1}+N_{t} \\
Y_{t} &=&c+\sum_{j=1}^{P}\beta
_{j}Y_{t-j}+\sum_{k=1}^{K}b_{k}\beta_{t-k}+A^{-1}(\tilde{d}_{t}\odot \tau_{t})+\tilde{e}_{t}  \notag \\
var\left( \tilde{e}_{t}\right)  &=&\Sigma_t= A^{-1} H_{t} A^{-1\prime}  \notag
\end{eqnarray}

where: 
\begin{equation}
\digamma _{t}=\left( 
\begin{array}{c}
\Theta_{t} \\
\beta_{t} \\
\vdots \\ 
\beta_{t-k}%
\end{array}%
\right) 
\end{equation}

\begin{equation}
C=\left( 
\begin{array}{c}
0 \\ 
\alpha +\sum_{j=1}^{Q}d_{j}Y_{t-j} \\ 
0 \\ 
\vdots\\
0
\end{array}%
\right) 
\end{equation}

\begin{equation}
\Psi_{1} =\left( 
\begin{array}{cc}
\theta & 0_{K \times (NS-K)}\\
I_{(NS-K)\times (NS-K)} & 0_{(NS-K) \times K} 
\end{array}%
\right)
\end{equation}%

\begin{equation}
\Psi =BLKDIAG\left(0_{N\times N},
\Psi_{1} \right)
\end{equation}

\begin{equation}
N_{1t}=\left( 
\begin{array}{c}
\eta_{t}\\ 
0 \\ 
0 \\ 
\vdots\\ 
0 
\end{array}%
\right)
\end{equation}

\begin{equation}
    N_{t}=BLKDIAG\left( 0_{N\times N},N_{1t}  \right)
\end{equation}

Moreover: 
\begin{equation*}
var\left( N_{t}\right) =\tilde{Q}=BLKDIAG\left(0_{N\times N},\left( 
\begin{array}{ccccc}
Q & . & . & . & 0 \\ 
0 & . & . &  & . \\ 
0 & . & . &  & . \\ 
0 & . &  &  & . \\ 
0 & . & 0 &  & 0%
\end{array}%
\right) \right)
\end{equation*}

Following recent developments in the seminal paper by \cite%
{RePEc:bla:jorssb:v:72:y:2010:i:3:p:269-342}, we employ a particle Gibbs
step to sample from the conditional posterior of $\digamma _{t}$. \cite%
{RePEc:bla:jorssb:v:72:y:2010:i:3:p:269-342} show how a version of the
particle filter, conditioned on a fixed trajectory for one of the particles
can be used to produce draws that result in a Markov Kernel with a target
distribution that is invariant. However, the usual problem of path
degeneracy in the particle filter can result in poor mixing in the original
version of particle Gibbs. Recent development, however, suggest that small
modifications of this algorithm can largely alleviate this problem. In
particular, \cite{JMLR:v15:lindsten14a} propose the addition of a step that
involves sampling the `ancestors' or indices associated with the particle
that is being conditioned on. They show that this results in a substantial
improvement in the mixing of the algorithm even with a few particles. As
explained in \cite{JMLR:v15:lindsten14a}, ancestor sampling breaks the
reference path into pieces and this causes the particle system to collapse
towards something different than the reference path. In the absence of this
step, the particle system tends to collapse to the conditioning path. We
employ particle Gibbs with ancestor sampling in this step.

Let $\digamma _{t}^{\left( i-1\right) }$ denote the fixed the fixed
trajectory, for $t=1,2,..T$ obtained in the previous draw of the Gibbs
algorithm. We denote all the parameters of the model by $\Xi $, and $j=1,2,..%
\tilde{M}$ indexes the particles. The conditional particle filter with
ancestor sampling proceeds in the following steps:

\begin{enumerate}
\item For $t=1$

\begin{enumerate}
\item Draw $\digamma _{1}^{(j)}\backslash \digamma _{0}^{(j)}$,$\Xi $ for $%
j=1,2,..\tilde{M}-1$. Fix $\digamma _{1}^{(\tilde{M})}=\digamma _{1}^{\left(
i-1\right) }$

\item Compute the normalised weights $p_{1}^{(j)}=\frac{w_{1}^{(j)}}{%
\sum_{j=1}^{\tilde{M}}w_{1}^{(j)}}$ where $w_{1}^{(j)}$ denotes the
conditional likelihood: $\left\vert \Sigma _{1}^{(j)}\right\vert
^{-0.5}-0.5\exp \left( e_{1}\left( \Sigma _{1}^{(j)}\right) ^{-1}%
e_{1}^{\prime }\right) $ where $\Sigma _{1}^{(j)}=A^{-1} H_{1}^{(j)}A^{(-1)\prime }$ with $H_{1}^{(j)}=diag\left( \exp
\left( \tilde{h}_{1,[0]}^{(j)}\right) \right) $ and:
\begin{equation}
e_{1}=Y_{t} -\left(c+\sum_{j=1}^{P}\beta _{j}Y_{t-j}+\sum_{l=1}^{L}b_{l}\tilde{h}^{j}_{1,[-l]}+\sum_{l=1}^{L}a_{L}\tilde{d}^{j}_{1,[-l]}+A^{-1}(\tilde{d}^{j}_{1,[0]}\odot \tau_{1,[0]})\right)
\end{equation}
The subscript $[0]$
denotes the contemporaneous value in the state vector while $[-l]$ denote
the $l$ lagged states.
\end{enumerate}

\item For $t=2$ to $T$

\begin{enumerate}
\item Resample $\digamma _{t-1}^{(j)}$ for $j=1,2,..\tilde{M}-1$ using
indices $a_{t}^{(j)}$ with $\Pr \left( a_{t}^{(j)}=j\right) \propto $ $%
p_{t-1}^{(j)}$

\item Draw $\digamma _{t}^{(j)}\backslash \digamma _{t-1}^{(a_{t}^{(j)})}$,$%
\Xi $ for $j=1,2,..\tilde{M}-1$ using the transition equation of the model.
Note that $\digamma _{t-1}^{(a_{t}^{(j)})}$ denotes the resampled particles
in step (a) above.

\item Fix $\digamma _{t}^{(\tilde{M})}=\digamma _{t}^{\left( i-1\right) }$

\item Sample $a_{t}^{(\tilde{M})}$ with probability defined by weights:
\begin{equation}
w_{t-1}^{(j)}\prod_{s=t}^{t-1+\mathcal{L}}g(F_{s}|\digamma^{(j)}_{1:t-1},\digamma^{(i-1)}_{t:s})f(\digamma^{(i-1)}_{s}|\digamma^{(j)}_{1:t-1},\digamma^{(i-1)}_{t:s-1})
\end{equation}
As discussed in \cite{JMLR:v15:lindsten14a}, this formula is an approximation that can be used when, as in our case, the transition density is degenerate. The number of factors in the approximation $\mathcal{L}$ is set to $5$ in our application.  

\item Update the weights $p_{t}^{(j)}=\frac{w_{t}^{(j)}}{\sum_{j=1}^{\tilde{M%
}}w_{t}^{(j)}}$ where $w_{1}^{(j)}$ denotes the conditional likelihood: 
$\left\vert \Sigma _{t}^{(j)}\right\vert
^{-0.5}-0.5\exp \left( e_{t}\left( \Sigma _{t}^{(j)}\right) ^{-1}%
e_{t}^{\prime }\right)$
\end{enumerate}

\item End

\item Sample $\digamma _{t}^{\left( i\right) }$ with $\Pr \left( \digamma
_{t}^{\left( i\right) }=\digamma _{t}^{\left( j\right) }\right) \propto $ $%
p_{T}^{(j)}$ to obtain a draw from the conditional posterior distribution
\end{enumerate}
We use $\tilde{M}=20$ particles in our application. The initial values $\mu
_{0}$ defined above are used to initialise step 1 of the filter.

\section{Forecast evaluation}
\subsection{Scoring rules}
Log scores evaluate the precision of the predictive density by measuring the logarithm of the likelihood assigned to the realised outcome based on the information available at the time of the forecast \citep{mitchell2011evaluating, Alessandri2017}. For a predictive density $p_{t}(y_{t+h} | \mathcal{I}_t)$ at forecast horizon $h$ and information set $\mathcal{I}_t$ at time $t$, the log-score is defined as $\ln p_t(y_{t+h}^{*} | \mathcal{I}_t)$, where $y_{t+h}^{*}$ is the observed outcome. 
The predictive score is commonly viewed as the broadest measure of density accuracy \citep{GEWEKE2010216}. Higher log scores indicate better predictive performance, as they reflect a higher probability mass assigned to the realized value. 

In addition to log scores, we evaluate density forecast accuracy using the CRPS which measures the difference between the cumulative distribution function (CDF) of the predictive density and the empirical CDF of the observed outcome, providing a comprehensive assessment of forecast accuracy across the entire distribution \citep{Gneiting2007, Gneiting2011}. As indicated in Gneiting and Raftery (2007) and Gneiting and Ranjan (2011), some researchers view the continuous ranked probability score as having advantages over log scores. In particular, the CRPS does a better job of rewarding values from the predictive density that are close to but not equal to the outcome, and it is less sensitive to outlier outcomes. 
For a predictive density $p_t(y | \mathcal{I}t)$ and observed outcome $y_{t+h}^{*}$, the CRPS is defined as $\int_{-\infty}^{\infty} \left[ F_t(y | \mathcal{I}t) - \mathbf{1}{y \geq y_{t+h}^{*}} \right]^2 dy$, where $F_t(y | \mathcal{I}t)$ is the CDF of the predictive density and $\mathbf{1}{y \geq y_{t+h}^{*}}$ is the indicator function for the observed outcome.  

Weighted scoring rules extend the standard log scores and CRPS by applying a weighting function that emphasises specific regions of the predictive distribution (\cite{Amisano-Giacomini-07, Gneiting2011}). For instance, for a predictive density $p_t(y | \mathcal{I}_t)$, the weighted log-score is computed as $\int w(y) \ln p_t(y | \mathcal{I}_t) dy$, where $w(y)$ is a non-negative weight function that integrates to unity.
Common choices for weight function include functions that focus on the tails of the distribution (e.g., to evaluate forecasts of extreme events) or specific quantiles relevant to macroeconomic variables, such as inflation or output growth. For both the weighted log scores and CRPS, we employ a naïve weighting scheme that attaches constant equal weights to all models \cite{GEWEKE2010216}, where the weights are updated recursively over time. This approach allows for a more targeted evaluation of forecast accuracy in regions of economic interest, such as tail risks, by assigning higher weights to those areas of the predictive density. Specifically, to focus on both tails of the predictive distribution, we calculate $w(y)=1-\frac{\phi(y)}{\phi(0)}$, with $\phi$ being a standard normal pdf.

\subsection{US Average Forecast Performance Results}

\subsubsection{Baseline results with Conditional Giacomini-White tests }

\begin{table}[H]
\centering
\caption{US Forecast Performance Comparison Across Multiple Metrics}
\label{tab:combined_forecast_comparison_conditional}
\begin{threeparttable}
\scriptsize
\begin{tabular}{l*{8}{c}}
\toprule
\textbf{Variable} & \multicolumn{8}{c}{\textbf{Forecast Horizon}} \\
\cmidrule(lr){2-9}
 & \textbf{H1} & \textbf{H2} & \textbf{H3} & \textbf{H4} & \textbf{H5} & \textbf{H6} & \textbf{H7} & \textbf{H8} \\
\midrule
\multicolumn{9}{c}{\cellcolor{gray!15}\textbf{Panel A: RMSE}} \\
\midrule
\multicolumn{9}{c}{\textbf{vs. $VAR^{\sigma,\kappa}_{restricted}$}} \\
\midrule
\textit{GNP} & 1.009 & \textbf{\textcolor{blue}{0.996}} & \textbf{\textcolor{blue}{0.982}} & \textbf{\textcolor{blue}{0.983}} & \textbf{\textcolor{blue}{0.987}} & \textbf{\textcolor{blue}{0.990}} & \textbf{\textcolor{blue}{0.997}} & 1.008 \\
\textit{GNP def} & 1.033 & 1.045 & 1.057 & 1.064 & 1.072 & 1.071 & 1.072 & 1.075$^{*}$ \\
\textit{SPREAD} & 1.004 & 1.009$^{**}$ & 1.011$^{***}$ & 1.024$^{***}$ & 1.029$^{***}$ & 1.037$^{***}$ & 1.052$^{***}$ & 1.073$^{***}$ \\
\midrule
\multicolumn{9}{c}{\textbf{vs. $VAR^{\sigma}$}} \\
\midrule
\textit{GNP} & 1.011 & \textbf{\textcolor{blue}{0.984}} & \textbf{\textcolor{blue}{0.970}} & \textbf{\textcolor{blue}{0.968}} & \textbf{\textcolor{blue}{0.963}} & \textbf{\textcolor{blue}{0.961}} & \textbf{\textcolor{blue}{0.960}} & \textbf{\textcolor{blue}{0.964}} \\
\textit{GNP def} & \textbf{\textcolor{blue}{0.962}} & \textbf{\textcolor{blue}{0.966$^{*}$}} & \textbf{\textcolor{blue}{0.964$^{*}$}} & \textbf{\textcolor{blue}{0.968$^{*}$}} & \textbf{\textcolor{blue}{0.975}} & \textbf{\textcolor{blue}{0.975}} & \textbf{\textcolor{blue}{0.976}} & \textbf{\textcolor{blue}{0.980}} \\
\textit{SPREAD} & \textbf{\textcolor{blue}{1.000}} & 1.010 & 1.011 & 1.010 & 1.025$^{*}$ & 1.026$^{**}$ & 1.031$^{**}$ & 1.035$^{**}$ \\
\midrule
\multicolumn{9}{c}{\cellcolor{gray!15}\textbf{Panel B: Weighted Both Tails Log Score}} \\
\midrule
\multicolumn{9}{c}{\textbf{vs. $VAR^{\sigma,\kappa}_{restricted}$}} \\
\midrule
\textit{GNP} & -29.9\% & \textbf{\textcolor{blue}{66.7\%}} & \textbf{\textcolor{blue}{14.4\%}} & \textbf{\textcolor{blue}{0.7\%}} & -4.1\% & -0.3\% & -1.1\% & -0.5\% \\
\textit{GNP def} & \textbf{\textcolor{blue}{0.1\%}} & \textbf{\textcolor{blue}{0.2\%}} & \textbf{\textcolor{blue}{0.3\%}} & \textbf{\textcolor{blue}{0.3\%}} & \textbf{\textcolor{blue}{0.3\%}} & \textbf{\textcolor{blue}{0.5\%}} & \textbf{\textcolor{blue}{0.6\%}} & \textbf{\textcolor{blue}{0.8\%}} \\
\textit{SPREAD} & -127.0\% & -1.2\%$^{*}$ & -0.9\%$^{**}$ & -0.1\% & -3.7\% & -7.1\% & \textbf{\textcolor{blue}{26.8\%}} & -11.1\% \\
\midrule
\multicolumn{9}{c}{\textbf{vs. $VAR^{\sigma}$}} \\
\midrule
\textit{GNP} & \textbf{\textcolor{blue}{30.8\%}} & -161.8\% & -55.5\% & -0.5\% & -3.9\% & -0.5\% & -1.4\% & -1.0\% \\
\textit{GNP def} & \textbf{\textcolor{blue}{1.3\%$^{***}$}} & \textbf{\textcolor{blue}{1.1\%$^{***}$}} & \textbf{\textcolor{blue}{0.9\%$^{***}$}} & \textbf{\textcolor{blue}{0.7\%$^{***}$}} & \textbf{\textcolor{blue}{0.5\%$^{***}$}} & \textbf{\textcolor{blue}{0.4\%$^{***}$}} & \textbf{\textcolor{blue}{0.3\%$^{***}$}} & \textbf{\textcolor{blue}{0.4\%$^{***}$}} \\
\textit{SPREAD} & -136.8\% & \textbf{\textcolor{blue}{6.9\%$^{***}$}} & \textbf{\textcolor{blue}{6.1\%$^{***}$}} & \textbf{\textcolor{blue}{5.0\%$^{***}$}} & -1.4\%$^{**}$ & -9.1\% & \textbf{\textcolor{blue}{1.4\%$^{***}$}} & -7.8\% \\
\midrule
\multicolumn{9}{c}{\cellcolor{gray!15}\textbf{Panel C: Weighted Both Tails CRPS}} \\
\midrule
\multicolumn{9}{c}{\textbf{vs. $VAR^{\sigma,\kappa}_{restricted}$}} \\
\midrule
\textit{GNP} & 1.026 & 1.006 & 1.005 & \textbf{\textcolor{blue}{0.996}} & 1.025$^{*}$ & \textbf{\textcolor{blue}{0.983}} & 1.026 & 1.027 \\
\textit{GNP def} & \textbf{\textcolor{blue}{0.966}} & \textbf{\textcolor{blue}{0.942}} & \textbf{\textcolor{blue}{0.944}} & \textbf{\textcolor{blue}{0.941}} & \textbf{\textcolor{blue}{0.945}} & \textbf{\textcolor{blue}{0.899}} & \textbf{\textcolor{blue}{0.889}} & \textbf{\textcolor{blue}{0.865}} \\
\textit{SPREAD} & 1.011 & 1.036 & 1.034$^{*}$ & 1.043$^{**}$ & 1.040$^{**}$ & 1.035$^{**}$ & 1.041$^{***}$ & 1.054$^{***}$ \\
\midrule
\multicolumn{9}{c}{\textbf{vs. $VAR^{\sigma}$}} \\
\midrule
\textit{GNP} & 1.031 & \textbf{\textcolor{blue}{0.999}} & 1.042 & 1.053 & 1.071 & 1.046 & 1.082 & 1.090 \\
\textit{GNP def} & \textbf{\textcolor{blue}{0.819$^{***}$}} & \textbf{\textcolor{blue}{0.826$^{**}$}} & \textbf{\textcolor{blue}{0.852}} & \textbf{\textcolor{blue}{0.880}} & \textbf{\textcolor{blue}{0.914}} & \textbf{\textcolor{blue}{0.894}} & \textbf{\textcolor{blue}{0.907}} & \textbf{\textcolor{blue}{0.892}} \\
\textit{SPREAD} & \textbf{\textcolor{blue}{0.843$^{***}$}} & \textbf{\textcolor{blue}{0.860$^{***}$}} & \textbf{\textcolor{blue}{0.867$^{***}$}} & \textbf{\textcolor{blue}{0.888$^{***}$}} & \textbf{\textcolor{blue}{0.935$^{***}$}} & \textbf{\textcolor{blue}{0.947$^{***}$}} & \textbf{\textcolor{blue}{0.955$^{***}$}} & \textbf{\textcolor{blue}{0.950$^{***}$}} \\
\bottomrule
\end{tabular}
\begin{tablenotes}
\scriptsize
\item \textbf{Notes:} For RMSE and CRPS, values shown are ratios ($VAR^{\sigma,\kappa}$ model/Competitor); values less than 1 (\textcolor{blue}{\textbf{bold blue}}) indicate the proposed model outperforms the competitor. For Log Score, values shown are percentage differences ($VAR^{\sigma,\kappa}$ - Competitor); positive values (\textcolor{blue}{\textbf{bold blue}}) indicate superior performance. Significance levels from Giacomini-White tests: *** p$<$0.01, ** p$<$0.05, * p$<$0.1.
\end{tablenotes}
\end{threeparttable}
\end{table}

\subsubsection{Additional results - standard LS, CRPS and other weighted tails}

\begin{table}[H]
\centering
\caption{Log Score Comparison}
\label{tab:log_score_comparison_combined}
\begin{threeparttable}
\scriptsize
\begin{tabular}{l*{8}{c}}
\toprule
\textbf{Variable} & \multicolumn{8}{c}{\textbf{Forecast Horizon}} \\
\cmidrule(lr){2-9}
 & \textbf{H1} & \textbf{H2} & \textbf{H3} & \textbf{H4} & \textbf{H5} & \textbf{H6} & \textbf{H7} & \textbf{H8} \\
\midrule
\multicolumn{9}{c}{\cellcolor{gray!15}\textbf{Panel A: GW Unconditional}} \\
\midrule
\multicolumn{9}{c}{\textbf{vs. $VAR^{\sigma,\kappa}_{restricted}$}} \\
\midrule
\textit{GNP} & -34.1\% & \textbf{\textcolor{blue}{63.7\%}} & \textbf{\textcolor{blue}{12.2\%}} & -1.1\% & -4.8\% & -0.6\% & -1.2\% & \textbf{\textcolor{blue}{0.0\%}} \\
\textit{GNP def} & -7.9\%$^{***}$ & -8.3\%$^{***}$ & -7.8\%$^{***}$ & -6.5\%$^{***}$ & -5.9\%$^{**}$ & -5.8\%$^{**}$ & -5.4\%$^{*}$ & -5.1\% \\
\textit{SPREAD} & -161.5\% & -2.1\% & -2.7\% & -2.2\% & -5.3\% & -9.5\% & \textbf{\textcolor{blue}{23.8\%}} & -15.3\% \\
\midrule
\multicolumn{9}{c}{\textbf{vs. $VAR^{\sigma}$}} \\
\midrule
\textit{GNP} & \textbf{\textcolor{blue}{43.1\%}} & -149.8\% & -45.4\% & \textbf{\textcolor{blue}{7.8\%$^{***}$}} & \textbf{\textcolor{blue}{3.5\%}} & \textbf{\textcolor{blue}{6.0\%$^{**}$}} & \textbf{\textcolor{blue}{4.3\%}} & \textbf{\textcolor{blue}{5.1\%$^{*}$}} \\
\textit{GNP def} & \textbf{\textcolor{blue}{92.8\%$^{***}$}} & \textbf{\textcolor{blue}{82.3\%$^{***}$}} & \textbf{\textcolor{blue}{74.8\%$^{***}$}} & \textbf{\textcolor{blue}{68.0\%$^{***}$}} & \textbf{\textcolor{blue}{62.2\%$^{***}$}} & \textbf{\textcolor{blue}{57.4\%$^{***}$}} & \textbf{\textcolor{blue}{53.0\%$^{***}$}} & \textbf{\textcolor{blue}{49.3\%$^{***}$}} \\
\textit{SPREAD} & -126.7\% & \textbf{\textcolor{blue}{54.8\%$^{***}$}} & \textbf{\textcolor{blue}{49.2\%$^{***}$}} & \textbf{\textcolor{blue}{48.2\%$^{***}$}} & \textbf{\textcolor{blue}{40.0\%$^{***}$}} & \textbf{\textcolor{blue}{33.4\%$^{*}$}} & \textbf{\textcolor{blue}{42.8\%$^{***}$}} & \textbf{\textcolor{blue}{32.0\%$^{*}$}} \\
\midrule
\multicolumn{9}{c}{\cellcolor{gray!15}\textbf{Panel B: GW Conditional}} \\
\midrule
\multicolumn{9}{c}{\textbf{vs. $VAR^{\sigma,\kappa}_{restricted}$}} \\
\midrule
\textit{GNP} & -34.1\% & \textbf{\textcolor{blue}{63.7\%}} & \textbf{\textcolor{blue}{12.2\%}} & -1.1\% & -4.8\% & -0.6\%$^{**}$ & -1.2\% & \textbf{\textcolor{blue}{0.0\%$^{***}$}} \\
\textit{GNP def} & -7.9\%$^{***}$ & -8.3\%$^{***}$ & -7.8\%$^{***}$ & -6.5\%$^{***}$ & -5.9\%$^{***}$ & -5.8\%$^{***}$ & -5.4\%$^{**}$ & -5.1\%$^{***}$ \\
\textit{SPREAD} & -161.5\% & -2.1\%$^{***}$ & -2.7\%$^{***}$ & -2.2\%$^{***}$ & -5.3\%$^{*}$ & -9.5\%$^{***}$ & \textbf{\textcolor{blue}{23.8\%}} & -15.3\%$^{***}$ \\
\midrule
\multicolumn{9}{c}{\textbf{vs. $VAR^{\sigma}$}} \\
\midrule
\textit{GNP} & \textbf{\textcolor{blue}{43.1\%}} & -149.8\%$^{***}$ & -45.4\% & \textbf{\textcolor{blue}{7.8\%$^{***}$}} & \textbf{\textcolor{blue}{3.5\%$^{***}$}} & \textbf{\textcolor{blue}{6.0\%$^{***}$}} & \textbf{\textcolor{blue}{4.3\%$^{***}$}} & \textbf{\textcolor{blue}{5.1\%$^{***}$}} \\
\textit{GNP def} & \textbf{\textcolor{blue}{92.8\%$^{***}$}} & \textbf{\textcolor{blue}{82.3\%$^{***}$}} & \textbf{\textcolor{blue}{74.8\%$^{***}$}} & \textbf{\textcolor{blue}{68.0\%$^{***}$}} & \textbf{\textcolor{blue}{62.2\%$^{***}$}} & \textbf{\textcolor{blue}{57.4\%$^{***}$}} & \textbf{\textcolor{blue}{53.0\%$^{***}$}} & \textbf{\textcolor{blue}{49.3\%$^{***}$}} \\
\textit{SPREAD} & -126.7\% & \textbf{\textcolor{blue}{54.8\%$^{***}$}} & \textbf{\textcolor{blue}{49.2\%$^{***}$}} & \textbf{\textcolor{blue}{48.2\%$^{***}$}} & \textbf{\textcolor{blue}{40.0\%$^{***}$}} & \textbf{\textcolor{blue}{33.4\%$^{***}$}} & \textbf{\textcolor{blue}{42.8\%$^{***}$}} & \textbf{\textcolor{blue}{32.0\%$^{***}$}} \\
\bottomrule
\end{tabular}
\begin{tablenotes}
\scriptsize
\item \textbf{Notes:} Values shown are percentage differences ($VAR^{\sigma,\kappa}$ model - Competitor); positive values (\textcolor{blue}{\textbf{bold blue}}) indicate the proposed model ($VAR^{\sigma,\kappa}$) outperforms the competitor. Panel A shows results from Giacomini-White unconditional test. Panel B shows results from Giacomini-White conditional test. Significance levels from these tests: *** p$<$0.01, ** p$<$0.05, * p$<$0.1.
\end{tablenotes}
\end{threeparttable}
\end{table}

\begin{table}[H]
\centering
\caption{Tail Weighted Log Score Comparison}
\label{tab:tail_weighted_log_score_comparison_combined}
\begin{threeparttable}
\scriptsize
\begin{tabular}{l*{8}{c}}
\toprule
\textbf{Variable} & \multicolumn{8}{c}{\textbf{Forecast Horizon}} \\
\cmidrule(lr){2-9}
 & \textbf{H1} & \textbf{H2} & \textbf{H3} & \textbf{H4} & \textbf{H5} & \textbf{H6} & \textbf{H7} & \textbf{H8} \\
\midrule
\multicolumn{9}{c}{\cellcolor{gray!15}\textbf{Panel A: Left Tail - GW Unconditional}} \\
\midrule
\multicolumn{9}{c}{\textbf{vs. $VAR^{\sigma,\kappa}_{restricted}$}} \\
\midrule
\textit{GNP} & -36.6\% & \textbf{\textcolor{blue}{65.3\%}} & \textbf{\textcolor{blue}{13.7\%}} & \textbf{\textcolor{blue}{0.3\%}} & -4.1\% & -0.5\% & -1.1\% & -0.1\% \\
\textit{GNP def} & -4.2\%$^{***}$ & -4.4\%$^{***}$ & -4.2\%$^{***}$ & -3.6\%$^{***}$ & -3.5\%$^{***}$ & -3.6\%$^{***}$ & -3.4\%$^{***}$ & -3.6\%$^{***}$ \\
\textit{SPREAD} & -4.8\% & \textbf{\textcolor{blue}{1.3\%}} & \textbf{\textcolor{blue}{1.7\%}} & \textbf{\textcolor{blue}{1.7\%}} & \textbf{\textcolor{blue}{2.1\%}} & \textbf{\textcolor{blue}{2.0\%}} & \textbf{\textcolor{blue}{2.0\%}} & \textbf{\textcolor{blue}{1.6\%}} \\
\midrule
\multicolumn{9}{c}{\textbf{vs. $VAR^{\sigma}$}} \\
\midrule
\textit{GNP} & \textbf{\textcolor{blue}{40.7\%}} & -156.1\% & -49.5\% & \textbf{\textcolor{blue}{4.3\%$^{***}$}} & \textbf{\textcolor{blue}{0.3\%}} & \textbf{\textcolor{blue}{2.9\%$^{*}$}} & \textbf{\textcolor{blue}{1.4\%}} & \textbf{\textcolor{blue}{2.2\%}} \\
\textit{GNP def} & \textbf{\textcolor{blue}{45.1\%$^{***}$}} & \textbf{\textcolor{blue}{40.2\%$^{***}$}} & \textbf{\textcolor{blue}{37.0\%$^{***}$}} & \textbf{\textcolor{blue}{33.9\%$^{***}$}} & \textbf{\textcolor{blue}{31.2\%$^{***}$}} & \textbf{\textcolor{blue}{28.9\%$^{***}$}} & \textbf{\textcolor{blue}{26.8\%$^{***}$}} & \textbf{\textcolor{blue}{24.7\%$^{***}$}} \\
\textit{SPREAD} & \textbf{\textcolor{blue}{34.6\%$^{***}$}} & \textbf{\textcolor{blue}{41.1\%$^{***}$}} & \textbf{\textcolor{blue}{37.6\%$^{***}$}} & \textbf{\textcolor{blue}{36.7\%$^{***}$}} & \textbf{\textcolor{blue}{35.3\%$^{***}$}} & \textbf{\textcolor{blue}{36.0\%$^{***}$}} & \textbf{\textcolor{blue}{35.3\%$^{***}$}} & \textbf{\textcolor{blue}{34.5\%$^{***}$}} \\
\midrule
\multicolumn{9}{c}{\cellcolor{gray!15}\textbf{Panel B: Left Tail - GW Conditional}} \\
\midrule
\multicolumn{9}{c}{\textbf{vs. $VAR^{\sigma,\kappa}_{restricted}$}} \\
\midrule
\textit{GNP} & -36.6\%$^{**}$ & \textbf{\textcolor{blue}{65.3\%}} & \textbf{\textcolor{blue}{13.7\%}} & \textbf{\textcolor{blue}{0.3\%$^{**}$}} & -4.1\% & -0.5\%$^{**}$ & -1.1\% & -0.1\% \\
\textit{GNP def} & -4.2\%$^{***}$ & -4.4\%$^{***}$ & -4.2\%$^{***}$ & -3.6\%$^{***}$ & -3.5\%$^{***}$ & -3.6\%$^{***}$ & -3.4\%$^{***}$ & -3.6\%$^{**}$ \\
\textit{SPREAD} & -4.8\% & \textbf{\textcolor{blue}{1.3\%$^{***}$}} & \textbf{\textcolor{blue}{1.7\%$^{***}$}} & \textbf{\textcolor{blue}{1.7\%$^{***}$}} & \textbf{\textcolor{blue}{2.1\%$^{***}$}} & \textbf{\textcolor{blue}{2.0\%$^{***}$}} & \textbf{\textcolor{blue}{2.0\%$^{***}$}} & \textbf{\textcolor{blue}{1.6\%$^{***}$}} \\
\midrule
\multicolumn{9}{c}{\textbf{vs. $VAR^{\sigma}$}} \\
\midrule
\textit{GNP} & \textbf{\textcolor{blue}{40.7\%$^{***}$}} & -156.1\%$^{***}$ & -49.5\%$^{***}$ & \textbf{\textcolor{blue}{4.3\%$^{***}$}} & \textbf{\textcolor{blue}{0.3\%}} & \textbf{\textcolor{blue}{2.9\%$^{***}$}} & \textbf{\textcolor{blue}{1.4\%}} & \textbf{\textcolor{blue}{2.2\%$^{***}$}} \\
\textit{GNP def} & \textbf{\textcolor{blue}{45.1\%$^{***}$}} & \textbf{\textcolor{blue}{40.2\%$^{***}$}} & \textbf{\textcolor{blue}{37.0\%$^{***}$}} & \textbf{\textcolor{blue}{33.9\%$^{***}$}} & \textbf{\textcolor{blue}{31.2\%$^{***}$}} & \textbf{\textcolor{blue}{28.9\%$^{***}$}} & \textbf{\textcolor{blue}{26.8\%$^{***}$}} & \textbf{\textcolor{blue}{24.7\%$^{***}$}} \\
\textit{SPREAD} & \textbf{\textcolor{blue}{34.6\%$^{***}$}} & \textbf{\textcolor{blue}{41.1\%$^{***}$}} & \textbf{\textcolor{blue}{37.6\%$^{***}$}} & \textbf{\textcolor{blue}{36.7\%$^{***}$}} & \textbf{\textcolor{blue}{35.3\%$^{***}$}} & \textbf{\textcolor{blue}{36.0\%$^{***}$}} & \textbf{\textcolor{blue}{35.3\%$^{***}$}} & \textbf{\textcolor{blue}{34.5\%$^{***}$}} \\
\midrule
\multicolumn{9}{c}{\cellcolor{gray!15}\textbf{Panel C: Right Tail - GW Unconditional}} \\
\midrule
\multicolumn{9}{c}{\textbf{vs. $VAR^{\sigma,\kappa}_{restricted}$}} \\
\midrule
\textit{GNP} & \textbf{\textcolor{blue}{2.5\%}} & -1.6\%$^{**}$ & -1.5\%$^{*}$ & -1.4\% & -0.7\% & -0.1\% & -0.0\% & \textbf{\textcolor{blue}{0.1\%}} \\
\textit{GNP def} & -3.7\%$^{***}$ & -3.9\%$^{***}$ & -3.6\%$^{***}$ & -2.9\%$^{**}$ & -2.5\%$^{*}$ & -2.2\% & -1.9\% & -1.5\% \\
\textit{SPREAD} & -156.7\% & -3.4\%$^{**}$ & -4.4\%$^{**}$ & -3.9\%$^{**}$ & -7.4\% & -11.5\% & \textbf{\textcolor{blue}{21.7\%}} & -17.0\% \\
\midrule
\multicolumn{9}{c}{\textbf{vs. $VAR^{\sigma}$}} \\
\midrule
\textit{GNP} & \textbf{\textcolor{blue}{2.4\%}} & \textbf{\textcolor{blue}{6.3\%$^{***}$}} & \textbf{\textcolor{blue}{4.1\%$^{**}$}} & \textbf{\textcolor{blue}{3.5\%$^{**}$}} & \textbf{\textcolor{blue}{3.1\%$^{**}$}} & \textbf{\textcolor{blue}{3.2\%$^{**}$}} & \textbf{\textcolor{blue}{2.9\%$^{**}$}} & \textbf{\textcolor{blue}{2.9\%$^{**}$}} \\
\textit{GNP def} & \textbf{\textcolor{blue}{47.7\%$^{***}$}} & \textbf{\textcolor{blue}{42.1\%$^{***}$}} & \textbf{\textcolor{blue}{37.9\%$^{***}$}} & \textbf{\textcolor{blue}{34.1\%$^{***}$}} & \textbf{\textcolor{blue}{31.0\%$^{***}$}} & \textbf{\textcolor{blue}{28.5\%$^{***}$}} & \textbf{\textcolor{blue}{26.2\%$^{***}$}} & \textbf{\textcolor{blue}{24.6\%$^{***}$}} \\
\textit{SPREAD} & -161.3\% & \textbf{\textcolor{blue}{13.8\%$^{***}$}} & \textbf{\textcolor{blue}{11.6\%$^{**}$}} & \textbf{\textcolor{blue}{11.5\%$^{*}$}} & \textbf{\textcolor{blue}{4.7\%}} & -2.6\% & \textbf{\textcolor{blue}{7.5\%}} & -2.5\% \\
\midrule
\multicolumn{9}{c}{\cellcolor{gray!15}\textbf{Panel D: Right Tail - GW Conditional}} \\
\midrule
\multicolumn{9}{c}{\textbf{vs. $VAR^{\sigma,\kappa}_{restricted}$}} \\
\midrule
\textit{GNP} & \textbf{\textcolor{blue}{2.5\%}} & -1.6\%$^{**}$ & -1.5\%$^{*}$ & -1.4\% & -0.7\% & -0.1\%$^{*}$ & -0.0\%$^{*}$ & \textbf{\textcolor{blue}{0.1\%$^{***}$}} \\
\textit{GNP def} & -3.7\%$^{**}$ & -3.9\%$^{***}$ & -3.6\%$^{***}$ & -2.9\%$^{***}$ & -2.5\%$^{***}$ & -2.2\%$^{***}$ & -1.9\%$^{**}$ & -1.5\%$^{***}$ \\
\textit{SPREAD} & -156.7\% & -3.4\%$^{**}$ & -4.4\%$^{***}$ & -3.9\% & -7.4\% & -11.5\% & \textbf{\textcolor{blue}{21.7\%}} & -17.0\% \\
\midrule
\multicolumn{9}{c}{\textbf{vs. $VAR^{\sigma}$}} \\
\midrule
\textit{GNP} & \textbf{\textcolor{blue}{2.4\%$^{***}$}} & \textbf{\textcolor{blue}{6.3\%$^{***}$}} & \textbf{\textcolor{blue}{4.1\%$^{***}$}} & \textbf{\textcolor{blue}{3.5\%$^{***}$}} & \textbf{\textcolor{blue}{3.1\%$^{***}$}} & \textbf{\textcolor{blue}{3.2\%$^{***}$}} & \textbf{\textcolor{blue}{2.9\%$^{***}$}} & \textbf{\textcolor{blue}{2.9\%$^{***}$}} \\
\textit{GNP def} & \textbf{\textcolor{blue}{47.7\%$^{***}$}} & \textbf{\textcolor{blue}{42.1\%$^{***}$}} & \textbf{\textcolor{blue}{37.9\%$^{***}$}} & \textbf{\textcolor{blue}{34.1\%$^{***}$}} & \textbf{\textcolor{blue}{31.0\%$^{***}$}} & \textbf{\textcolor{blue}{28.5\%$^{***}$}} & \textbf{\textcolor{blue}{26.2\%$^{***}$}} & \textbf{\textcolor{blue}{24.6\%$^{***}$}} \\
\textit{SPREAD} & -161.3\% & \textbf{\textcolor{blue}{13.8\%$^{***}$}} & \textbf{\textcolor{blue}{11.6\%$^{***}$}} & \textbf{\textcolor{blue}{11.5\%$^{***}$}} & \textbf{\textcolor{blue}{4.7\%$^{***}$}} & -2.6\%$^{**}$ & \textbf{\textcolor{blue}{7.5\%$^{***}$}} & -2.5\%$^{***}$ \\
\bottomrule
\end{tabular}
\begin{tablenotes}
\scriptsize
\item \textbf{Notes:} Values shown are percentage differences ($VAR^{\sigma,\kappa}$ model - Competitor); positive values (\textcolor{blue}{\textbf{bold blue}}) indicate the proposed model ($VAR^{\sigma,\kappa}$) outperforms the competitor. Panels A and B show left tail weighted log score for unconditional and conditional Giacomini-White tests. Panels C and D show right tail weighted log score for unconditional and conditional Giacomini-White tests. Significance levels: *** p$<$0.01, ** p$<$0.05, * p$<$0.1.
\end{tablenotes}
\end{threeparttable}
\end{table}

\begin{table}[H]
\centering
\caption{CRPS Comparison}
\label{tab:crps_comparison_combined}
\begin{threeparttable}
\scriptsize
\begin{tabular}{l*{8}{c}}
\toprule
\textbf{Variable} & \multicolumn{8}{c}{\textbf{Forecast Horizon}} \\
\cmidrule(lr){2-9}
 & \textbf{H1} & \textbf{H2} & \textbf{H3} & \textbf{H4} & \textbf{H5} & \textbf{H6} & \textbf{H7} & \textbf{H8} \\
\midrule
\multicolumn{9}{c}{\cellcolor{gray!15}\textbf{Panel A: GW Unconditional}} \\
\midrule
\multicolumn{9}{c}{\textbf{vs. $VAR^{\sigma,\kappa}_{restricted}$}} \\
\midrule
\textit{GNP} & 1.019$^{*}$ & 1.019 & 1.010 & 1.009 & 1.010 & 1.012 & 1.012 & 1.014 \\
\textit{GNP def} & 1.042$^{*}$ & 1.049$^{*}$ & 1.061$^{*}$ & 1.067$^{*}$ & 1.074$^{**}$ & 1.077$^{*}$ & 1.079 & 1.083 \\
\textit{SPREAD} & 1.000 & 1.013 & 1.018 & 1.023 & 1.030 & 1.031 & 1.043 & 1.057 \\
\midrule
\multicolumn{9}{c}{\textbf{vs. $VAR^{\sigma}$}} \\
\midrule
\textit{GNP} & \textbf{\textcolor{blue}{0.964$^{***}$}} & \textbf{\textcolor{blue}{0.938$^{***}$}} & \textbf{\textcolor{blue}{0.947$^{***}$}} & \textbf{\textcolor{blue}{0.953$^{**}$}} & \textbf{\textcolor{blue}{0.959$^{*}$}} & \textbf{\textcolor{blue}{0.963}} & \textbf{\textcolor{blue}{0.963}} & \textbf{\textcolor{blue}{0.964}} \\
\textit{GNP def} & \textbf{\textcolor{blue}{0.554$^{***}$}} & \textbf{\textcolor{blue}{0.601$^{***}$}} & \textbf{\textcolor{blue}{0.636$^{***}$}} & \textbf{\textcolor{blue}{0.668$^{***}$}} & \textbf{\textcolor{blue}{0.698$^{***}$}} & \textbf{\textcolor{blue}{0.722$^{***}$}} & \textbf{\textcolor{blue}{0.744$^{***}$}} & \textbf{\textcolor{blue}{0.766$^{***}$}} \\
\textit{SPREAD} & \textbf{\textcolor{blue}{0.733$^{***}$}} & \textbf{\textcolor{blue}{0.752$^{***}$}} & \textbf{\textcolor{blue}{0.768$^{***}$}} & \textbf{\textcolor{blue}{0.773$^{***}$}} & \textbf{\textcolor{blue}{0.791$^{***}$}} & \textbf{\textcolor{blue}{0.788$^{***}$}} & \textbf{\textcolor{blue}{0.794$^{***}$}} & \textbf{\textcolor{blue}{0.797$^{***}$}} \\
\midrule
\multicolumn{9}{c}{\cellcolor{gray!15}\textbf{Panel B: GW Conditional}} \\
\midrule
\multicolumn{9}{c}{\textbf{vs. $VAR^{\sigma,\kappa}_{restricted}$}} \\
\midrule
\textit{GNP} & 1.019 & 1.019 & 1.010 & 1.009 & 1.010$^{*}$ & 1.012 & 1.012 & 1.014 \\
\textit{GNP def} & 1.042 & 1.049$^{*}$ & 1.061$^{**}$ & 1.067$^{**}$ & 1.074$^{**}$ & 1.077$^{**}$ & 1.079$^{***}$ & 1.083$^{***}$ \\
\textit{SPREAD} & 1.000 & 1.013$^{***}$ & 1.018$^{***}$ & 1.023$^{***}$ & 1.030$^{***}$ & 1.031$^{***}$ & 1.043$^{***}$ & 1.057$^{***}$ \\
\midrule
\multicolumn{9}{c}{\textbf{vs. $VAR^{\sigma}$}} \\
\midrule
\textit{GNP} & \textbf{\textcolor{blue}{0.964$^{**}$}} & \textbf{\textcolor{blue}{0.938$^{***}$}} & \textbf{\textcolor{blue}{0.947$^{***}$}} & \textbf{\textcolor{blue}{0.953$^{***}$}} & \textbf{\textcolor{blue}{0.959$^{***}$}} & \textbf{\textcolor{blue}{0.963$^{***}$}} & \textbf{\textcolor{blue}{0.963$^{***}$}} & \textbf{\textcolor{blue}{0.964$^{***}$}} \\
\textit{GNP def} & \textbf{\textcolor{blue}{0.554$^{***}$}} & \textbf{\textcolor{blue}{0.601$^{***}$}} & \textbf{\textcolor{blue}{0.636$^{***}$}} & \textbf{\textcolor{blue}{0.668$^{***}$}} & \textbf{\textcolor{blue}{0.698$^{***}$}} & \textbf{\textcolor{blue}{0.722$^{***}$}} & \textbf{\textcolor{blue}{0.744$^{***}$}} & \textbf{\textcolor{blue}{0.766$^{***}$}} \\
\textit{SPREAD} & \textbf{\textcolor{blue}{0.733$^{***}$}} & \textbf{\textcolor{blue}{0.752$^{***}$}} & \textbf{\textcolor{blue}{0.768$^{***}$}} & \textbf{\textcolor{blue}{0.773$^{***}$}} & \textbf{\textcolor{blue}{0.791$^{***}$}} & \textbf{\textcolor{blue}{0.788$^{***}$}} & \textbf{\textcolor{blue}{0.794$^{***}$}} & \textbf{\textcolor{blue}{0.797$^{***}$}} \\
\bottomrule
\end{tabular}
\begin{tablenotes}
\scriptsize
\item \textbf{Notes:} Values shown are CRPS ratios ($VAR^{\sigma,\kappa}$ model/Competitor); values less than 1 (\textcolor{blue}{\textbf{bold blue}}) indicate the proposed model ($VAR^{\sigma,\kappa}$) outperforms the competitor. Panel A shows results from Giacomini-White unconditional test. Panel B shows results from Giacomini-White conditional test. Significance levels from the tests: *** p$<$0.01, ** p$<$0.05, * p$<$0.1.
\end{tablenotes}
\end{threeparttable}
\end{table}

\begin{table}[H]
\centering
\caption{Tail Weighted CRPS Comparison}
\label{tab:tail_weighted_crps_comparison_combined}
\begin{threeparttable}
\scriptsize
\begin{tabular}{l*{8}{c}}
\toprule
\textbf{Variable} & \multicolumn{8}{c}{\textbf{Forecast Horizon}} \\
\cmidrule(lr){2-9}
 & \textbf{H1} & \textbf{H2} & \textbf{H3} & \textbf{H4} & \textbf{H5} & \textbf{H6} & \textbf{H7} & \textbf{H8} \\
\midrule
\multicolumn{9}{c}{\cellcolor{gray!15}\textbf{Panel A: Left Tail - GW Unconditional}} \\
\midrule
\multicolumn{9}{c}{\textbf{vs. $VAR^{\sigma,\kappa}_{restricted}$}} \\
\midrule
\textit{GNP} & 1.024$^{*}$ & 1.017 & 1.010 & 1.010 & 1.012 & 1.015 & 1.011 & 1.013 \\
\textit{GNP def} & 1.056$^{***}$ & 1.070$^{***}$ & 1.085$^{***}$ & 1.096$^{***}$ & 1.110$^{***}$ & 1.127$^{***}$ & 1.131$^{***}$ & 1.145$^{**}$ \\
\textit{SPREAD} & \textbf{\textcolor{blue}{0.977}} & \textbf{\textcolor{blue}{0.973}} & \textbf{\textcolor{blue}{0.965}} & \textbf{\textcolor{blue}{0.960}} & \textbf{\textcolor{blue}{0.959}} & \textbf{\textcolor{blue}{0.959}} & \textbf{\textcolor{blue}{0.963}} & \textbf{\textcolor{blue}{0.969}} \\
\midrule
\multicolumn{9}{c}{\textbf{vs. $VAR^{\sigma}$}} \\
\midrule
\textit{GNP} & \textbf{\textcolor{blue}{0.968$^{**}$}} & \textbf{\textcolor{blue}{0.954$^{***}$}} & \textbf{\textcolor{blue}{0.954$^{***}$}} & \textbf{\textcolor{blue}{0.963$^{*}$}} & \textbf{\textcolor{blue}{0.963}} & \textbf{\textcolor{blue}{0.969}} & \textbf{\textcolor{blue}{0.974}} & \textbf{\textcolor{blue}{0.971}} \\
\textit{GNP def} & \textbf{\textcolor{blue}{0.517$^{***}$}} & \textbf{\textcolor{blue}{0.569$^{***}$}} & \textbf{\textcolor{blue}{0.598$^{***}$}} & \textbf{\textcolor{blue}{0.629$^{***}$}} & \textbf{\textcolor{blue}{0.658$^{***}$}} & \textbf{\textcolor{blue}{0.686$^{***}$}} & \textbf{\textcolor{blue}{0.714$^{***}$}} & \textbf{\textcolor{blue}{0.743$^{***}$}} \\
\textit{SPREAD} & \textbf{\textcolor{blue}{0.626$^{***}$}} & \textbf{\textcolor{blue}{0.637$^{***}$}} & \textbf{\textcolor{blue}{0.662$^{***}$}} & \textbf{\textcolor{blue}{0.666$^{***}$}} & \textbf{\textcolor{blue}{0.676$^{***}$}} & \textbf{\textcolor{blue}{0.666$^{***}$}} & \textbf{\textcolor{blue}{0.670$^{***}$}} & \textbf{\textcolor{blue}{0.672$^{***}$}} \\
\midrule
\multicolumn{9}{c}{\cellcolor{gray!15}\textbf{Panel B: Left Tail - GW Conditional}} \\
\midrule
\multicolumn{9}{c}{\textbf{vs. $VAR^{\sigma,\kappa}_{restricted}$}} \\
\midrule
\textit{GNP} & 1.024 & 1.017 & 1.010 & 1.010 & 1.012$^{*}$ & 1.015$^{*}$ & 1.011 & 1.013 \\
\textit{GNP def} & 1.056$^{**}$ & 1.070$^{**}$ & 1.085$^{***}$ & 1.096$^{***}$ & 1.110$^{***}$ & 1.127$^{***}$ & 1.131$^{**}$ & 1.145$^{**}$ \\
\textit{SPREAD} & \textbf{\textcolor{blue}{0.977}} & \textbf{\textcolor{blue}{0.973$^{***}$}} & \textbf{\textcolor{blue}{0.965$^{***}$}} & \textbf{\textcolor{blue}{0.960$^{***}$}} & \textbf{\textcolor{blue}{0.959$^{***}$}} & \textbf{\textcolor{blue}{0.959$^{***}$}} & \textbf{\textcolor{blue}{0.963$^{***}$}} & \textbf{\textcolor{blue}{0.969$^{***}$}} \\
\midrule
\multicolumn{9}{c}{\textbf{vs. $VAR^{\sigma}$}} \\
\midrule
\textit{GNP} & \textbf{\textcolor{blue}{0.968}} & \textbf{\textcolor{blue}{0.954$^{***}$}} & \textbf{\textcolor{blue}{0.954$^{**}$}} & \textbf{\textcolor{blue}{0.963$^{**}$}} & \textbf{\textcolor{blue}{0.963$^{**}$}} & \textbf{\textcolor{blue}{0.969$^{***}$}} & \textbf{\textcolor{blue}{0.974$^{**}$}} & \textbf{\textcolor{blue}{0.971$^{***}$}} \\
\textit{GNP def} & \textbf{\textcolor{blue}{0.517$^{***}$}} & \textbf{\textcolor{blue}{0.569$^{***}$}} & \textbf{\textcolor{blue}{0.598$^{***}$}} & \textbf{\textcolor{blue}{0.629$^{***}$}} & \textbf{\textcolor{blue}{0.658$^{***}$}} & \textbf{\textcolor{blue}{0.686$^{***}$}} & \textbf{\textcolor{blue}{0.714$^{***}$}} & \textbf{\textcolor{blue}{0.743$^{***}$}} \\
\textit{SPREAD} & \textbf{\textcolor{blue}{0.626$^{***}$}} & \textbf{\textcolor{blue}{0.637$^{***}$}} & \textbf{\textcolor{blue}{0.662$^{***}$}} & \textbf{\textcolor{blue}{0.666$^{***}$}} & \textbf{\textcolor{blue}{0.676$^{***}$}} & \textbf{\textcolor{blue}{0.666$^{***}$}} & \textbf{\textcolor{blue}{0.670$^{***}$}} & \textbf{\textcolor{blue}{0.672$^{***}$}} \\
\midrule
\multicolumn{9}{c}{\cellcolor{gray!15}\textbf{Panel C: Right Tail - GW Unconditional}} \\
\midrule
\multicolumn{9}{c}{\textbf{vs. $VAR^{\sigma,\kappa}_{restricted}$}} \\
\midrule
\textit{GNP} & 1.014 & 1.020 & 1.010 & 1.007 & 1.007 & 1.008 & 1.014 & 1.015 \\
\textit{GNP def} & 1.033 & 1.034 & 1.044 & 1.048 & 1.050 & 1.045 & 1.044 & 1.042 \\
\textit{SPREAD} & 1.017 & 1.042$^{**}$ & 1.061$^{**}$ & 1.073$^{**}$ & 1.088$^{**}$ & 1.089$^{**}$ & 1.108$^{**}$ & 1.129$^{***}$ \\
\midrule
\multicolumn{9}{c}{\textbf{vs. $VAR^{\sigma}$}} \\
\midrule
\textit{GNP} & \textbf{\textcolor{blue}{0.959$^{***}$}} & \textbf{\textcolor{blue}{0.916$^{***}$}} & \textbf{\textcolor{blue}{0.938$^{***}$}} & \textbf{\textcolor{blue}{0.942$^{***}$}} & \textbf{\textcolor{blue}{0.955$^{*}$}} & \textbf{\textcolor{blue}{0.955$^{*}$}} & \textbf{\textcolor{blue}{0.951$^{**}$}} & \textbf{\textcolor{blue}{0.955$^{*}$}} \\
\textit{GNP def} & \textbf{\textcolor{blue}{0.584$^{***}$}} & \textbf{\textcolor{blue}{0.626$^{***}$}} & \textbf{\textcolor{blue}{0.666$^{***}$}} & \textbf{\textcolor{blue}{0.698$^{***}$}} & \textbf{\textcolor{blue}{0.728$^{***}$}} & \textbf{\textcolor{blue}{0.749$^{***}$}} & \textbf{\textcolor{blue}{0.768$^{***}$}} & \textbf{\textcolor{blue}{0.784$^{***}$}} \\
\textit{SPREAD} & \textbf{\textcolor{blue}{0.834$^{***}$}} & \textbf{\textcolor{blue}{0.857$^{***}$}} & \textbf{\textcolor{blue}{0.869$^{***}$}} & \textbf{\textcolor{blue}{0.875$^{***}$}} & \textbf{\textcolor{blue}{0.900$^{**}$}} & \textbf{\textcolor{blue}{0.904$^{*}$}} & \textbf{\textcolor{blue}{0.913}} & \textbf{\textcolor{blue}{0.917}} \\
\midrule
\multicolumn{9}{c}{\cellcolor{gray!15}\textbf{Panel D: Right Tail - GW Conditional}} \\
\midrule
\multicolumn{9}{c}{\textbf{vs. $VAR^{\sigma,\kappa}_{restricted}$}} \\
\midrule
\textit{GNP} & 1.014 & 1.020 & 1.010 & 1.007 & 1.007 & 1.008 & 1.014 & 1.015 \\
\textit{GNP def} & 1.033 & 1.034 & 1.044$^{*}$ & 1.048$^{*}$ & 1.050$^{*}$ & 1.045$^{**}$ & 1.044$^{**}$ & 1.042$^{***}$ \\
\textit{SPREAD} & 1.017 & 1.042$^{***}$ & 1.061$^{***}$ & 1.073$^{***}$ & 1.088$^{***}$ & 1.089$^{***}$ & 1.108$^{***}$ & 1.129$^{***}$ \\
\midrule
\multicolumn{9}{c}{\textbf{vs. $VAR^{\sigma}$}} \\
\midrule
\textit{GNP} & \textbf{\textcolor{blue}{0.959$^{**}$}} & \textbf{\textcolor{blue}{0.916$^{***}$}} & \textbf{\textcolor{blue}{0.938$^{***}$}} & \textbf{\textcolor{blue}{0.942$^{***}$}} & \textbf{\textcolor{blue}{0.955$^{***}$}} & \textbf{\textcolor{blue}{0.955$^{***}$}} & \textbf{\textcolor{blue}{0.951$^{***}$}} & \textbf{\textcolor{blue}{0.955$^{***}$}} \\
\textit{GNP def} & \textbf{\textcolor{blue}{0.584$^{***}$}} & \textbf{\textcolor{blue}{0.626$^{***}$}} & \textbf{\textcolor{blue}{0.666$^{***}$}} & \textbf{\textcolor{blue}{0.698$^{***}$}} & \textbf{\textcolor{blue}{0.728$^{***}$}} & \textbf{\textcolor{blue}{0.749$^{***}$}} & \textbf{\textcolor{blue}{0.768$^{***}$}} & \textbf{\textcolor{blue}{0.784$^{***}$}} \\
\textit{SPREAD} & \textbf{\textcolor{blue}{0.834$^{***}$}} & \textbf{\textcolor{blue}{0.857$^{***}$}} & \textbf{\textcolor{blue}{0.869$^{***}$}} & \textbf{\textcolor{blue}{0.875$^{***}$}} & \textbf{\textcolor{blue}{0.900$^{***}$}} & \textbf{\textcolor{blue}{0.904$^{***}$}} & \textbf{\textcolor{blue}{0.913$^{***}$}} & \textbf{\textcolor{blue}{0.917$^{***}$}} \\
\bottomrule
\end{tabular}
\begin{tablenotes}
\scriptsize
\item \textbf{Notes:} Values shown are CRPS ratios ($VAR^{\sigma,\kappa}$ model/Competitor); values less than 1 (\textcolor{blue}{\textbf{bold blue}}) indicate the proposed model ($VAR^{\sigma,\kappa}$) outperforms the competitor. Panels A and B show left tail weighted CRPS for unconditional and conditional Giacomini-White tests. Panels C and D show right tail weighted CRPS for unconditional and conditional Giacomini-White tests. Significance levels from the tests: *** p$<$0.01, ** p$<$0.05, * p$<$0.1.
\end{tablenotes}
\end{threeparttable}
\end{table}

\subsection{US Forecast Performance Over Time}

\subsubsection{Baseline results with Conditional Giacomini-White test}

\begin{table}[H]
\centering
\caption{Forecast Performance by Period: Log Score and CRPS Comparison}
\label{tab:combined_period_comparison_conditional}
\begin{threeparttable}
\scriptsize
\begin{tabular}{@{}l*{6}{c}@{}}
\toprule
 & \multicolumn{6}{c}{\textbf{Variables}} \\
\cmidrule(lr){2-7}
\textbf{Period} & \multicolumn{2}{c}{\textbf{GNP}} & \multicolumn{2}{c}{\textbf{GNP Deflator}} & \multicolumn{2}{c}{\textbf{SPREAD}} \\
\cmidrule(lr){2-3} \cmidrule(lr){4-5} \cmidrule(lr){6-7}
 & \textbf{$VAR^{\sigma,\kappa}_{rest}$} & \textbf{$VAR^{\sigma}$} & \textbf{$VAR^{\sigma,\kappa}_{rest}$} & \textbf{$VAR^{\sigma}$} & \textbf{$VAR^{\sigma,\kappa}_{rest}$} & \textbf{$VAR^{\sigma}$} \\
\midrule
\multicolumn{7}{c}{\cellcolor{gray!15}\textbf{Panel A: Cumulative Weighted Both Tails Log Score Differences}} \\
\midrule
Full Sample & \textbf{\textcolor{blue}{1085.0}} & -4577.3 & \textbf{\textcolor{blue}{77.9}} & \textbf{\textcolor{blue}{130.6}}$^{***}$ & -2936.1$^{**}$ & -3236.5$^{***}$ \\
1975Q2--2023Q1 & (\textbf{\textcolor{blue}{5.7}}) & (-24.2) & (\textbf{\textcolor{blue}{0.4}}) & (\textbf{\textcolor{blue}{0.7}}) & (-15.5) & (-17.2) \\
\addlinespace
Great Inflation & \textbf{\textcolor{blue}{9.8}} & -28.3$^{*}$ & \textbf{\textcolor{blue}{69.0}}$^{**}$ & \textbf{\textcolor{blue}{108.1}}$^{***}$ & \textbf{\textcolor{blue}{108.0}}$^{***}$ & \textbf{\textcolor{blue}{440.2}}$^{*}$ \\
1975Q2--1984Q4 & (\textbf{\textcolor{blue}{0.3}}) & (-0.8) & (\textbf{\textcolor{blue}{2.0}}) & (\textbf{\textcolor{blue}{3.0}}) & (\textbf{\textcolor{blue}{3.0}}) & (\textbf{\textcolor{blue}{12.3}}) \\
\addlinespace
Volcker Disinflation & \textbf{\textcolor{blue}{4.9}}$^{***}$ & -16.2$^{***}$ & \textbf{\textcolor{blue}{1.9}}$^{***}$ & \textbf{\textcolor{blue}{10.9}}$^{**}$ & -24.4$^{***}$ & -51.6$^{***}$ \\
1981Q3--1983Q2 & (\textbf{\textcolor{blue}{0.6}}) & (-2.0) & (\textbf{\textcolor{blue}{0.2}}) & (\textbf{\textcolor{blue}{1.4}}) & (-3.1) & (-6.5) \\
\addlinespace
Great Moderation & -6.6$^{**}$ & \textbf{\textcolor{blue}{19.7}}$^{***}$ & -4.0$^{***}$ & \textbf{\textcolor{blue}{45.0}}$^{***}$ & -16.4$^{***}$ & \textbf{\textcolor{blue}{1114.3}}$^{***}$ \\
1985Q1--2007Q4 & (-0.1) & (\textbf{\textcolor{blue}{0.2}}) & (-0.0) & (\textbf{\textcolor{blue}{0.5}}) & (-0.2) & (\textbf{\textcolor{blue}{12.1}}) \\
\addlinespace
Housing Boom & \textbf{\textcolor{blue}{1.7}}$^{***}$ & \textbf{\textcolor{blue}{6.6}}$^{***}$ & -0.4$^{***}$ & \textbf{\textcolor{blue}{11.8}}$^{***}$ & \textbf{\textcolor{blue}{15.9}}$^{***}$ & \textbf{\textcolor{blue}{261.7}}$^{***}$ \\
2004Q1--2007Q4 & (\textbf{\textcolor{blue}{0.1}}) & (\textbf{\textcolor{blue}{0.4}}) & (-0.0) & (\textbf{\textcolor{blue}{0.7}}) & (\textbf{\textcolor{blue}{1.0}}) & (\textbf{\textcolor{blue}{16.4}}) \\
\addlinespace
GFC & -29.8 & -29.8 & -0.2$^{*}$ & \textbf{\textcolor{blue}{20.4}}$^{***}$ & \textbf{\textcolor{blue}{51.3}} & -1277.2 \\
2008Q1--2014Q4 & (-1.1) & (-1.1) & (-0.0) & (\textbf{\textcolor{blue}{0.7}}) & (\textbf{\textcolor{blue}{1.8}}) & (-45.6) \\
\addlinespace
COVID-19 & \textbf{\textcolor{blue}{1108.2}}$^{***}$ & -4530.3$^{***}$ & \textbf{\textcolor{blue}{1.5}}$^{**}$ & \textbf{\textcolor{blue}{6.1}}$^{**}$ & -8.2$^{***}$ & \textbf{\textcolor{blue}{4.8}}$^{***}$ \\
2020Q1--2021Q2 & (\textbf{\textcolor{blue}{221.6}}) & (-906.1) & (\textbf{\textcolor{blue}{0.3}}) & (\textbf{\textcolor{blue}{1.2}}) & (-1.6) & (\textbf{\textcolor{blue}{1.0}}) \\
\addlinespace
Since 2021Q3 & \textbf{\textcolor{blue}{7.3}}$^{**}$ & \textbf{\textcolor{blue}{1.8}}$^{*}$ & \textbf{\textcolor{blue}{10.3}}$^{*}$ & -60.7$^{***}$ & \textbf{\textcolor{blue}{1.2}}$^{***}$ & \textbf{\textcolor{blue}{58.4}}$^{***}$ \\
2021Q3--2024Q1 & (\textbf{\textcolor{blue}{1.1}}) & (\textbf{\textcolor{blue}{0.1}}) & (\textbf{\textcolor{blue}{0.4}}) & (-8.4) & (\textbf{\textcolor{blue}{0.1}}) & (\textbf{\textcolor{blue}{8.8}}) \\
\midrule
\multicolumn{7}{c}{\cellcolor{gray!15}\textbf{Panel B: Weighted CRPS Both Tails Ratios}} \\
\midrule
Full Sample & 1.012$^{*}$ & 1.052 & \textbf{\textcolor{blue}{0.924}} & \textbf{\textcolor{blue}{0.873}}$^{***}$ & 1.037$^{***}$ & \textbf{\textcolor{blue}{0.909}}$^{***}$ \\
1975Q2--2023Q1 &  & &  & &  & \\
\addlinespace
Great Inflation & \textbf{\textcolor{blue}{0.939}}$^{*}$ & 1.090$^{**}$ & \textbf{\textcolor{blue}{0.892}}$^{***}$ & \textbf{\textcolor{blue}{0.812}}$^{***}$ & \textbf{\textcolor{blue}{0.946}}$^{***}$ & \textbf{\textcolor{blue}{0.825}}$^{**}$ \\
1975Q2--1984Q4 &  & &  & &  & \\
\addlinespace
Volcker Disinflation & \textbf{\textcolor{blue}{0.962}}$^{***}$ & 1.188$^{***}$ & 1.031$^{***}$ & \textbf{\textcolor{blue}{0.913}}$^{***}$ & 1.096$^{***}$ & 1.053$^{***}$ \\
1981Q3--1983Q2 &  & &  & &  & \\
\addlinespace
Great Moderation & 1.045$^{*}$ & \textbf{\textcolor{blue}{0.914}}$^{***}$ & 1.053$^{**}$ & \textbf{\textcolor{blue}{0.583}}$^{***}$ & 1.039$^{**}$ & \textbf{\textcolor{blue}{0.631}}$^{***}$ \\
1985Q1--2007Q4 &  & &  & &  & \\
\addlinespace
Housing Boom & \textbf{\textcolor{blue}{0.936}}$^{***}$ & \textbf{\textcolor{blue}{0.809}}$^{***}$ & \textbf{\textcolor{blue}{0.992}}$^{***}$ & \textbf{\textcolor{blue}{0.431}}$^{***}$ & \textbf{\textcolor{blue}{0.879}}$^{***}$ & \textbf{\textcolor{blue}{0.398}}$^{***}$ \\
2004Q1--2007Q4 &  & &  & &  & \\
\addlinespace
GFC & 1.060 & 1.066 & 1.055 & \textbf{\textcolor{blue}{0.654}}$^{***}$ & 1.069 & 1.155$^{*}$ \\
2008Q1--2014Q4 &  & &  & &  & \\
\addlinespace
COVID-19 & 1.044$^{***}$ & 1.070$^{***}$ & \textbf{\textcolor{blue}{0.991}}$^{**}$ & \textbf{\textcolor{blue}{0.917}}$^{***}$ & 1.102$^{***}$ & \textbf{\textcolor{blue}{0.934}}$^{***}$ \\
2020Q1--2021Q2 &  & &  & &  & \\
\addlinespace
Since 2021Q3 & \textbf{\textcolor{blue}{0.851}}$^{***}$ & \textbf{\textcolor{blue}{0.986}}$^{***}$ & \textbf{\textcolor{blue}{0.996}}$^{*}$ & 1.257 & \textbf{\textcolor{blue}{0.942}}$^{***}$ & \textbf{\textcolor{blue}{0.451}}$^{***}$ \\
2021Q3--2024Q1 &  & &  & &  & \\
\bottomrule
\end{tabular}
\begin{tablenotes}
\tiny
\item \textbf{Notes:} Panel A shows cumulative weighted both tails log score differences ($VAR^{\sigma,\kappa}$ - Competitor) over specified periods. Values in parentheses show average score differences per quarter within each period. Positive values (\textcolor{blue}{\textbf{bold blue}}) indicate superior performance. Panel B shows weighted CRPS both tails ratios ($VAR^{\sigma,\kappa}$ WCRPS-LR / Competitor WCRPS-LR). Values less than 1 (\textcolor{blue}{\textbf{bold blue}}) indicate superior performance. All values are averaged across forecast horizons 1-8. Significance levels from Giacomini-White conditional test: *** p$<$0.01, ** p$<$0.05, * p$<$0.1.
\end{tablenotes}
\end{threeparttable}
\end{table}

\subsubsection{Additional results - standard LS, CRPS and other weighted tails}

\begin{table}[H]
\centering
\caption{RMSE Ratios by Period}
\label{tab:rmse_ratios_by_period_combined}
\begin{threeparttable}
\scriptsize
\begin{tabular}{@{}l*{6}{c}@{}}
\toprule
 & \multicolumn{6}{c}{\textbf{Variables}} \\
\cmidrule(lr){2-7}
\textbf{Period} & \multicolumn{2}{c}{\textbf{GNP}} & \multicolumn{2}{c}{\textbf{GNP Deflator}} & \multicolumn{2}{c}{\textbf{SPREAD}} \\
\cmidrule(lr){2-3} \cmidrule(lr){4-5} \cmidrule(lr){6-7}
 & \textbf{$VAR^{\sigma,\kappa}_{rest}$} & \textbf{$VAR^{\sigma}$} & \textbf{$VAR^{\sigma,\kappa}_{rest}$} & \textbf{$VAR^{\sigma}$} & \textbf{$VAR^{\sigma,\kappa}_{rest}$} & \textbf{$VAR^{\sigma}$} \\
\midrule
\multicolumn{7}{c}{\cellcolor{gray!15}\textbf{Panel A: GW Unconditional}} \\
\midrule
Full Sample & \textbf{\textcolor{blue}{0.995}} & \textbf{\textcolor{blue}{0.973}} & 1.062 & \textbf{\textcolor{blue}{0.971}} & 1.030 & 1.019$^{***}$ \\
1975Q2--2023Q1 &  & &  & &  & \\
\addlinespace
Great Inflation & \textbf{\textcolor{blue}{0.929}}$^{*}$ & \textbf{\textcolor{blue}{0.980}} & 1.052 & \textbf{\textcolor{blue}{0.894}} & \textbf{\textcolor{blue}{0.994}} & 1.042$^{***}$ \\
1975Q2--1984Q4 &  & &  & &  & \\
\addlinespace
Volcker Disinflation & \textbf{\textcolor{blue}{0.918}}$^{**}$ & 1.129$^{***}$ & 1.219$^{*}$ & 1.104 & 1.116$^{***}$ & 1.029$^{**}$ \\
1981Q3--1983Q2 &  & &  & &  & \\
\addlinespace
Great Moderation & 1.079$^{*}$ & \textbf{\textcolor{blue}{0.908}}$^{**}$ & 1.238$^{***}$ & 1.079$^{***}$ & 1.112 & \textbf{\textcolor{blue}{0.998}}$^{*}$ \\
1985Q1--2007Q4 &  & &  & &  & \\
\addlinespace
Housing Boom & \textbf{\textcolor{blue}{0.988}}$^{***}$ & \textbf{\textcolor{blue}{0.885}}$^{**}$ & \textbf{\textcolor{blue}{0.872}}$^{***}$ & 1.126$^{*}$ & \textbf{\textcolor{blue}{0.861}} & 1.038$^{**}$ \\
2004Q1--2007Q4 &  & &  & &  & \\
\addlinespace
GFC & 1.044$^{*}$ & \textbf{\textcolor{blue}{0.994}} & 1.042$^{*}$ & \textbf{\textcolor{blue}{0.915}}$^{**}$ & 1.014 & 1.023$^{**}$ \\
2008Q1--2014Q4 &  & &  & &  & \\
\addlinespace
COVID-19 & 1.006$^{***}$ & 1.013$^{***}$ & 1.006 & 1.030$^{***}$ & 1.052$^{***}$ & 1.034$^{***}$ \\
2020Q1--2021Q2 &  & &  & &  & \\
\addlinespace
Since 2021Q3 & \textbf{\textcolor{blue}{0.856}}$^{***}$ & \textbf{\textcolor{blue}{0.937}}$^{***}$ & 1.002$^{***}$ & 1.056$^{**}$ & \textbf{\textcolor{blue}{0.866}}$^{***}$ & 1.012 \\
2021Q3--2024Q1 &  & &  & &  & \\
\midrule
\multicolumn{7}{c}{\cellcolor{gray!15}\textbf{Panel B: GW Conditional}} \\
\midrule
Full Sample & \textbf{\textcolor{blue}{0.995}} & \textbf{\textcolor{blue}{0.973}} & 1.062$^{*}$ & \textbf{\textcolor{blue}{0.971}}$^{*}$ & 1.030$^{***}$ & 1.019$^{**}$ \\
1975Q2--2023Q1 &  & &  & &  & \\
\addlinespace
Great Inflation & \textbf{\textcolor{blue}{0.929}}$^{**}$ & \textbf{\textcolor{blue}{0.980}}$^{**}$ & 1.052$^{***}$ & \textbf{\textcolor{blue}{0.894}}$^{**}$ & \textbf{\textcolor{blue}{0.994}}$^{***}$ & 1.042$^{***}$ \\
1975Q2--1984Q4 &  & &  & &  & \\
\addlinespace
Volcker Disinflation & \textbf{\textcolor{blue}{0.918}}$^{**}$ & 1.129$^{***}$ & 1.219$^{***}$ & 1.104$^{***}$ & 1.116$^{***}$ & 1.029$^{***}$ \\
1981Q3--1983Q2 &  & &  & &  & \\
\addlinespace
Great Moderation & 1.079$^{**}$ & \textbf{\textcolor{blue}{0.908}} & 1.238$^{**}$ & 1.079$^{***}$ & 1.112$^{***}$ & \textbf{\textcolor{blue}{0.998}} \\
1985Q1--2007Q4 &  & &  & &  & \\
\addlinespace
Housing Boom & \textbf{\textcolor{blue}{0.988}}$^{***}$ & \textbf{\textcolor{blue}{0.885}}$^{**}$ & \textbf{\textcolor{blue}{0.872}}$^{***}$ & 1.126$^{*}$ & \textbf{\textcolor{blue}{0.861}}$^{**}$ & 1.038$^{**}$ \\
2004Q1--2007Q4 &  & &  & &  & \\
\addlinespace
GFC & 1.044 & \textbf{\textcolor{blue}{0.994}} & 1.042$^{**}$ & \textbf{\textcolor{blue}{0.915}}$^{**}$ & 1.014$^{***}$ & 1.023 \\
2008Q1--2014Q4 &  & &  & &  & \\
\addlinespace
COVID-19 & 1.006$^{***}$ & 1.013$^{***}$ & 1.006 & 1.030$^{***}$ & 1.052$^{***}$ & 1.034$^{***}$ \\
2020Q1--2021Q2 &  & &  & &  & \\
\addlinespace
Since 2021Q3 & \textbf{\textcolor{blue}{0.856}}$^{***}$ & \textbf{\textcolor{blue}{0.937}}$^{***}$ & 1.002$^{***}$ & 1.056$^{***}$ & \textbf{\textcolor{blue}{0.866}}$^{**}$ & 1.012$^{***}$ \\
2021Q3--2024Q1 &  & &  & &  & \\
\bottomrule
\end{tabular}
\begin{tablenotes}
\tiny
\item \textbf{Notes:} RMSE ratios ($VAR^{\sigma,\kappa}$ RMSE / Competitor RMSE) over specified periods. Values less than 1 (\textcolor{blue}{\textbf{bold blue}}) indicate the proposed model outperforms the competitor. Ratios are averaged across forecast horizons 1-8. Panel A: Significance levels from Giacomini-White unconditional test. Panel B: Significance levels from Giacomini-White conditional test. *** p$<$0.01, ** p$<$0.05, * p$<$0.1.
\end{tablenotes}
\end{threeparttable}
\end{table}

\begin{table}[H]
\centering
\caption{Cumulative Log Score Differences: GW Tests Comparison}
\label{tab:cumulative_logscore_gw_comparison}
\begin{threeparttable}
\scriptsize
\begin{tabular}{@{}l*{6}{c}@{}}
\toprule
 & \multicolumn{6}{c}{\textbf{Variables}} \\
\cmidrule(lr){2-7}
\textbf{Period} & \multicolumn{2}{c}{\textbf{GNP}} & \multicolumn{2}{c}{\textbf{GNP Deflator}} & \multicolumn{2}{c}{\textbf{SPREAD}} \\
\cmidrule(lr){2-3} \cmidrule(lr){4-5} \cmidrule(lr){6-7}
 & \textbf{$VAR^{\sigma,\kappa}_{rest}$} & \textbf{$VAR^{\sigma}$} & \textbf{$VAR^{\sigma,\kappa}_{rest}$} & \textbf{$VAR^{\sigma}$} & \textbf{$VAR^{\sigma,\kappa}_{rest}$} & \textbf{$VAR^{\sigma}$} \\
\midrule
\multicolumn{7}{c}{\cellcolor{gray!15}\textbf{Panel A: GW Unconditional Test}} \\
\midrule
Full Sample & \textbf{\textcolor{blue}{787.0}} & -2964.2$^{***}$ & -1247.4$^{***}$ & \textbf{\textcolor{blue}{12671.4}}$^{***}$ & -4130.0 & \textbf{\textcolor{blue}{3946.3}}$^{***}$ \\
1975Q2--2023Q1 & (\textbf{\textcolor{blue}{4.2}}) & (-15.7) & (-6.6) & (\textbf{\textcolor{blue}{67.4}}) & (-21.9) & (\textbf{\textcolor{blue}{21.1}}) \\
\addlinespace
Great Inflation & \textbf{\textcolor{blue}{31.8}}$^{**}$ & -177.9$^{*}$ & \textbf{\textcolor{blue}{80.3}} & \textbf{\textcolor{blue}{865.2}}$^{***}$ & \textbf{\textcolor{blue}{138.2}} & \textbf{\textcolor{blue}{1123.7}}$^{***}$ \\
1975Q2--1984Q4 & (\textbf{\textcolor{blue}{1.1}}) & (-5.0) & (\textbf{\textcolor{blue}{2.4}}) & (\textbf{\textcolor{blue}{23.6}}) & (\textbf{\textcolor{blue}{3.8}}) & (\textbf{\textcolor{blue}{31.2}}) \\
\addlinespace
Volcker Disinflation & \textbf{\textcolor{blue}{33.5}}$^{*}$ & -121.9$^{***}$ & -71.1 & \textbf{\textcolor{blue}{118.3}}$^{**}$ & -80.1$^{***}$ & -36.8$^{***}$ \\
1981Q3--1983Q2 & (\textbf{\textcolor{blue}{4.2}}) & (-15.2) & (-8.9) & (\textbf{\textcolor{blue}{14.8}}) & (-10.0) & (-4.6) \\
\addlinespace
Great Moderation & -403.0$^{***}$ & \textbf{\textcolor{blue}{1290.6}}$^{***}$ & -1359.5$^{***}$ & \textbf{\textcolor{blue}{7543.4}}$^{***}$ & -456.7 & \textbf{\textcolor{blue}{5966.7}}$^{***}$ \\
1985Q1--2007Q4 & (-4.4) & (\textbf{\textcolor{blue}{14.0}}) & (-14.8) & (\textbf{\textcolor{blue}{82.0}}) & (-5.0) & (\textbf{\textcolor{blue}{64.9}}) \\
\addlinespace
Housing Boom & \textbf{\textcolor{blue}{94.7}}$^{***}$ & \textbf{\textcolor{blue}{442.4}}$^{***}$ & -50.0$^{***}$ & \textbf{\textcolor{blue}{1620.4}}$^{***}$ & \textbf{\textcolor{blue}{112.8}} & \textbf{\textcolor{blue}{1564.7}}$^{***}$ \\
2004Q1--2007Q4 & (\textbf{\textcolor{blue}{5.9}}) & (\textbf{\textcolor{blue}{27.7}}) & (-3.1) & (\textbf{\textcolor{blue}{101.3}}) & (\textbf{\textcolor{blue}{7.0}}) & (\textbf{\textcolor{blue}{97.8}}) \\
\addlinespace
GFC & -127.7 & \textbf{\textcolor{blue}{88.1}}$^{***}$ & \textbf{\textcolor{blue}{18.1}} & \textbf{\textcolor{blue}{2427.5}}$^{***}$ & \textbf{\textcolor{blue}{164.7}} & -402.9$^{**}$ \\
2008Q1--2014Q4 & (-4.6) & (\textbf{\textcolor{blue}{3.1}}) & (\textbf{\textcolor{blue}{0.6}}) & (\textbf{\textcolor{blue}{86.7}}) & (\textbf{\textcolor{blue}{5.9}}) & (-14.4) \\
\addlinespace
COVID-19 & \textbf{\textcolor{blue}{1114.0}}$^{***}$ & -4625.0$^{***}$ & \textbf{\textcolor{blue}{16.4}} & \textbf{\textcolor{blue}{207.1}}$^{***}$ & -53.7$^{***}$ & \textbf{\textcolor{blue}{146.8}}$^{***}$ \\
2020Q1--2021Q2 & (\textbf{\textcolor{blue}{222.8}}) & (-925.0) & (\textbf{\textcolor{blue}{3.3}}) & (\textbf{\textcolor{blue}{41.4}}) & (-10.7) & (\textbf{\textcolor{blue}{29.4}}) \\
\addlinespace
Since 2021Q3 & \textbf{\textcolor{blue}{103.0}}$^{***}$ & \textbf{\textcolor{blue}{13.7}}$^{***}$ & \textbf{\textcolor{blue}{28.3}}$^{***}$ & -127.3 & \textbf{\textcolor{blue}{10.2}}$^{**}$ & \textbf{\textcolor{blue}{544.9}}$^{***}$ \\
2021Q3--2024Q1 & (\textbf{\textcolor{blue}{14.7}}) & (\textbf{\textcolor{blue}{0.5}}) & (-0.1) & (-16.5) & (\textbf{\textcolor{blue}{0.3}}) & (\textbf{\textcolor{blue}{83.9}}) \\
\midrule
\multicolumn{7}{c}{\cellcolor{gray!15}\textbf{Panel B: GW Conditional Test}} \\
\midrule
Full Sample & \textbf{\textcolor{blue}{787.0}}$^{***}$ & -2964.2$^{***}$ & -1247.4$^{***}$ & \textbf{\textcolor{blue}{12671.4}}$^{***}$ & -4130.0$^{***}$ & \textbf{\textcolor{blue}{3946.3}}$^{***}$ \\
1975Q2--2023Q1 & (\textbf{\textcolor{blue}{4.2}}) & (-15.7) & (-6.6) & (\textbf{\textcolor{blue}{67.4}}) & (-21.9) & (\textbf{\textcolor{blue}{21.1}}) \\
\addlinespace
Great Inflation & \textbf{\textcolor{blue}{31.8}}$^{**}$ & -177.9$^{***}$ & \textbf{\textcolor{blue}{80.3}}$^{***}$ & \textbf{\textcolor{blue}{865.2}}$^{***}$ & \textbf{\textcolor{blue}{138.2}}$^{***}$ & \textbf{\textcolor{blue}{1123.7}}$^{***}$ \\
1975Q2--1984Q4 & (\textbf{\textcolor{blue}{1.1}}) & (-5.0) & (\textbf{\textcolor{blue}{2.4}}) & (\textbf{\textcolor{blue}{23.6}}) & (\textbf{\textcolor{blue}{3.8}}) & (\textbf{\textcolor{blue}{31.2}}) \\
\addlinespace
Volcker Disinflation & \textbf{\textcolor{blue}{33.5}}$^{**}$ & -121.9$^{***}$ & -71.1$^{***}$ & \textbf{\textcolor{blue}{118.3}}$^{***}$ & -80.1$^{***}$ & -36.8$^{***}$ \\
1981Q3--1983Q2 & (\textbf{\textcolor{blue}{4.2}}) & (-15.2) & (-8.9) & (\textbf{\textcolor{blue}{14.8}}) & (-10.0) & (-4.6) \\
\addlinespace
Great Moderation & -403.0$^{***}$ & \textbf{\textcolor{blue}{1290.6}}$^{***}$ & -1359.5$^{***}$ & \textbf{\textcolor{blue}{7543.4}}$^{***}$ & -456.7$^{***}$ & \textbf{\textcolor{blue}{5966.7}}$^{***}$ \\
1985Q1--2007Q4 & (-4.4) & (\textbf{\textcolor{blue}{14.0}}) & (-14.8) & (\textbf{\textcolor{blue}{82.0}}) & (-5.0) & (\textbf{\textcolor{blue}{64.9}}) \\
\addlinespace
Housing Boom & \textbf{\textcolor{blue}{94.7}}$^{***}$ & \textbf{\textcolor{blue}{442.4}}$^{***}$ & -50.0$^{***}$ & \textbf{\textcolor{blue}{1620.4}}$^{***}$ & \textbf{\textcolor{blue}{112.8}}$^{***}$ & \textbf{\textcolor{blue}{1564.7}}$^{***}$ \\
2004Q1--2007Q4 & (\textbf{\textcolor{blue}{5.9}}) & (\textbf{\textcolor{blue}{27.7}}) & (-3.1) & (\textbf{\textcolor{blue}{101.3}}) & (\textbf{\textcolor{blue}{7.0}}) & (\textbf{\textcolor{blue}{97.8}}) \\
\addlinespace
GFC & -127.7$^{*}$ & \textbf{\textcolor{blue}{88.1}}$^{**}$ & \textbf{\textcolor{blue}{18.1}} & \textbf{\textcolor{blue}{2427.5}}$^{***}$ & \textbf{\textcolor{blue}{164.7}}$^{*}$ & -402.9$^{***}$ \\
2008Q1--2014Q4 & (-4.6) & (\textbf{\textcolor{blue}{3.1}}) & (\textbf{\textcolor{blue}{0.6}}) & (\textbf{\textcolor{blue}{86.7}}) & (\textbf{\textcolor{blue}{5.9}}) & (-14.4) \\
\addlinespace
COVID-19 & \textbf{\textcolor{blue}{1114.0}}$^{***}$ & -4625.0$^{***}$ & \textbf{\textcolor{blue}{16.4}}$^{***}$ & \textbf{\textcolor{blue}{207.1}}$^{***}$ & -53.7$^{***}$ & \textbf{\textcolor{blue}{146.8}}$^{***}$ \\
2020Q1--2021Q2 & (\textbf{\textcolor{blue}{222.8}}) & (-925.0) & (\textbf{\textcolor{blue}{3.3}}) & (\textbf{\textcolor{blue}{41.4}}) & (-10.7) & (\textbf{\textcolor{blue}{29.4}}) \\
\addlinespace
Since 2021Q3 & \textbf{\textcolor{blue}{103.0}}$^{***}$ & \textbf{\textcolor{blue}{13.7}}$^{***}$ & \textbf{\textcolor{blue}{28.3}}$^{***}$ & -127.3$^{***}$ & \textbf{\textcolor{blue}{10.2}}$^{**}$ & \textbf{\textcolor{blue}{544.9}}$^{***}$ \\
2021Q3--2024Q1 & (\textbf{\textcolor{blue}{14.7}}) & (\textbf{\textcolor{blue}{0.5}}) & (-0.1) & (-16.5) & (\textbf{\textcolor{blue}{0.3}}) & (\textbf{\textcolor{blue}{83.9}}) \\
\bottomrule
\end{tabular}
\begin{tablenotes}
\tiny
\item \textbf{Notes:} Panel A shows cumulative log score differences using Giacomini-White unconditional test. Panel B shows cumulative log score differences using Giacomini-White conditional test. Values represent ($VAR^{\sigma,\kappa}$ model - Competitor) over specified periods. Values in parentheses show average score differences per quarter within each period. Positive values (\textcolor{blue}{\textbf{bold blue}}) indicate the proposed model outperforms the competitor. All values are averaged across forecast horizons 1-8. Significance levels: *** p$<$0.01, ** p$<$0.05, * p$<$0.1.
\end{tablenotes}
\end{threeparttable}
\end{table}

\begin{table}[H]
\centering
\caption{Cumulative Weighted Left Tail Log Score Differences}
\label{tab:cumulative_weighted_left_tail_gw_comparison}
\begin{threeparttable}
\scriptsize
\begin{tabular}{@{}l*{6}{c}@{}}
\toprule
 & \multicolumn{6}{c}{\textbf{Variables}} \\
\cmidrule(lr){2-7}
\textbf{Period} & \multicolumn{2}{c}{\textbf{GNP}} & \multicolumn{2}{c}{\textbf{GNP Deflator}} & \multicolumn{2}{c}{\textbf{SPREAD}} \\
\cmidrule(lr){2-3} \cmidrule(lr){4-5} \cmidrule(lr){6-7}
 & \textbf{$VAR^{\sigma,\kappa}_{rest}$} & \textbf{$VAR^{\sigma}$} & \textbf{$VAR^{\sigma,\kappa}_{rest}$} & \textbf{$VAR^{\sigma}$} & \textbf{$VAR^{\sigma,\kappa}_{rest}$} & \textbf{$VAR^{\sigma}$} \\
\midrule
\multicolumn{7}{c}{\cellcolor{gray!15}\textbf{Panel A: GW Unconditional Test}} \\
\midrule
Full Sample & \textbf{\textcolor{blue}{861.0}} & -3637.0$^{***}$ & -722.5$^{***}$ & \textbf{\textcolor{blue}{6284.9}}$^{***}$ & \textbf{\textcolor{blue}{181.3}} & \textbf{\textcolor{blue}{6760.7}}$^{***}$ \\
1975Q2--2023Q1 & (\textbf{\textcolor{blue}{4.5}}) & (-19.2) & (-3.8) & (\textbf{\textcolor{blue}{33.4}}) & (\textbf{\textcolor{blue}{1.0}}) & (\textbf{\textcolor{blue}{36.0}}) \\
\addlinespace
Great Inflation & \textbf{\textcolor{blue}{33.0}}$^{**}$ & -113.1$^{**}$ & -24.3 & \textbf{\textcolor{blue}{283.5}}$^{***}$ & \textbf{\textcolor{blue}{216.6}}$^{*}$ & \textbf{\textcolor{blue}{1062.3}}$^{***}$ \\
1975Q2--1984Q4 & (\textbf{\textcolor{blue}{1.0}}) & (-3.2) & (-0.6) & (\textbf{\textcolor{blue}{7.7}}) & (\textbf{\textcolor{blue}{6.1}}) & (\textbf{\textcolor{blue}{29.6}}) \\
\addlinespace
Volcker Disinflation & \textbf{\textcolor{blue}{22.1}} & -71.4$^{***}$ & -33.6 & \textbf{\textcolor{blue}{41.4}}$^{**}$ & -4.2$^{***}$ & \textbf{\textcolor{blue}{36.3}}$^{***}$ \\
1981Q3--1983Q2 & (\textbf{\textcolor{blue}{2.8}}) & (-8.9) & (-4.2) & (\textbf{\textcolor{blue}{5.2}}) & (-0.5) & (\textbf{\textcolor{blue}{4.5}}) \\
\addlinespace
Great Moderation & -210.9$^{***}$ & \textbf{\textcolor{blue}{638.8}}$^{***}$ & -680.1$^{***}$ & \textbf{\textcolor{blue}{3686.7}}$^{***}$ & -10.7 & \textbf{\textcolor{blue}{4273.1}}$^{***}$ \\
1985Q1--2007Q4 & (-2.3) & (\textbf{\textcolor{blue}{6.9}}) & (-7.4) & (\textbf{\textcolor{blue}{40.1}}) & (-0.1) & (\textbf{\textcolor{blue}{46.4}}) \\
\addlinespace
Housing Boom & \textbf{\textcolor{blue}{50.7}}$^{***}$ & \textbf{\textcolor{blue}{230.2}}$^{***}$ & -21.6$^{***}$ & \textbf{\textcolor{blue}{752.5}}$^{***}$ & \textbf{\textcolor{blue}{82.5}} & \textbf{\textcolor{blue}{1134.5}}$^{***}$ \\
2004Q1--2007Q4 & (\textbf{\textcolor{blue}{3.2}}) & (\textbf{\textcolor{blue}{14.4}}) & (-1.4) & (\textbf{\textcolor{blue}{47.0}}) & (\textbf{\textcolor{blue}{5.2}}) & (\textbf{\textcolor{blue}{70.9}}) \\
\addlinespace
GFC & -91.2 & \textbf{\textcolor{blue}{18.9}}$^{***}$ & \textbf{\textcolor{blue}{6.9}} & \textbf{\textcolor{blue}{1286.1}}$^{***}$ & \textbf{\textcolor{blue}{71.0}}$^{**}$ & \textbf{\textcolor{blue}{415.8}}$^{***}$ \\
2008Q1--2014Q4 & (-3.3) & (\textbf{\textcolor{blue}{0.7}}) & (\textbf{\textcolor{blue}{0.2}}) & (\textbf{\textcolor{blue}{45.9}}) & (\textbf{\textcolor{blue}{2.5}}) & (\textbf{\textcolor{blue}{14.9}}) \\
\addlinespace
COVID-19 & \textbf{\textcolor{blue}{1040.4}}$^{***}$ & -4413.6$^{*}$ & -3.1 & \textbf{\textcolor{blue}{93.1}}$^{***}$ & -10.1$^{***}$ & \textbf{\textcolor{blue}{113.1}}$^{***}$ \\
2020Q1--2021Q2 & (\textbf{\textcolor{blue}{208.1}}) & (-882.7) & (-0.6) & (\textbf{\textcolor{blue}{18.6}}) & (-2.0) & (\textbf{\textcolor{blue}{22.6}}) \\
\addlinespace
Since 2021Q3 & \textbf{\textcolor{blue}{57.2}}$^{***}$ & \textbf{\textcolor{blue}{9.0}}$^{***}$ & \textbf{\textcolor{blue}{4.2}}$^{***}$ & -1.7 & \textbf{\textcolor{blue}{7.1}}$^{**}$ & \textbf{\textcolor{blue}{362.6}}$^{***}$ \\
2021Q3--2024Q1 & (\textbf{\textcolor{blue}{8.1}}) & (\textbf{\textcolor{blue}{0.5}}) & (-0.3) & (\textbf{\textcolor{blue}{0.1}}) & (\textbf{\textcolor{blue}{0.3}}) & (\textbf{\textcolor{blue}{55.9}}) \\
\midrule
\multicolumn{7}{c}{\cellcolor{gray!15}\textbf{Panel B: GW Conditional Test}} \\
\midrule
Full Sample & \textbf{\textcolor{blue}{861.0}}$^{**}$ & -3637.0$^{***}$ & -722.5$^{***}$ & \textbf{\textcolor{blue}{6284.9}}$^{***}$ & \textbf{\textcolor{blue}{181.3}}$^{***}$ & \textbf{\textcolor{blue}{6760.7}}$^{***}$ \\
1975Q2--2023Q1 & (\textbf{\textcolor{blue}{4.5}}) & (-19.2) & (-3.8) & (\textbf{\textcolor{blue}{33.4}}) & (\textbf{\textcolor{blue}{1.0}}) & (\textbf{\textcolor{blue}{36.0}}) \\
\addlinespace
Great Inflation & \textbf{\textcolor{blue}{33.0}}$^{**}$ & -113.1$^{**}$ & -24.3$^{***}$ & \textbf{\textcolor{blue}{283.5}}$^{***}$ & \textbf{\textcolor{blue}{216.6}}$^{***}$ & \textbf{\textcolor{blue}{1062.3}}$^{***}$ \\
1975Q2--1984Q4 & (\textbf{\textcolor{blue}{1.0}}) & (-3.2) & (-0.6) & (\textbf{\textcolor{blue}{7.7}}) & (\textbf{\textcolor{blue}{6.1}}) & (\textbf{\textcolor{blue}{29.6}}) \\
\addlinespace
Volcker Disinflation & \textbf{\textcolor{blue}{22.1}}$^{***}$ & -71.4$^{***}$ & -33.6$^{**}$ & \textbf{\textcolor{blue}{41.4}}$^{***}$ & -4.2$^{***}$ & \textbf{\textcolor{blue}{36.3}}$^{***}$ \\
1981Q3--1983Q2 & (\textbf{\textcolor{blue}{2.8}}) & (-8.9) & (-4.2) & (\textbf{\textcolor{blue}{5.2}}) & (-0.5) & (\textbf{\textcolor{blue}{4.5}}) \\
\addlinespace
Great Moderation & -210.9$^{***}$ & \textbf{\textcolor{blue}{638.8}}$^{***}$ & -680.1$^{***}$ & \textbf{\textcolor{blue}{3686.7}}$^{***}$ & -10.7$^{***}$ & \textbf{\textcolor{blue}{4273.1}}$^{***}$ \\
1985Q1--2007Q4 & (-2.3) & (\textbf{\textcolor{blue}{6.9}}) & (-7.4) & (\textbf{\textcolor{blue}{40.1}}) & (-0.1) & (\textbf{\textcolor{blue}{46.4}}) \\
\addlinespace
Housing Boom & \textbf{\textcolor{blue}{50.7}}$^{***}$ & \textbf{\textcolor{blue}{230.2}}$^{***}$ & -21.6$^{***}$ & \textbf{\textcolor{blue}{752.5}}$^{***}$ & \textbf{\textcolor{blue}{82.5}}$^{***}$ & \textbf{\textcolor{blue}{1134.5}}$^{***}$ \\
2004Q1--2007Q4 & (\textbf{\textcolor{blue}{3.2}}) & (\textbf{\textcolor{blue}{14.4}}) & (-1.4) & (\textbf{\textcolor{blue}{47.0}}) & (\textbf{\textcolor{blue}{5.2}}) & (\textbf{\textcolor{blue}{70.9}}) \\
\addlinespace
GFC & -91.2 & \textbf{\textcolor{blue}{18.9}}$^{**}$ & \textbf{\textcolor{blue}{6.9}}$^{*}$ & \textbf{\textcolor{blue}{1286.1}}$^{***}$ & \textbf{\textcolor{blue}{71.0}}$^{**}$ & \textbf{\textcolor{blue}{415.8}}$^{***}$ \\
2008Q1--2014Q4 & (-3.3) & (\textbf{\textcolor{blue}{0.7}}) & (\textbf{\textcolor{blue}{0.2}}) & (\textbf{\textcolor{blue}{45.9}}) & (\textbf{\textcolor{blue}{2.5}}) & (\textbf{\textcolor{blue}{14.9}}) \\
\addlinespace
COVID-19 & \textbf{\textcolor{blue}{1040.4}}$^{***}$ & -4413.6$^{***}$ & -3.1$^{***}$ & \textbf{\textcolor{blue}{93.1}}$^{***}$ & -10.1$^{***}$ & \textbf{\textcolor{blue}{113.1}}$^{***}$ \\
2020Q1--2021Q2 & (\textbf{\textcolor{blue}{208.1}}) & (-882.7) & (-0.6) & (\textbf{\textcolor{blue}{18.6}}) & (-2.0) & (\textbf{\textcolor{blue}{22.6}}) \\
\addlinespace
Since 2021Q3 & \textbf{\textcolor{blue}{57.2}}$^{***}$ & \textbf{\textcolor{blue}{9.0}}$^{***}$ & \textbf{\textcolor{blue}{4.2}}$^{***}$ & -1.7$^{***}$ & \textbf{\textcolor{blue}{7.1}}$^{**}$ & \textbf{\textcolor{blue}{362.6}}$^{***}$ \\
2021Q3--2024Q1 & (\textbf{\textcolor{blue}{8.1}}) & (\textbf{\textcolor{blue}{0.5}}) & (-0.3) & (\textbf{\textcolor{blue}{0.1}}) & (\textbf{\textcolor{blue}{0.3}}) & (\textbf{\textcolor{blue}{55.9}}) \\
\bottomrule
\end{tabular}
\begin{tablenotes}
\tiny
\item \textbf{Notes:} Panel A shows cumulative weighted left tail log score differences using Giacomini-White unconditional test. Panel B shows cumulative weighted left tail log score differences using Giacomini-White conditional test. Values represent ($VAR^{\sigma,\kappa}$ model - Competitor) over specified periods. Values in parentheses show average score differences per quarter within each period. Positive values (\textcolor{blue}{\textbf{bold blue}}) indicate the proposed model outperforms the competitor on left tail forecasting. All values are averaged across forecast horizons 1-8. Significance levels: *** p$<$0.01, ** p$<$0.05, * p$<$0.1.
\end{tablenotes}
\end{threeparttable}
\end{table}

\begin{table}[H]
\centering
\caption{Cumulative Weighted Right Tail Log Score Differences}
\label{tab:cumulative_weighted_right_tail_gw_comparison}
\begin{threeparttable}
\scriptsize
\begin{tabular}{@{}l*{6}{c}@{}}
\toprule
 & \multicolumn{6}{c}{\textbf{Variables}} \\
\cmidrule(lr){2-7}
\textbf{Period} & \multicolumn{2}{c}{\textbf{GNP}} & \multicolumn{2}{c}{\textbf{GNP Deflator}} & \multicolumn{2}{c}{\textbf{SPREAD}} \\
\cmidrule(lr){2-3} \cmidrule(lr){4-5} \cmidrule(lr){6-7}
 & \textbf{$VAR^{\sigma,\kappa}_{rest}$} & \textbf{$VAR^{\sigma}$} & \textbf{$VAR^{\sigma,\kappa}_{rest}$} & \textbf{$VAR^{\sigma}$} & \textbf{$VAR^{\sigma,\kappa}_{rest}$} & \textbf{$VAR^{\sigma}$} \\
\midrule
\multicolumn{7}{c}{\cellcolor{gray!15}\textbf{Panel A: GW Unconditional Test}} \\
\midrule
Full Sample & -74.0$^{**}$ & \textbf{\textcolor{blue}{672.8}}$^{***}$ & -524.9$^{***}$ & \textbf{\textcolor{blue}{6386.6}}$^{***}$ & -4311.3$^{**}$ & -2814.4$^{***}$ \\
1975Q2--2023Q1 & (-0.4) & (\textbf{\textcolor{blue}{3.6}}) & (-2.8) & (\textbf{\textcolor{blue}{34.0}}) & (-22.8) & (-14.9) \\
\addlinespace
Great Inflation & -1.2$^{*}$ & -64.8$^{*}$ & \textbf{\textcolor{blue}{104.6}} & \textbf{\textcolor{blue}{581.8}}$^{***}$ & -78.4 & \textbf{\textcolor{blue}{61.4}} \\
1975Q2--1984Q4 & (\textbf{\textcolor{blue}{0.1}}) & (-1.8) & (\textbf{\textcolor{blue}{3.1}}) & (\textbf{\textcolor{blue}{15.9}}) & (-2.3) & (\textbf{\textcolor{blue}{1.6}}) \\
\addlinespace
Volcker Disinflation & \textbf{\textcolor{blue}{11.3}}$^{**}$ & -50.5$^{***}$ & -37.5 & \textbf{\textcolor{blue}{77.0}}$^{**}$ & -75.9$^{***}$ & -73.1$^{***}$ \\
1981Q3--1983Q2 & (\textbf{\textcolor{blue}{1.4}}) & (-6.3) & (-4.7) & (\textbf{\textcolor{blue}{9.6}}) & (-9.5) & (-9.1) \\
\addlinespace
Great Moderation & -192.1$^{***}$ & \textbf{\textcolor{blue}{651.8}}$^{***}$ & -679.5$^{***}$ & \textbf{\textcolor{blue}{3856.7}}$^{***}$ & -445.9$^{**}$ & \textbf{\textcolor{blue}{1693.6}}$^{***}$ \\
1985Q1--2007Q4 & (-2.1) & (\textbf{\textcolor{blue}{7.1}}) & (-7.4) & (\textbf{\textcolor{blue}{41.9}}) & (-4.8) & (\textbf{\textcolor{blue}{18.4}}) \\
\addlinespace
Housing Boom & \textbf{\textcolor{blue}{44.0}}$^{***}$ & \textbf{\textcolor{blue}{212.2}}$^{***}$ & -28.3$^{***}$ & \textbf{\textcolor{blue}{867.9}}$^{***}$ & \textbf{\textcolor{blue}{30.3}} & \textbf{\textcolor{blue}{430.2}}$^{***}$ \\
2004Q1--2007Q4 & (\textbf{\textcolor{blue}{2.8}}) & (\textbf{\textcolor{blue}{13.3}}) & (-1.8) & (\textbf{\textcolor{blue}{54.2}}) & (\textbf{\textcolor{blue}{1.9}}) & (\textbf{\textcolor{blue}{26.9}}) \\
\addlinespace
GFC & -36.5 & \textbf{\textcolor{blue}{69.2}}$^{**}$ & \textbf{\textcolor{blue}{11.2}} & \textbf{\textcolor{blue}{1141.3}}$^{***}$ & \textbf{\textcolor{blue}{93.7}} & -818.7$^{**}$ \\
2008Q1--2014Q4 & (-1.3) & (\textbf{\textcolor{blue}{2.5}}) & (\textbf{\textcolor{blue}{0.4}}) & (\textbf{\textcolor{blue}{40.8}}) & (\textbf{\textcolor{blue}{3.3}}) & (-29.2) \\
\addlinespace
COVID-19 & \textbf{\textcolor{blue}{73.6}}$^{***}$ & -211.4$^{***}$ & \textbf{\textcolor{blue}{19.5}} & \textbf{\textcolor{blue}{114.1}}$^{***}$ & -43.7$^{**}$ & \textbf{\textcolor{blue}{33.8}}$^{***}$ \\
2020Q1--2021Q2 & (\textbf{\textcolor{blue}{14.7}}) & (-42.3) & (\textbf{\textcolor{blue}{3.9}}) & (\textbf{\textcolor{blue}{22.8}}) & (-8.7) & (\textbf{\textcolor{blue}{6.8}}) \\
\addlinespace
Since 2021Q3 & \textbf{\textcolor{blue}{45.8}}$^{***}$ & \textbf{\textcolor{blue}{4.8}}$^{***}$ & \textbf{\textcolor{blue}{24.1}}$^{***}$ & -125.7 & \textbf{\textcolor{blue}{3.1}}$^{**}$ & \textbf{\textcolor{blue}{182.3}}$^{***}$ \\
2021Q3--2024Q1 & (\textbf{\textcolor{blue}{6.6}}) & (-0.1) & (\textbf{\textcolor{blue}{0.3}}) & (-16.6) & (-0.0) & (\textbf{\textcolor{blue}{28.0}}) \\
\midrule
\multicolumn{7}{c}{\cellcolor{gray!15}\textbf{Panel B: GW Conditional Test}} \\
\midrule
Full Sample & -74.0$^{***}$ & \textbf{\textcolor{blue}{672.8}}$^{***}$ & -524.9$^{***}$ & \textbf{\textcolor{blue}{6386.6}}$^{***}$ & -4311.3$^{***}$ & -2814.4$^{***}$ \\
1975Q2--2023Q1 & (-0.4) & (\textbf{\textcolor{blue}{3.6}}) & (-2.8) & (\textbf{\textcolor{blue}{34.0}}) & (-22.8) & (-14.9) \\
\addlinespace
Great Inflation & -1.2$^{***}$ & -64.8$^{***}$ & \textbf{\textcolor{blue}{104.6}}$^{***}$ & \textbf{\textcolor{blue}{581.8}}$^{***}$ & -78.4$^{***}$ & \textbf{\textcolor{blue}{61.4}}$^{***}$ \\
1975Q2--1984Q4 & (\textbf{\textcolor{blue}{0.1}}) & (-1.8) & (\textbf{\textcolor{blue}{3.1}}) & (\textbf{\textcolor{blue}{15.9}}) & (-2.3) & (\textbf{\textcolor{blue}{1.6}}) \\
\addlinespace
Volcker Disinflation & \textbf{\textcolor{blue}{11.3}}$^{***}$ & -50.5$^{***}$ & -37.5$^{***}$ & \textbf{\textcolor{blue}{77.0}}$^{***}$ & -75.9$^{***}$ & -73.1$^{***}$ \\
1981Q3--1983Q2 & (\textbf{\textcolor{blue}{1.4}}) & (-6.3) & (-4.7) & (\textbf{\textcolor{blue}{9.6}}) & (-9.5) & (-9.1) \\
\addlinespace
Great Moderation & -192.1$^{***}$ & \textbf{\textcolor{blue}{651.8}}$^{***}$ & -679.5$^{***}$ & \textbf{\textcolor{blue}{3856.7}}$^{***}$ & -445.9$^{***}$ & \textbf{\textcolor{blue}{1693.6}}$^{***}$ \\
1985Q1--2007Q4 & (-2.1) & (\textbf{\textcolor{blue}{7.1}}) & (-7.4) & (\textbf{\textcolor{blue}{41.9}}) & (-4.8) & (\textbf{\textcolor{blue}{18.4}}) \\
\addlinespace
Housing Boom & \textbf{\textcolor{blue}{44.0}}$^{***}$ & \textbf{\textcolor{blue}{212.2}}$^{***}$ & -28.3$^{***}$ & \textbf{\textcolor{blue}{867.9}}$^{***}$ & \textbf{\textcolor{blue}{30.3}}$^{**}$ & \textbf{\textcolor{blue}{430.2}}$^{***}$ \\
2004Q1--2007Q4 & (\textbf{\textcolor{blue}{2.8}}) & (\textbf{\textcolor{blue}{13.3}}) & (-1.8) & (\textbf{\textcolor{blue}{54.2}}) & (\textbf{\textcolor{blue}{1.9}}) & (\textbf{\textcolor{blue}{26.9}}) \\
\addlinespace
GFC & -36.5$^{*}$ & \textbf{\textcolor{blue}{69.2}}$^{**}$ & \textbf{\textcolor{blue}{11.2}} & \textbf{\textcolor{blue}{1141.3}}$^{***}$ & \textbf{\textcolor{blue}{93.7}} & -818.7$^{***}$ \\
2008Q1--2014Q4 & (-1.3) & (\textbf{\textcolor{blue}{2.5}}) & (\textbf{\textcolor{blue}{0.4}}) & (\textbf{\textcolor{blue}{40.8}}) & (\textbf{\textcolor{blue}{3.3}}) & (-29.2) \\
\addlinespace
COVID-19 & \textbf{\textcolor{blue}{73.6}}$^{***}$ & -211.4$^{***}$ & \textbf{\textcolor{blue}{19.5}}$^{***}$ & \textbf{\textcolor{blue}{114.1}}$^{***}$ & -43.7$^{**}$ & \textbf{\textcolor{blue}{33.8}}$^{***}$ \\
2020Q1--2021Q2 & (\textbf{\textcolor{blue}{14.7}}) & (-42.3) & (\textbf{\textcolor{blue}{3.9}}) & (\textbf{\textcolor{blue}{22.8}}) & (-8.7) & (\textbf{\textcolor{blue}{6.8}}) \\
\addlinespace
Since 2021Q3 & \textbf{\textcolor{blue}{45.8}}$^{***}$ & \textbf{\textcolor{blue}{4.8}}$^{***}$ & \textbf{\textcolor{blue}{24.1}}$^{***}$ & -125.7$^{***}$ & \textbf{\textcolor{blue}{3.1}}$^{**}$ & \textbf{\textcolor{blue}{182.3}}$^{***}$ \\
2021Q3--2024Q1 & (\textbf{\textcolor{blue}{6.6}}) & (-0.1) & (\textbf{\textcolor{blue}{0.3}}) & (-16.6) & (-0.0) & (\textbf{\textcolor{blue}{28.0}}) \\
\bottomrule
\end{tabular}
\begin{tablenotes}
\tiny
\item \textbf{Notes:} Panel A shows cumulative weighted right tail log score differences using Giacomini-White unconditional test. Panel B shows cumulative weighted right tail log score differences using Giacomini-White conditional test. Values represent ($VAR^{\sigma,\kappa}$ model - Competitor) over specified periods. Values in parentheses show average score differences per quarter within each period. Positive values (\textcolor{blue}{\textbf{bold blue}}) indicate the proposed model outperforms the competitor on right tail forecasting. All values are averaged across forecast horizons 1-8. Significance levels: *** p$<$0.01, ** p$<$0.05, * p$<$0.1.
\end{tablenotes}
\end{threeparttable}
\end{table}

\begin{table}[H]
\centering
\caption{CRPS Ratios by Period}
\label{tab:crps_ratios_gw_comparison}
\begin{threeparttable}
\scriptsize
\begin{tabular}{@{}l*{6}{c}@{}}
\toprule
 & \multicolumn{6}{c}{\textbf{Variables}} \\
\cmidrule(lr){2-7}
\textbf{Period} & \multicolumn{2}{c}{\textbf{GNP}} & \multicolumn{2}{c}{\textbf{GNP Deflator}} & \multicolumn{2}{c}{\textbf{SPREAD}} \\
\cmidrule(lr){2-3} \cmidrule(lr){4-5} \cmidrule(lr){6-7}
 & \textbf{$VAR^{\sigma,\kappa}_{rest}$} & \textbf{$VAR^{\sigma}$} & \textbf{$VAR^{\sigma,\kappa}_{rest}$} & \textbf{$VAR^{\sigma}$} & \textbf{$VAR^{\sigma,\kappa}_{rest}$} & \textbf{$VAR^{\sigma}$} \\
\midrule
\multicolumn{7}{c}{\cellcolor{gray!15}\textbf{Panel A: GW Unconditional Test}} \\
\midrule
Full Sample & 1.014$^{*}$ & \textbf{\textcolor{blue}{0.956}}$^{***}$ & 1.068$^{**}$ & \textbf{\textcolor{blue}{0.675}}$^{***}$ & 1.027 & \textbf{\textcolor{blue}{0.777}}$^{***}$ \\
1975Q2--2023Q1 &  & &  & &  & \\
\addlinespace
Great Inflation & \textbf{\textcolor{blue}{0.961}} & 1.017 & 1.010 & \textbf{\textcolor{blue}{0.853}}$^{***}$ & \textbf{\textcolor{blue}{0.974}} & \textbf{\textcolor{blue}{0.865}}$^{***}$ \\
1975Q2--1984Q4 &  & &  & &  & \\
\addlinespace
Volcker Disinflation & \textbf{\textcolor{blue}{0.946}}$^{*}$ & 1.147$^{***}$ & 1.167$^{*}$ & \textbf{\textcolor{blue}{0.981}}$^{*}$ & 1.098$^{***}$ & 1.013$^{***}$ \\
1981Q3--1983Q2 &  & &  & &  & \\
\addlinespace
Great Moderation & 1.057$^{***}$ & \textbf{\textcolor{blue}{0.897}}$^{***}$ & 1.179$^{***}$ & \textbf{\textcolor{blue}{0.562}}$^{***}$ & 1.073 & \textbf{\textcolor{blue}{0.665}}$^{***}$ \\
1985Q1--2007Q4 &  & &  & &  & \\
\addlinespace
Housing Boom & \textbf{\textcolor{blue}{0.945}}$^{***}$ & \textbf{\textcolor{blue}{0.804}}$^{***}$ & \textbf{\textcolor{blue}{0.982}} & \textbf{\textcolor{blue}{0.432}}$^{***}$ & \textbf{\textcolor{blue}{0.897}} & \textbf{\textcolor{blue}{0.503}}$^{***}$ \\
2004Q1--2007Q4 &  & &  & &  & \\
\addlinespace
GFC & 1.054 & \textbf{\textcolor{blue}{0.995}} & 1.020 & \textbf{\textcolor{blue}{0.542}}$^{***}$ & 1.009 & \textbf{\textcolor{blue}{0.958}}$^{**}$ \\
2008Q1--2014Q4 &  & &  & &  & \\
\addlinespace
COVID-19 & 1.032$^{***}$ & 1.066$^{***}$ & 1.007 & \textbf{\textcolor{blue}{0.810}}$^{***}$ & 1.072$^{**}$ & \textbf{\textcolor{blue}{0.862}}$^{***}$ \\
2020Q1--2021Q2 &  & &  & &  & \\
\addlinespace
Since 2021Q3 & \textbf{\textcolor{blue}{0.866}}$^{***}$ & \textbf{\textcolor{blue}{0.965}}$^{**}$ & 1.008$^{***}$ & 1.116 & \textbf{\textcolor{blue}{0.964}}$^{***}$ & \textbf{\textcolor{blue}{0.554}}$^{***}$ \\
2021Q3--2024Q1 &  & &  & &  & \\
\midrule
\multicolumn{7}{c}{\cellcolor{gray!15}\textbf{Panel B: GW Conditional Test}} \\
\midrule
Full Sample & 1.014$^{*}$ & \textbf{\textcolor{blue}{0.956}}$^{***}$ & 1.068$^{***}$ & \textbf{\textcolor{blue}{0.675}}$^{***}$ & 1.027$^{***}$ & \textbf{\textcolor{blue}{0.777}}$^{***}$ \\
1975Q2--2023Q1 &  & &  & &  & \\
\addlinespace
Great Inflation & \textbf{\textcolor{blue}{0.961}}$^{**}$ & 1.017$^{***}$ & 1.010$^{***}$ & \textbf{\textcolor{blue}{0.853}}$^{***}$ & \textbf{\textcolor{blue}{0.974}}$^{***}$ & \textbf{\textcolor{blue}{0.865}}$^{***}$ \\
1975Q2--1984Q4 &  & &  & &  & \\
\addlinespace
Volcker Disinflation & \textbf{\textcolor{blue}{0.946}}$^{*}$ & 1.147$^{***}$ & 1.167$^{***}$ & \textbf{\textcolor{blue}{0.981}}$^{***}$ & 1.098$^{***}$ & 1.013$^{***}$ \\
1981Q3--1983Q2 &  & &  & &  & \\
\addlinespace
Great Moderation & 1.057$^{***}$ & \textbf{\textcolor{blue}{0.897}}$^{***}$ & 1.179$^{***}$ & \textbf{\textcolor{blue}{0.562}}$^{***}$ & 1.073$^{***}$ & \textbf{\textcolor{blue}{0.665}}$^{***}$ \\
1985Q1--2007Q4 &  & &  & &  & \\
\addlinespace
Housing Boom & \textbf{\textcolor{blue}{0.945}}$^{***}$ & \textbf{\textcolor{blue}{0.804}}$^{***}$ & \textbf{\textcolor{blue}{0.982}}$^{***}$ & \textbf{\textcolor{blue}{0.432}}$^{***}$ & \textbf{\textcolor{blue}{0.897}}$^{***}$ & \textbf{\textcolor{blue}{0.503}}$^{***}$ \\
2004Q1--2007Q4 &  & &  & &  & \\
\addlinespace
GFC & 1.054 & \textbf{\textcolor{blue}{0.995}} & 1.020 & \textbf{\textcolor{blue}{0.542}}$^{***}$ & 1.009$^{***}$ & \textbf{\textcolor{blue}{0.958}}$^{***}$ \\
2008Q1--2014Q4 &  & &  & &  & \\
\addlinespace
COVID-19 & 1.032$^{***}$ & 1.066$^{***}$ & 1.007$^{**}$ & \textbf{\textcolor{blue}{0.810}}$^{***}$ & 1.072$^{***}$ & \textbf{\textcolor{blue}{0.862}}$^{***}$ \\
2020Q1--2021Q2 &  & &  & &  & \\
\addlinespace
Since 2021Q3 & \textbf{\textcolor{blue}{0.866}}$^{***}$ & \textbf{\textcolor{blue}{0.965}}$^{***}$ & 1.008$^{***}$ & 1.116$^{***}$ & \textbf{\textcolor{blue}{0.964}}$^{***}$ & \textbf{\textcolor{blue}{0.554}}$^{***}$ \\
2021Q3--2024Q1 &  & &  & &  & \\
\bottomrule
\end{tabular}
\begin{tablenotes}
\tiny
\item \textbf{Notes:} Panel A shows CRPS ratios ($VAR^{\sigma,\kappa}$ CRPS / Competitor CRPS) using Giacomini-White unconditional test. Panel B shows CRPS ratios using Giacomini-White conditional test. Values less than 1 (\textcolor{blue}{\textbf{bold blue}}) indicate the proposed model outperforms the competitor. All ratios are averaged across forecast horizons 1-8. Significance levels: *** p$<$0.01, ** p$<$0.05, * p$<$0.1.
\end{tablenotes}
\end{threeparttable}
\end{table}

\begin{table}[H]
\centering
\caption{Weighted CRPS Left Tail Ratios by Period}
\label{tab:wcrps_left_tail_ratios_gw_comparison}
\begin{threeparttable}
\scriptsize
\begin{tabular}{@{}l*{6}{c}@{}}
\toprule
 & \multicolumn{6}{c}{\textbf{Variables}} \\
\cmidrule(lr){2-7}
\textbf{Period} & \multicolumn{2}{c}{\textbf{GNP}} & \multicolumn{2}{c}{\textbf{GNP Deflator}} & \multicolumn{2}{c}{\textbf{SPREAD}} \\
\cmidrule(lr){2-3} \cmidrule(lr){4-5} \cmidrule(lr){6-7}
 & \textbf{$VAR^{\sigma,\kappa}_{rest}$} & \textbf{$VAR^{\sigma}$} & \textbf{$VAR^{\sigma,\kappa}_{rest}$} & \textbf{$VAR^{\sigma}$} & \textbf{$VAR^{\sigma,\kappa}_{rest}$} & \textbf{$VAR^{\sigma}$} \\
\midrule
\multicolumn{7}{c}{\cellcolor{gray!15}\textbf{Panel A: GW Unconditional Test}} \\
\midrule
Full Sample & 1.015$^{*}$ & \textbf{\textcolor{blue}{0.965}}$^{***}$ & 1.104$^{***}$ & \textbf{\textcolor{blue}{0.640}}$^{***}$ & \textbf{\textcolor{blue}{0.966}} & \textbf{\textcolor{blue}{0.662}}$^{***}$ \\
1975Q2--2023Q1 &  & &  & &  & \\
\addlinespace
Great Inflation & \textbf{\textcolor{blue}{0.955}}$^{*}$ & 1.039 & 1.075 & \textbf{\textcolor{blue}{0.871}}$^{***}$ & \textbf{\textcolor{blue}{0.914}}$^{*}$ & \textbf{\textcolor{blue}{0.785}}$^{***}$ \\
1975Q2--1984Q4 &  & &  & &  & \\
\addlinespace
Volcker Disinflation & \textbf{\textcolor{blue}{0.950}} & 1.155$^{***}$ & 1.206$^{*}$ & \textbf{\textcolor{blue}{0.998}}$^{**}$ & 1.028$^{***}$ & \textbf{\textcolor{blue}{0.934}}$^{**}$ \\
1981Q3--1983Q2 &  & &  & &  & \\
\addlinespace
Great Moderation & 1.060$^{***}$ & \textbf{\textcolor{blue}{0.897}}$^{***}$ & 1.188$^{***}$ & \textbf{\textcolor{blue}{0.563}}$^{***}$ & 1.014 & \textbf{\textcolor{blue}{0.594}}$^{***}$ \\
1985Q1--2007Q4 &  & &  & &  & \\
\addlinespace
Housing Boom & \textbf{\textcolor{blue}{0.945}}$^{***}$ & \textbf{\textcolor{blue}{0.806}}$^{***}$ & \textbf{\textcolor{blue}{0.981}} & \textbf{\textcolor{blue}{0.434}}$^{***}$ & \textbf{\textcolor{blue}{0.885}} & \textbf{\textcolor{blue}{0.469}}$^{***}$ \\
2004Q1--2007Q4 &  & &  & &  & \\
\addlinespace
GFC & 1.056$^{*}$ & 1.006 & 1.023 & \textbf{\textcolor{blue}{0.548}}$^{***}$ & \textbf{\textcolor{blue}{0.918}} & \textbf{\textcolor{blue}{0.758}}$^{***}$ \\
2008Q1--2014Q4 &  & &  & &  & \\
\addlinespace
COVID-19 & 1.032$^{***}$ & 1.055$^{***}$ & 1.049 & \textbf{\textcolor{blue}{0.791}}$^{***}$ & 1.015$^{**}$ & \textbf{\textcolor{blue}{0.797}}$^{***}$ \\
2020Q1--2021Q2 &  & &  & &  & \\
\addlinespace
Since 2021Q3 & \textbf{\textcolor{blue}{0.864}}$^{***}$ & \textbf{\textcolor{blue}{0.957}}$^{**}$ & 1.014$^{***}$ & 1.041 & \textbf{\textcolor{blue}{0.960}}$^{***}$ & \textbf{\textcolor{blue}{0.537}}$^{***}$ \\
2021Q3--2024Q1 &  & &  & &  & \\
\midrule
\multicolumn{7}{c}{\cellcolor{gray!15}\textbf{Panel B: GW Conditional Test}} \\
\midrule
Full Sample & 1.015$^{*}$ & \textbf{\textcolor{blue}{0.965}}$^{***}$ & 1.104$^{***}$ & \textbf{\textcolor{blue}{0.640}}$^{***}$ & \textbf{\textcolor{blue}{0.966}}$^{***}$ & \textbf{\textcolor{blue}{0.662}}$^{***}$ \\
1975Q2--2023Q1 &  & &  & &  & \\
\addlinespace
Great Inflation & \textbf{\textcolor{blue}{0.955}}$^{*}$ & 1.039$^{*}$ & 1.075$^{***}$ & \textbf{\textcolor{blue}{0.871}}$^{***}$ & \textbf{\textcolor{blue}{0.914}}$^{***}$ & \textbf{\textcolor{blue}{0.785}}$^{***}$ \\
1975Q2--1984Q4 &  & &  & &  & \\
\addlinespace
Volcker Disinflation & \textbf{\textcolor{blue}{0.950}}$^{**}$ & 1.155$^{***}$ & 1.206$^{***}$ & \textbf{\textcolor{blue}{0.998}}$^{***}$ & 1.028$^{***}$ & \textbf{\textcolor{blue}{0.934}}$^{***}$ \\
1981Q3--1983Q2 &  & &  & &  & \\
\addlinespace
Great Moderation & 1.060$^{***}$ & \textbf{\textcolor{blue}{0.897}}$^{***}$ & 1.188$^{***}$ & \textbf{\textcolor{blue}{0.563}}$^{***}$ & 1.014$^{***}$ & \textbf{\textcolor{blue}{0.594}}$^{***}$ \\
1985Q1--2007Q4 &  & &  & &  & \\
\addlinespace
Housing Boom & \textbf{\textcolor{blue}{0.945}}$^{***}$ & \textbf{\textcolor{blue}{0.806}}$^{***}$ & \textbf{\textcolor{blue}{0.981}}$^{***}$ & \textbf{\textcolor{blue}{0.434}}$^{***}$ & \textbf{\textcolor{blue}{0.885}}$^{***}$ & \textbf{\textcolor{blue}{0.469}}$^{***}$ \\
2004Q1--2007Q4 &  & &  & &  & \\
\addlinespace
GFC & 1.056 & 1.006 & 1.023 & \textbf{\textcolor{blue}{0.548}}$^{***}$ & \textbf{\textcolor{blue}{0.918}}$^{***}$ & \textbf{\textcolor{blue}{0.758}}$^{***}$ \\
2008Q1--2014Q4 &  & &  & &  & \\
\addlinespace
COVID-19 & 1.032$^{***}$ & 1.055$^{***}$ & 1.049$^{**}$ & \textbf{\textcolor{blue}{0.791}}$^{***}$ & 1.015$^{***}$ & \textbf{\textcolor{blue}{0.797}}$^{***}$ \\
2020Q1--2021Q2 &  & &  & &  & \\
\addlinespace
Since 2021Q3 & \textbf{\textcolor{blue}{0.864}}$^{***}$ & \textbf{\textcolor{blue}{0.957}}$^{***}$ & 1.014$^{***}$ & 1.041$^{***}$ & \textbf{\textcolor{blue}{0.960}}$^{***}$ & \textbf{\textcolor{blue}{0.537}}$^{***}$ \\
2021Q3--2024Q1 &  & &  & &  & \\
\bottomrule
\end{tabular}
\begin{tablenotes}
\tiny
\item \textbf{Notes:} Panel A shows weighted CRPS left tail ratios ($VAR^{\sigma,\kappa}$ WCRPS-L / Competitor WCRPS-L) using Giacomini-White unconditional test. Panel B shows weighted CRPS left tail ratios using Giacomini-White conditional test. Values less than 1 (\textcolor{blue}{\textbf{bold blue}}) indicate the proposed model outperforms the competitor on left tail forecasting. All ratios are averaged across forecast horizons 1-8. Significance levels: *** p$<$0.01, ** p$<$0.05, * p$<$0.1.
\end{tablenotes}
\end{threeparttable}
\end{table}

\begin{table}[H]
\centering
\caption{Weighted CRPS Right Tail Ratios by Period}
\label{tab:wcrps_right_tail_ratios_gw_comparison}
\begin{threeparttable}
\scriptsize
\begin{tabular}{@{}l*{6}{c}@{}}
\toprule
 & \multicolumn{6}{c}{\textbf{Variables}} \\
\cmidrule(lr){2-7}
\textbf{Period} & \multicolumn{2}{c}{\textbf{GNP}} & \multicolumn{2}{c}{\textbf{GNP Deflator}} & \multicolumn{2}{c}{\textbf{SPREAD}} \\
\cmidrule(lr){2-3} \cmidrule(lr){4-5} \cmidrule(lr){6-7}
 & \textbf{$VAR^{\sigma,\kappa}_{rest}$} & \textbf{$VAR^{\sigma}$} & \textbf{$VAR^{\sigma,\kappa}_{rest}$} & \textbf{$VAR^{\sigma}$} & \textbf{$VAR^{\sigma,\kappa}_{rest}$} & \textbf{$VAR^{\sigma}$} \\
\midrule
\multicolumn{7}{c}{\cellcolor{gray!15}\textbf{Panel A: GW Unconditional Test}} \\
\midrule
Full Sample & 1.013 & \textbf{\textcolor{blue}{0.946}}$^{***}$ & 1.043 & \textbf{\textcolor{blue}{0.701}}$^{***}$ & 1.076$^{***}$ & \textbf{\textcolor{blue}{0.885}}$^{***}$ \\
1975Q2--2023Q1 &  & &  & &  & \\
\addlinespace
Great Inflation & \textbf{\textcolor{blue}{0.968}} & \textbf{\textcolor{blue}{0.994}} & \textbf{\textcolor{blue}{0.985}} & \textbf{\textcolor{blue}{0.846}}$^{***}$ & 1.056 & \textbf{\textcolor{blue}{0.980}}$^{**}$ \\
1975Q2--1984Q4 &  & &  & &  & \\
\addlinespace
Volcker Disinflation & \textbf{\textcolor{blue}{0.941}}$^{**}$ & 1.139$^{***}$ & 1.146$^{*}$ & \textbf{\textcolor{blue}{0.972}}$^{*}$ & 1.128$^{***}$ & 1.044$^{**}$ \\
1981Q3--1983Q2 &  & &  & &  & \\
\addlinespace
Great Moderation & 1.055$^{***}$ & \textbf{\textcolor{blue}{0.897}}$^{***}$ & 1.171$^{***}$ & \textbf{\textcolor{blue}{0.561}}$^{***}$ & 1.143$^{**}$ & \textbf{\textcolor{blue}{0.758}}$^{***}$ \\
1985Q1--2007Q4 &  & &  & &  & \\
\addlinespace
Housing Boom & \textbf{\textcolor{blue}{0.946}}$^{***}$ & \textbf{\textcolor{blue}{0.803}}$^{***}$ & \textbf{\textcolor{blue}{0.982}} & \textbf{\textcolor{blue}{0.430}}$^{***}$ & \textbf{\textcolor{blue}{0.916}} & \textbf{\textcolor{blue}{0.572}}$^{***}$ \\
2004Q1--2007Q4 &  & &  & &  & \\
\addlinespace
GFC & 1.051 & \textbf{\textcolor{blue}{0.976}}$^{*}$ & 1.017 & \textbf{\textcolor{blue}{0.536}}$^{***}$ & 1.029 & 1.011$^{**}$ \\
2008Q1--2014Q4 &  & &  & &  & \\
\addlinespace
COVID-19 & 1.034$^{***}$ & 1.082$^{***}$ & \textbf{\textcolor{blue}{0.977}} & \textbf{\textcolor{blue}{0.821}}$^{***}$ & 1.120$^{**}$ & \textbf{\textcolor{blue}{0.921}}$^{***}$ \\
2020Q1--2021Q2 &  & &  & &  & \\
\addlinespace
Since 2021Q3 & \textbf{\textcolor{blue}{0.871}}$^{***}$ & \textbf{\textcolor{blue}{0.976}}$^{*}$ & 1.006$^{***}$ & 1.148 & \textbf{\textcolor{blue}{0.969}}$^{***}$ & \textbf{\textcolor{blue}{0.582}}$^{***}$ \\
2021Q3--2024Q1 &  & &  & &  & \\
\midrule
\multicolumn{7}{c}{\cellcolor{gray!15}\textbf{Panel B: GW Conditional Test}} \\
\midrule
Full Sample & 1.013 & \textbf{\textcolor{blue}{0.946}}$^{***}$ & 1.043$^{***}$ & \textbf{\textcolor{blue}{0.701}}$^{***}$ & 1.076$^{***}$ & \textbf{\textcolor{blue}{0.885}}$^{***}$ \\
1975Q2--2023Q1 &  & &  & &  & \\
\addlinespace
Great Inflation & \textbf{\textcolor{blue}{0.968}}$^{**}$ & \textbf{\textcolor{blue}{0.994}}$^{***}$ & \textbf{\textcolor{blue}{0.985}}$^{***}$ & \textbf{\textcolor{blue}{0.846}}$^{***}$ & 1.056$^{*}$ & \textbf{\textcolor{blue}{0.980}}$^{**}$ \\
1975Q2--1984Q4 &  & &  & &  & \\
\addlinespace
Volcker Disinflation & \textbf{\textcolor{blue}{0.941}}$^{***}$ & 1.139$^{***}$ & 1.146$^{***}$ & \textbf{\textcolor{blue}{0.972}}$^{***}$ & 1.128$^{***}$ & 1.044$^{***}$ \\
1981Q3--1983Q2 &  & &  & &  & \\
\addlinespace
Great Moderation & 1.055$^{***}$ & \textbf{\textcolor{blue}{0.897}}$^{***}$ & 1.171$^{***}$ & \textbf{\textcolor{blue}{0.561}}$^{***}$ & 1.143$^{***}$ & \textbf{\textcolor{blue}{0.758}}$^{***}$ \\
1985Q1--2007Q4 &  & &  & &  & \\
\addlinespace
Housing Boom & \textbf{\textcolor{blue}{0.946}}$^{***}$ & \textbf{\textcolor{blue}{0.803}}$^{***}$ & \textbf{\textcolor{blue}{0.982}}$^{***}$ & \textbf{\textcolor{blue}{0.430}}$^{***}$ & \textbf{\textcolor{blue}{0.916}}$^{**}$ & \textbf{\textcolor{blue}{0.572}}$^{***}$ \\
2004Q1--2007Q4 &  & &  & &  & \\
\addlinespace
GFC & 1.051$^{*}$ & \textbf{\textcolor{blue}{0.976}}$^{*}$ & 1.017 & \textbf{\textcolor{blue}{0.536}}$^{***}$ & 1.029$^{***}$ & 1.011$^{***}$ \\
2008Q1--2014Q4 &  & &  & &  & \\
\addlinespace
COVID-19 & 1.034$^{***}$ & 1.082$^{***}$ & \textbf{\textcolor{blue}{0.977}}$^{**}$ & \textbf{\textcolor{blue}{0.821}}$^{***}$ & 1.120$^{***}$ & \textbf{\textcolor{blue}{0.921}}$^{***}$ \\
2020Q1--2021Q2 &  & &  & &  & \\
\addlinespace
Since 2021Q3 & \textbf{\textcolor{blue}{0.871}}$^{***}$ & \textbf{\textcolor{blue}{0.976}}$^{***}$ & 1.006$^{***}$ & 1.148$^{***}$ & \textbf{\textcolor{blue}{0.969}}$^{***}$ & \textbf{\textcolor{blue}{0.582}}$^{***}$ \\
2021Q3--2024Q1 &  & &  & &  & \\
\bottomrule
\end{tabular}
\begin{tablenotes}
\tiny
\item \textbf{Notes:} Panel A shows weighted CRPS right tail ratios ($VAR^{\sigma,\kappa}$ WCRPS-R / Competitor WCRPS-R) using Giacomini-White unconditional test. Panel B shows weighted CRPS right tail ratios using Giacomini-White conditional test. Values less than 1 (\textcolor{blue}{\textbf{bold blue}}) indicate the proposed model outperforms the competitor on right tail forecasting. All ratios are averaged across forecast horizons 1-8. Significance levels: *** p$<$0.01, ** p$<$0.05, * p$<$0.1.
\end{tablenotes}
\end{threeparttable}
\end{table}

\subsection{UK Average Forecast Performance}

\subsubsection{Baseline results with Conditional Giacomini-White test}

\begin{table}[H]
\centering
\caption{UK Forecast Performance Comparison Across Multiple Metrics}
\label{tab:combined_forecast_comparison_uk_conditional}
\begin{threeparttable}
\scriptsize
\begin{tabular}{l*{8}{c}}
\toprule
\textbf{Variable} & \multicolumn{8}{c}{\textbf{Forecast Horizon}} \\
\cmidrule(lr){2-9}
 & \textbf{H1} & \textbf{H2} & \textbf{H3} & \textbf{H4} & \textbf{H5} & \textbf{H6} & \textbf{H7} & \textbf{H8} \\
\midrule
\multicolumn{9}{c}{\cellcolor{gray!15}\textbf{Panel A: RMSE}} \\
\midrule
\multicolumn{9}{c}{\textbf{vs. $VAR^{\sigma,\kappa}_{restricted}$}} \\
\midrule
\textit{GNP} & \textbf{\textcolor{blue}{0.936}} & \textbf{\textcolor{blue}{0.927}} & \textbf{\textcolor{blue}{0.951}} & \textbf{\textcolor{blue}{0.944}} & \textbf{\textcolor{blue}{0.951}} & \textbf{\textcolor{blue}{0.949}} & \textbf{\textcolor{blue}{0.960}} & \textbf{\textcolor{blue}{0.969}} \\
\textit{GNP def} & \textbf{\textcolor{blue}{0.954}} & \textbf{\textcolor{blue}{0.944}} & \textbf{\textcolor{blue}{0.947}} & \textbf{\textcolor{blue}{0.925}} & \textbf{\textcolor{blue}{0.917}} & \textbf{\textcolor{blue}{0.903}} & \textbf{\textcolor{blue}{0.885}} & \textbf{\textcolor{blue}{0.876}} \\
\textit{SPREAD} & \textbf{\textcolor{blue}{0.984}} & \textbf{\textcolor{blue}{0.989}} & \textbf{\textcolor{blue}{0.987}} & \textbf{\textcolor{blue}{0.999}} & 1.002 & \textbf{\textcolor{blue}{0.999}} & \textbf{\textcolor{blue}{0.999}} & \textbf{\textcolor{blue}{0.991}} \\
\midrule
\multicolumn{9}{c}{\textbf{vs. $VAR^{\sigma}$}} \\
\midrule
\textit{GNP} & \textbf{\textcolor{blue}{0.919}} & \textbf{\textcolor{blue}{0.885}} & \textbf{\textcolor{blue}{0.877}} & \textbf{\textcolor{blue}{0.864}} & \textbf{\textcolor{blue}{0.865}} & \textbf{\textcolor{blue}{0.854}} & \textbf{\textcolor{blue}{0.857}} & \textbf{\textcolor{blue}{0.857}} \\
\textit{GNP def} & \textbf{\textcolor{blue}{0.981}} & \textbf{\textcolor{blue}{0.964}} & \textbf{\textcolor{blue}{0.975}} & \textbf{\textcolor{blue}{0.961}} & \textbf{\textcolor{blue}{0.954}} & \textbf{\textcolor{blue}{0.939}} & \textbf{\textcolor{blue}{0.929}} & \textbf{\textcolor{blue}{0.933}} \\
\textit{SPREAD} & \textbf{\textcolor{blue}{0.969}} & \textbf{\textcolor{blue}{0.983}} & \textbf{\textcolor{blue}{0.977}} & \textbf{\textcolor{blue}{0.972}} & \textbf{\textcolor{blue}{0.991}} & \textbf{\textcolor{blue}{0.995}} & \textbf{\textcolor{blue}{0.991}} & \textbf{\textcolor{blue}{0.997}} \\
\midrule
\multicolumn{9}{c}{\cellcolor{gray!15}\textbf{Panel B: Weighted Both Tails Log Score}} \\
\midrule
\multicolumn{9}{c}{\textbf{vs. $VAR^{\sigma,\kappa}_{restricted}$}} \\
\midrule
\textit{GNP} & -0.3\%$^{*}$ & -6.3\% & \textbf{\textcolor{blue}{1.3\%}} & -1.0\% & \textbf{\textcolor{blue}{3.7\%}} & \textbf{\textcolor{blue}{3.6\%}} & \textbf{\textcolor{blue}{23.3\%}} & -3.7\% \\
\textit{GNP def} & -7.6\% & \textbf{\textcolor{blue}{1.6\%$^{*}$}} & \textbf{\textcolor{blue}{1.9\%}} & \textbf{\textcolor{blue}{2.3\%$^{*}$}} & \textbf{\textcolor{blue}{1.9\%}} & \textbf{\textcolor{blue}{2.0\%}} & \textbf{\textcolor{blue}{2.3\%}} & \textbf{\textcolor{blue}{2.5\%}} \\
\textit{SPREAD} & -3.6\%$^{**}$ & -14.3\% & -0.9\% & -30.2\% & \textbf{\textcolor{blue}{11.9\%}} & -10.5\% & -44.8\% & \textbf{\textcolor{blue}{5.8\%}} \\
\midrule
\multicolumn{9}{c}{\textbf{vs. $VAR^{\sigma}$}} \\
\midrule
\textit{GNP} & -1.3\% & -7.8\% & -2.1\% & -177.0\% & -90.3\% & -35.9\% & -37.9\% & -15.4\% \\
\textit{GNP def} & -3.2\% & \textbf{\textcolor{blue}{3.2\%$^{***}$}} & \textbf{\textcolor{blue}{2.5\%$^{**}$}} & \textbf{\textcolor{blue}{2.3\%$^{**}$}} & \textbf{\textcolor{blue}{2.0\%$^{*}$}} & \textbf{\textcolor{blue}{1.6\%$^{*}$}} & \textbf{\textcolor{blue}{1.5\%$^{**}$}} & \textbf{\textcolor{blue}{1.2\%$^{**}$}} \\
\textit{SPREAD} & -4.8\% & -18.4\% & -6.4\% & -36.5\% & -15.9\% & -106.5\% & -79.5\% & -9.4\% \\
\midrule
\multicolumn{9}{c}{\cellcolor{gray!15}\textbf{Panel C: Weighted Both Tails CRPS}} \\
\midrule
\multicolumn{9}{c}{\textbf{vs. $VAR^{\sigma,\kappa}_{restricted}$}} \\
\midrule
\textit{GNP} & \textbf{\textcolor{blue}{0.904}} & 1.025 & \textbf{\textcolor{blue}{0.983}} & \textbf{\textcolor{blue}{0.895}} & \textbf{\textcolor{blue}{0.950}} & \textbf{\textcolor{blue}{0.959}} & \textbf{\textcolor{blue}{0.998}} & 1.002 \\
\textit{GNP def} & \textbf{\textcolor{blue}{0.934}} & \textbf{\textcolor{blue}{0.875}} & \textbf{\textcolor{blue}{0.884}} & \textbf{\textcolor{blue}{0.799$^{*}$}} & \textbf{\textcolor{blue}{0.806}} & \textbf{\textcolor{blue}{0.771}} & \textbf{\textcolor{blue}{0.725}} & \textbf{\textcolor{blue}{0.717}} \\
\textit{SPREAD} & 1.008 & 1.011$^{**}$ & 1.002 & 1.020 & 1.004 & \textbf{\textcolor{blue}{0.999}} & 1.004 & \textbf{\textcolor{blue}{0.992}} \\
\midrule
\multicolumn{9}{c}{\textbf{vs. $VAR^{\sigma}$}} \\
\midrule
\textit{GNP} & \textbf{\textcolor{blue}{0.934}} & \textbf{\textcolor{blue}{0.982}} & \textbf{\textcolor{blue}{0.932}} & \textbf{\textcolor{blue}{0.942}} & \textbf{\textcolor{blue}{0.966}} & \textbf{\textcolor{blue}{0.991$^{*}$}} & 1.056 & 1.034 \\
\textit{GNP def} & \textbf{\textcolor{blue}{0.924}} & \textbf{\textcolor{blue}{0.875$^{*}$}} & \textbf{\textcolor{blue}{0.909}} & \textbf{\textcolor{blue}{0.848}} & \textbf{\textcolor{blue}{0.851}} & \textbf{\textcolor{blue}{0.848}} & \textbf{\textcolor{blue}{0.867}} & \textbf{\textcolor{blue}{0.926}} \\
\textit{SPREAD} & 1.012 & 1.042 & 1.046 & 1.047 & 1.058 & 1.048 & 1.047 & 1.041 \\
\bottomrule
\end{tabular}
\begin{tablenotes}
\tiny
\item \textbf{Notes:} For RMSE and CRPS, values shown are ratios ($VAR^{\sigma,\kappa}$ model/Competitor); values less than 1 (\textcolor{blue}{\textbf{bold blue}}) indicate the proposed model outperforms the competitor. For Log Score, values shown are percentage differences ($VAR^{\sigma,\kappa}$ model - Competitor); positive values (\textcolor{blue}{\textbf{bold blue}}) indicate superior performance. Significance levels: *** p$<$0.01, ** p$<$0.05, * p$<$0.1.
\end{tablenotes}
\end{threeparttable}
\end{table}

\subsubsection{Additional results - standard LS, CRPS, other weighted tails}

\begin{table}[H]
\centering
\caption{Log Score Comparison}
\label{tab:log_score_comparison_uk_combined}
\begin{threeparttable}
\scriptsize
\begin{tabular}{l*{8}{c}}
\toprule
\textbf{Variable} & \multicolumn{8}{c}{\textbf{Forecast Horizon}} \\
\cmidrule(lr){2-9}
 & \textbf{H1} & \textbf{H2} & \textbf{H3} & \textbf{H4} & \textbf{H5} & \textbf{H6} & \textbf{H7} & \textbf{H8} \\
\midrule
\multicolumn{9}{c}{\cellcolor{gray!15}\textbf{Panel A: GW Unconditional}} \\
\midrule
\multicolumn{9}{c}{\textbf{vs. $VAR^{\sigma,\kappa}_{restricted}$}} \\
\midrule
\textit{GNP} & -9.9\%$^{***}$ & -16.2\%$^{***}$ & -8.6\%$^{***}$ & -7.9\% & \textbf{\textcolor{blue}{0.1\%}} & -1.5\% & \textbf{\textcolor{blue}{17.7\%}} & -8.8\%$^{**}$ \\
\textit{GNP def} & -10.5\% & \textbf{\textcolor{blue}{0.4\%}} & \textbf{\textcolor{blue}{2.8\%}} & \textbf{\textcolor{blue}{4.0\%}} & \textbf{\textcolor{blue}{5.2\%$^{**}$}} & \textbf{\textcolor{blue}{5.8\%$^{**}$}} & \textbf{\textcolor{blue}{6.6\%$^{**}$}} & \textbf{\textcolor{blue}{7.1\%$^{**}$}} \\
\textit{SPREAD} & -2.3\% & -11.8\% & \textbf{\textcolor{blue}{2.7\%}} & -26.6\% & \textbf{\textcolor{blue}{15.7\%}} & -6.5\% & -39.9\% & \textbf{\textcolor{blue}{10.0\%}} \\
\midrule
\multicolumn{9}{c}{\textbf{vs. $VAR^{\sigma}$}} \\
\midrule
\textit{GNP} & \textbf{\textcolor{blue}{29.9\%$^{***}$}} & \textbf{\textcolor{blue}{15.6\%$^{***}$}} & \textbf{\textcolor{blue}{15.1\%$^{***}$}} & -163.3\%$^{**}$ & -81.3\% & -26.8\% & -28.9\% & -7.0\% \\
\textit{GNP def} & \textbf{\textcolor{blue}{44.6\%$^{***}$}} & \textbf{\textcolor{blue}{45.4\%$^{***}$}} & \textbf{\textcolor{blue}{42.0\%$^{***}$}} & \textbf{\textcolor{blue}{38.1\%$^{***}$}} & \textbf{\textcolor{blue}{35.7\%$^{***}$}} & \textbf{\textcolor{blue}{33.4\%$^{***}$}} & \textbf{\textcolor{blue}{31.2\%$^{***}$}} & \textbf{\textcolor{blue}{29.3\%$^{***}$}} \\
\textit{SPREAD} & \textbf{\textcolor{blue}{32.0\%$^{***}$}} & \textbf{\textcolor{blue}{23.4\%$^{*}$}} & \textbf{\textcolor{blue}{37.7\%$^{***}$}} & \textbf{\textcolor{blue}{9.5\%}} & \textbf{\textcolor{blue}{30.8\%$^{*}$}} & -58.8\% & -30.4\% & \textbf{\textcolor{blue}{39.9\%$^{***}$}} \\
\midrule
\multicolumn{9}{c}{\cellcolor{gray!15}\textbf{Panel B: GW Conditional}} \\
\midrule
\multicolumn{9}{c}{\textbf{vs. $VAR^{\sigma,\kappa}_{restricted}$}} \\
\midrule
\textit{GNP} & -9.9\%$^{***}$ & -16.2\%$^{***}$ & -8.6\%$^{***}$ & -7.9\% & \textbf{\textcolor{blue}{0.1\%$^{***}$}} & -1.5\%$^{**}$ & \textbf{\textcolor{blue}{17.7\%}} & -8.8\%$^{***}$ \\
\textit{GNP def} & -10.5\% & \textbf{\textcolor{blue}{0.4\%}} & \textbf{\textcolor{blue}{2.8\%$^{*}$}} & \textbf{\textcolor{blue}{4.0\%}} & \textbf{\textcolor{blue}{5.2\%}} & \textbf{\textcolor{blue}{5.8\%$^{**}$}} & \textbf{\textcolor{blue}{6.6\%$^{**}$}} & \textbf{\textcolor{blue}{7.1\%$^{**}$}} \\
\textit{SPREAD} & -2.3\%$^{*}$ & -11.8\% & \textbf{\textcolor{blue}{2.7\%}} & -26.6\% & \textbf{\textcolor{blue}{15.7\%}} & -6.5\% & -39.9\% & \textbf{\textcolor{blue}{10.0\%}} \\
\midrule
\multicolumn{9}{c}{\textbf{vs. $VAR^{\sigma}$}} \\
\midrule
\textit{GNP} & \textbf{\textcolor{blue}{29.9\%$^{***}$}} & \textbf{\textcolor{blue}{15.6\%$^{***}$}} & \textbf{\textcolor{blue}{15.1\%$^{***}$}} & -163.3\%$^{***}$ & -81.3\% & -26.8\% & -28.9\% & -7.0\% \\
\textit{GNP def} & \textbf{\textcolor{blue}{44.6\%$^{***}$}} & \textbf{\textcolor{blue}{45.4\%$^{***}$}} & \textbf{\textcolor{blue}{42.0\%$^{***}$}} & \textbf{\textcolor{blue}{38.1\%$^{***}$}} & \textbf{\textcolor{blue}{35.7\%$^{***}$}} & \textbf{\textcolor{blue}{33.4\%$^{***}$}} & \textbf{\textcolor{blue}{31.2\%$^{***}$}} & \textbf{\textcolor{blue}{29.3\%$^{***}$}} \\
\textit{SPREAD} & \textbf{\textcolor{blue}{32.0\%$^{***}$}} & \textbf{\textcolor{blue}{23.4\%$^{***}$}} & \textbf{\textcolor{blue}{37.7\%$^{***}$}} & \textbf{\textcolor{blue}{9.5\%$^{***}$}} & \textbf{\textcolor{blue}{30.8\%$^{***}$}} & -58.8\%$^{**}$ & -30.4\% & \textbf{\textcolor{blue}{39.9\%$^{***}$}} \\
\bottomrule
\end{tabular}
\begin{tablenotes}
\tiny
\item \textbf{Notes:} Values shown are percentage differences ($VAR^{\sigma,\kappa}$ model - Competitor); positive values (\textcolor{blue}{\textbf{bold blue}}) indicate the proposed model ($VAR^{\sigma,\kappa}$) outperforms the competitor. Panel A shows results from Giacomini-White unconditional test. Panel B shows results from Giacomini-White conditional test. Significance levels: *** p$<$0.01, ** p$<$0.05, * p$<$0.1.
\end{tablenotes}
\end{threeparttable}
\end{table}

\begin{table}[H]
\centering
\caption{Tail Weighted Log Score Comparison}
\label{tab:tail_weighted_log_score_comparison_uk_combined}
\begin{threeparttable}
\scriptsize
\begin{tabular}{l*{8}{c}}
\toprule
\textbf{Variable} & \multicolumn{8}{c}{\textbf{Forecast Horizon}} \\
\cmidrule(lr){2-9}
 & \textbf{H1} & \textbf{H2} & \textbf{H3} & \textbf{H4} & \textbf{H5} & \textbf{H6} & \textbf{H7} & \textbf{H8} \\
\midrule
\multicolumn{9}{c}{\cellcolor{gray!15}\textbf{Panel A: Left Tail - GW Unconditional}} \\
\midrule
\multicolumn{9}{c}{\textbf{vs. $VAR^{\sigma,\kappa}_{restricted}$}} \\
\midrule
\textit{GNP} & -6.8\%$^{***}$ & -11.5\%$^{***}$ & -4.0\%$^{**}$ & \textbf{\textcolor{blue}{2.7\%}} & \textbf{\textcolor{blue}{1.8\%}} & \textbf{\textcolor{blue}{0.6\%}} & \textbf{\textcolor{blue}{20.6\%}} & -5.9\% \\
\textit{GNP def} & -1.7\%$^{**}$ & -0.9\% & \textbf{\textcolor{blue}{0.1\%}} & \textbf{\textcolor{blue}{0.4\%}} & \textbf{\textcolor{blue}{1.2\%}} & \textbf{\textcolor{blue}{1.4\%}} & \textbf{\textcolor{blue}{1.8\%}} & \textbf{\textcolor{blue}{1.9\%}} \\
\textit{SPREAD} & \textbf{\textcolor{blue}{1.0\%}} & \textbf{\textcolor{blue}{1.6\%$^{**}$}} & \textbf{\textcolor{blue}{2.8\%$^{***}$}} & \textbf{\textcolor{blue}{2.5\%$^{***}$}} & \textbf{\textcolor{blue}{2.9\%$^{***}$}} & \textbf{\textcolor{blue}{2.8\%$^{***}$}} & \textbf{\textcolor{blue}{3.4\%$^{***}$}} & \textbf{\textcolor{blue}{2.9\%$^{***}$}} \\
\midrule
\multicolumn{9}{c}{\textbf{vs. $VAR^{\sigma}$}} \\
\midrule
\textit{GNP} & \textbf{\textcolor{blue}{14.7\%$^{***}$}} & \textbf{\textcolor{blue}{4.1\%$^{***}$}} & \textbf{\textcolor{blue}{8.7\%$^{***}$}} & -161.1\% & -84.6\% & -30.6\% & -32.1\% & -10.3\% \\
\textit{GNP def} & \textbf{\textcolor{blue}{28.6\%$^{***}$}} & \textbf{\textcolor{blue}{25.4\%$^{***}$}} & \textbf{\textcolor{blue}{23.7\%$^{***}$}} & \textbf{\textcolor{blue}{21.4\%$^{***}$}} & \textbf{\textcolor{blue}{19.9\%$^{***}$}} & \textbf{\textcolor{blue}{18.8\%$^{***}$}} & \textbf{\textcolor{blue}{17.5\%$^{***}$}} & \textbf{\textcolor{blue}{16.3\%$^{***}$}} \\
\textit{SPREAD} & \textbf{\textcolor{blue}{17.9\%$^{***}$}} & \textbf{\textcolor{blue}{22.4\%$^{***}$}} & \textbf{\textcolor{blue}{23.7\%$^{***}$}} & \textbf{\textcolor{blue}{25.4\%$^{***}$}} & \textbf{\textcolor{blue}{25.6\%$^{***}$}} & \textbf{\textcolor{blue}{26.4\%$^{***}$}} & \textbf{\textcolor{blue}{26.9\%$^{***}$}} & \textbf{\textcolor{blue}{27.6\%$^{***}$}} \\
\midrule
\multicolumn{9}{c}{\cellcolor{gray!15}\textbf{Panel B: Left Tail - GW Conditional}} \\
\midrule
\multicolumn{9}{c}{\textbf{vs. $VAR^{\sigma,\kappa}_{restricted}$}} \\
\midrule
\textit{GNP} & -6.8\%$^{***}$ & -11.5\%$^{***}$ & -4.0\%$^{*}$ & \textbf{\textcolor{blue}{2.7\%}} & \textbf{\textcolor{blue}{1.8\%$^{***}$}} & \textbf{\textcolor{blue}{0.6\%}} & \textbf{\textcolor{blue}{20.6\%$^{**}$}} & -5.9\%$^{***}$ \\
\textit{GNP def} & -1.7\%$^{**}$ & -0.9\%$^{**}$ & \textbf{\textcolor{blue}{0.1\%$^{**}$}} & \textbf{\textcolor{blue}{0.4\%$^{***}$}} & \textbf{\textcolor{blue}{1.2\%$^{***}$}} & \textbf{\textcolor{blue}{1.4\%$^{***}$}} & \textbf{\textcolor{blue}{1.8\%$^{***}$}} & \textbf{\textcolor{blue}{1.9\%$^{***}$}} \\
\textit{SPREAD} & \textbf{\textcolor{blue}{1.0\%$^{*}$}} & \textbf{\textcolor{blue}{1.6\%$^{**}$}} & \textbf{\textcolor{blue}{2.8\%$^{***}$}} & \textbf{\textcolor{blue}{2.5\%$^{***}$}} & \textbf{\textcolor{blue}{2.9\%$^{***}$}} & \textbf{\textcolor{blue}{2.8\%$^{***}$}} & \textbf{\textcolor{blue}{3.4\%$^{***}$}} & \textbf{\textcolor{blue}{2.9\%$^{***}$}} \\
\midrule
\multicolumn{9}{c}{\textbf{vs. $VAR^{\sigma}$}} \\
\midrule
\textit{GNP} & \textbf{\textcolor{blue}{14.7\%$^{***}$}} & \textbf{\textcolor{blue}{4.1\%$^{***}$}} & \textbf{\textcolor{blue}{8.7\%$^{***}$}} & -161.1\%$^{***}$ & -84.6\%$^{*}$ & -30.6\%$^{**}$ & -32.1\%$^{**}$ & -10.3\%$^{**}$ \\
\textit{GNP def} & \textbf{\textcolor{blue}{28.6\%$^{***}$}} & \textbf{\textcolor{blue}{25.4\%$^{***}$}} & \textbf{\textcolor{blue}{23.7\%$^{***}$}} & \textbf{\textcolor{blue}{21.4\%$^{***}$}} & \textbf{\textcolor{blue}{19.9\%$^{***}$}} & \textbf{\textcolor{blue}{18.8\%$^{***}$}} & \textbf{\textcolor{blue}{17.5\%$^{***}$}} & \textbf{\textcolor{blue}{16.3\%$^{***}$}} \\
\textit{SPREAD} & \textbf{\textcolor{blue}{17.9\%$^{***}$}} & \textbf{\textcolor{blue}{22.4\%$^{***}$}} & \textbf{\textcolor{blue}{23.7\%$^{***}$}} & \textbf{\textcolor{blue}{25.4\%$^{***}$}} & \textbf{\textcolor{blue}{25.6\%$^{***}$}} & \textbf{\textcolor{blue}{26.4\%$^{***}$}} & \textbf{\textcolor{blue}{26.9\%$^{***}$}} & \textbf{\textcolor{blue}{27.6\%$^{***}$}} \\
\midrule
\multicolumn{9}{c}{\cellcolor{gray!15}\textbf{Panel C: Right Tail - GW Unconditional}} \\
\midrule
\multicolumn{9}{c}{\textbf{vs. $VAR^{\sigma,\kappa}_{restricted}$}} \\
\midrule
\textit{GNP} & -3.1\%$^{**}$ & -4.7\%$^{***}$ & -4.6\%$^{***}$ & -10.7\% & -1.6\% & -2.0\% & -2.8\%$^{**}$ & -2.9\%$^{*}$ \\
\textit{GNP def} & -8.8\% & \textbf{\textcolor{blue}{1.2\%}} & \textbf{\textcolor{blue}{2.7\%}} & \textbf{\textcolor{blue}{3.5\%$^{**}$}} & \textbf{\textcolor{blue}{4.0\%$^{**}$}} & \textbf{\textcolor{blue}{4.3\%$^{**}$}} & \textbf{\textcolor{blue}{4.8\%$^{**}$}} & \textbf{\textcolor{blue}{5.2\%$^{**}$}} \\
\textit{SPREAD} & -3.3\%$^{*}$ & -13.4\% & -0.0\% & -29.1\% & \textbf{\textcolor{blue}{12.7\%}} & -9.2\% & -43.3\% & \textbf{\textcolor{blue}{7.1\%}} \\
\midrule
\multicolumn{9}{c}{\textbf{vs. $VAR^{\sigma}$}} \\
\midrule
\textit{GNP} & \textbf{\textcolor{blue}{15.1\%$^{***}$}} & \textbf{\textcolor{blue}{11.4\%$^{***}$}} & \textbf{\textcolor{blue}{6.5\%$^{***}$}} & -2.1\%$^{***}$ & \textbf{\textcolor{blue}{3.3\%$^{**}$}} & \textbf{\textcolor{blue}{3.8\%$^{**}$}} & \textbf{\textcolor{blue}{3.3\%$^{*}$}} & \textbf{\textcolor{blue}{3.4\%$^{*}$}} \\
\textit{GNP def} & \textbf{\textcolor{blue}{16.0\%$^{**}$}} & \textbf{\textcolor{blue}{20.0\%$^{***}$}} & \textbf{\textcolor{blue}{18.3\%$^{***}$}} & \textbf{\textcolor{blue}{16.7\%$^{***}$}} & \textbf{\textcolor{blue}{15.8\%$^{***}$}} & \textbf{\textcolor{blue}{14.7\%$^{***}$}} & \textbf{\textcolor{blue}{13.7\%$^{***}$}} & \textbf{\textcolor{blue}{13.0\%$^{***}$}} \\
\textit{SPREAD} & \textbf{\textcolor{blue}{14.1\%$^{***}$}} & \textbf{\textcolor{blue}{1.1\%}} & \textbf{\textcolor{blue}{14.0\%$^{**}$}} & -15.9\% & \textbf{\textcolor{blue}{5.2\%}} & -85.2\% & -57.3\% & \textbf{\textcolor{blue}{12.3\%}} \\
\midrule
\multicolumn{9}{c}{\cellcolor{gray!15}\textbf{Panel D: Right Tail - GW Conditional}} \\
\midrule
\multicolumn{9}{c}{\textbf{vs. $VAR^{\sigma,\kappa}_{restricted}$}} \\
\midrule
\textit{GNP} & -3.1\%$^{*}$ & -4.7\%$^{***}$ & -4.6\%$^{***}$ & -10.7\% & -1.6\% & -2.0\%$^{***}$ & -2.8\%$^{***}$ & -2.9\%$^{***}$ \\
\textit{GNP def} & -8.8\% & \textbf{\textcolor{blue}{1.2\%}} & \textbf{\textcolor{blue}{2.7\%}} & \textbf{\textcolor{blue}{3.5\%}} & \textbf{\textcolor{blue}{4.0\%}} & \textbf{\textcolor{blue}{4.3\%$^{*}$}} & \textbf{\textcolor{blue}{4.8\%$^{*}$}} & \textbf{\textcolor{blue}{5.2\%$^{*}$}} \\
\textit{SPREAD} & -3.3\%$^{*}$ & -13.4\% & -0.0\% & -29.1\% & \textbf{\textcolor{blue}{12.7\%}} & -9.2\% & -43.3\% & \textbf{\textcolor{blue}{7.1\%}} \\
\midrule
\multicolumn{9}{c}{\textbf{vs. $VAR^{\sigma}$}} \\
\midrule
\textit{GNP} & \textbf{\textcolor{blue}{15.1\%$^{***}$}} & \textbf{\textcolor{blue}{11.4\%$^{***}$}} & \textbf{\textcolor{blue}{6.5\%$^{***}$}} & -2.1\%$^{***}$ & \textbf{\textcolor{blue}{3.3\%$^{***}$}} & \textbf{\textcolor{blue}{3.8\%$^{**}$}} & \textbf{\textcolor{blue}{3.3\%$^{**}$}} & \textbf{\textcolor{blue}{3.4\%$^{*}$}} \\
\textit{GNP def} & \textbf{\textcolor{blue}{16.0\%$^{***}$}} & \textbf{\textcolor{blue}{20.0\%$^{***}$}} & \textbf{\textcolor{blue}{18.3\%$^{***}$}} & \textbf{\textcolor{blue}{16.7\%$^{***}$}} & \textbf{\textcolor{blue}{15.8\%$^{***}$}} & \textbf{\textcolor{blue}{14.7\%$^{***}$}} & \textbf{\textcolor{blue}{13.7\%$^{***}$}} & \textbf{\textcolor{blue}{13.0\%$^{***}$}} \\
\textit{SPREAD} & \textbf{\textcolor{blue}{14.1\%$^{***}$}} & \textbf{\textcolor{blue}{1.1\%$^{**}$}} & \textbf{\textcolor{blue}{14.0\%$^{***}$}} & -15.9\% & \textbf{\textcolor{blue}{5.2\%$^{***}$}} & -85.2\% & -57.3\% & \textbf{\textcolor{blue}{12.3\%$^{***}$}} \\
\bottomrule
\end{tabular}
\begin{tablenotes}
\tiny
\item \textbf{Notes:} Values shown are percentage differences ($VAR^{\sigma,\kappa}$ model - Competitor); positive values (\textcolor{blue}{\textbf{bold blue}}) indicate the benchmark model ($VAR^{\sigma,\kappa}$) outperforms the competitor. Panels A and B show left tail weighted log score for unconditional and conditional Giacomini-White tests. Panels C and D show right tail weighted log score for unconditional and conditional Giacomini-White tests. Significance levels: *** p$<$0.01, ** p$<$0.05, * p$<$0.1.
\end{tablenotes}
\end{threeparttable}
\end{table}

\begin{table}[H]
\centering
\caption{CRPS Comparison}
\label{tab:crps_comparison_uk_combined}
\begin{threeparttable}
\scriptsize
\begin{tabular}{l*{8}{c}}
\toprule
\textbf{Variable} & \multicolumn{8}{c}{\textbf{Forecast Horizon}} \\
\cmidrule(lr){2-9}
 & \textbf{H1} & \textbf{H2} & \textbf{H3} & \textbf{H4} & \textbf{H5} & \textbf{H6} & \textbf{H7} & \textbf{H8} \\
\midrule
\multicolumn{9}{c}{\cellcolor{gray!15}\textbf{Panel A: GW Unconditional}} \\
\midrule
\multicolumn{9}{c}{\textbf{vs. $VAR^{\sigma,\kappa}_{restricted}$}} \\
\midrule
\textit{GNP} & \textbf{\textcolor{blue}{0.992}} & \textbf{\textcolor{blue}{0.998}} & 1.011 & 1.011 & 1.009 & 1.008 & 1.014 & 1.020 \\
\textit{GNP def} & \textbf{\textcolor{blue}{0.990}} & \textbf{\textcolor{blue}{0.972}} & \textbf{\textcolor{blue}{0.965}} & \textbf{\textcolor{blue}{0.946}} & \textbf{\textcolor{blue}{0.939}} & \textbf{\textcolor{blue}{0.933}} & \textbf{\textcolor{blue}{0.924}} & \textbf{\textcolor{blue}{0.918}} \\
\textit{SPREAD} & \textbf{\textcolor{blue}{0.998}} & \textbf{\textcolor{blue}{0.991}} & \textbf{\textcolor{blue}{0.981}} & \textbf{\textcolor{blue}{0.990}} & \textbf{\textcolor{blue}{0.983}} & \textbf{\textcolor{blue}{0.980$^{*}$}} & \textbf{\textcolor{blue}{0.979}} & \textbf{\textcolor{blue}{0.974$^{*}$}} \\
\midrule
\multicolumn{9}{c}{\textbf{vs. $VAR^{\sigma}$}} \\
\midrule
\textit{GNP} & \textbf{\textcolor{blue}{0.881$^{***}$}} & \textbf{\textcolor{blue}{0.871$^{***}$}} & \textbf{\textcolor{blue}{0.879$^{***}$}} & \textbf{\textcolor{blue}{0.890$^{**}$}} & \textbf{\textcolor{blue}{0.897$^{**}$}} & \textbf{\textcolor{blue}{0.895$^{**}$}} & \textbf{\textcolor{blue}{0.897$^{*}$}} & \textbf{\textcolor{blue}{0.898$^{*}$}} \\
\textit{GNP def} & \textbf{\textcolor{blue}{0.808$^{***}$}} & \textbf{\textcolor{blue}{0.813$^{***}$}} & \textbf{\textcolor{blue}{0.825$^{***}$}} & \textbf{\textcolor{blue}{0.830$^{***}$}} & \textbf{\textcolor{blue}{0.837$^{***}$}} & \textbf{\textcolor{blue}{0.839$^{***}$}} & \textbf{\textcolor{blue}{0.841$^{***}$}} & \textbf{\textcolor{blue}{0.848$^{***}$}} \\
\textit{SPREAD} & \textbf{\textcolor{blue}{0.851$^{***}$}} & \textbf{\textcolor{blue}{0.834$^{***}$}} & \textbf{\textcolor{blue}{0.825$^{***}$}} & \textbf{\textcolor{blue}{0.816$^{***}$}} & \textbf{\textcolor{blue}{0.823$^{***}$}} & \textbf{\textcolor{blue}{0.817$^{***}$}} & \textbf{\textcolor{blue}{0.811$^{***}$}} & \textbf{\textcolor{blue}{0.806$^{***}$}} \\
\midrule
\multicolumn{9}{c}{\cellcolor{gray!15}\textbf{Panel B: GW Conditional}} \\
\midrule
\multicolumn{9}{c}{\textbf{vs. $VAR^{\sigma,\kappa}_{restricted}$}} \\
\midrule
\textit{GNP} & \textbf{\textcolor{blue}{0.992}} & \textbf{\textcolor{blue}{0.998}} & 1.011 & 1.011 & 1.009 & 1.008 & 1.014$^{*}$ & 1.020$^{**}$ \\
\textit{GNP def} & \textbf{\textcolor{blue}{0.990}} & \textbf{\textcolor{blue}{0.972}} & \textbf{\textcolor{blue}{0.965}} & \textbf{\textcolor{blue}{0.946}} & \textbf{\textcolor{blue}{0.939$^{*}$}} & \textbf{\textcolor{blue}{0.933$^{*}$}} & \textbf{\textcolor{blue}{0.924}} & \textbf{\textcolor{blue}{0.918}} \\
\textit{SPREAD} & \textbf{\textcolor{blue}{0.998}} & \textbf{\textcolor{blue}{0.991$^{**}$}} & \textbf{\textcolor{blue}{0.981}} & \textbf{\textcolor{blue}{0.990}} & \textbf{\textcolor{blue}{0.983}} & \textbf{\textcolor{blue}{0.980}} & \textbf{\textcolor{blue}{0.979$^{**}$}} & \textbf{\textcolor{blue}{0.974$^{*}$}} \\
\midrule
\multicolumn{9}{c}{\textbf{vs. $VAR^{\sigma}$}} \\
\midrule
\textit{GNP} & \textbf{\textcolor{blue}{0.881$^{***}$}} & \textbf{\textcolor{blue}{0.871$^{**}$}} & \textbf{\textcolor{blue}{0.879$^{**}$}} & \textbf{\textcolor{blue}{0.890$^{**}$}} & \textbf{\textcolor{blue}{0.897$^{*}$}} & \textbf{\textcolor{blue}{0.895$^{*}$}} & \textbf{\textcolor{blue}{0.897$^{*}$}} & \textbf{\textcolor{blue}{0.898$^{*}$}} \\
\textit{GNP def} & \textbf{\textcolor{blue}{0.808$^{***}$}} & \textbf{\textcolor{blue}{0.813$^{***}$}} & \textbf{\textcolor{blue}{0.825$^{***}$}} & \textbf{\textcolor{blue}{0.830$^{***}$}} & \textbf{\textcolor{blue}{0.837$^{***}$}} & \textbf{\textcolor{blue}{0.839$^{***}$}} & \textbf{\textcolor{blue}{0.841$^{***}$}} & \textbf{\textcolor{blue}{0.848$^{***}$}} \\
\textit{SPREAD} & \textbf{\textcolor{blue}{0.851$^{***}$}} & \textbf{\textcolor{blue}{0.834$^{***}$}} & \textbf{\textcolor{blue}{0.825$^{***}$}} & \textbf{\textcolor{blue}{0.816$^{***}$}} & \textbf{\textcolor{blue}{0.823$^{***}$}} & \textbf{\textcolor{blue}{0.817$^{***}$}} & \textbf{\textcolor{blue}{0.811$^{***}$}} & \textbf{\textcolor{blue}{0.806$^{***}$}} \\
\bottomrule
\end{tabular}
\begin{tablenotes}
\tiny
\item \textbf{Notes:} Values shown are CRPS ratios ($VAR^{\sigma,\kappa}$ model/Competitor); values less than 1 (\textcolor{blue}{\textbf{bold blue}}) indicate the proposed model ($VAR^{\sigma,\kappa}$) outperforms the competitor. Panel A shows results from Giacomini-White unconditional test. Panel B shows results from Giacomini-White conditional test. Significance levels: *** p$<$0.01, ** p$<$0.05, * p$<$0.1.
\end{tablenotes}
\end{threeparttable}
\end{table}

\begin{table}[H]
\centering
\caption{Tail Weighted CRPS Comparison}
\label{tab:tail_weighted_crps_comparison_uk_combined}
\begin{threeparttable}
\scriptsize
\begin{tabular}{l*{8}{c}}
\toprule
\textbf{Variable} & \multicolumn{8}{c}{\textbf{Forecast Horizon}} \\
\cmidrule(lr){2-9}
 & \textbf{H1} & \textbf{H2} & \textbf{H3} & \textbf{H4} & \textbf{H5} & \textbf{H6} & \textbf{H7} & \textbf{H8} \\
\midrule
\multicolumn{9}{c}{\cellcolor{gray!15}\textbf{Panel A: Left Tail - GW Unconditional}} \\
\midrule
\multicolumn{9}{c}{\textbf{vs. $VAR^{\sigma,\kappa}_{restricted}$}} \\
\midrule
\textit{GNP} & 1.066$^{***}$ & 1.021 & \textbf{\textcolor{blue}{0.992}} & 1.058$^{**}$ & 1.034 & 1.027 & 1.011 & 1.013 \\
\textit{GNP def} & 1.019 & 1.022 & 1.010 & 1.016 & 1.004 & 1.004 & 1.003 & 1.003 \\
\textit{SPREAD} & \textbf{\textcolor{blue}{0.973$^{*}$}} & \textbf{\textcolor{blue}{0.948$^{**}$}} & \textbf{\textcolor{blue}{0.926$^{***}$}} & \textbf{\textcolor{blue}{0.935$^{***}$}} & \textbf{\textcolor{blue}{0.928$^{***}$}} & \textbf{\textcolor{blue}{0.941$^{***}$}} & \textbf{\textcolor{blue}{0.918$^{***}$}} & \textbf{\textcolor{blue}{0.924$^{***}$}} \\
\midrule
\multicolumn{9}{c}{\textbf{vs. $VAR^{\sigma}$}} \\
\midrule
\textit{GNP} & \textbf{\textcolor{blue}{0.888$^{***}$}} & \textbf{\textcolor{blue}{0.882$^{***}$}} & \textbf{\textcolor{blue}{0.879$^{**}$}} & \textbf{\textcolor{blue}{0.898$^{**}$}} & \textbf{\textcolor{blue}{0.899$^{**}$}} & \textbf{\textcolor{blue}{0.904$^{*}$}} & \textbf{\textcolor{blue}{0.887$^{**}$}} & \textbf{\textcolor{blue}{0.893$^{*}$}} \\
\textit{GNP def} & \textbf{\textcolor{blue}{0.726$^{***}$}} & \textbf{\textcolor{blue}{0.754$^{***}$}} & \textbf{\textcolor{blue}{0.766$^{***}$}} & \textbf{\textcolor{blue}{0.797$^{***}$}} & \textbf{\textcolor{blue}{0.813$^{***}$}} & \textbf{\textcolor{blue}{0.817$^{***}$}} & \textbf{\textcolor{blue}{0.820$^{***}$}} & \textbf{\textcolor{blue}{0.828$^{***}$}} \\
\textit{SPREAD} & \textbf{\textcolor{blue}{0.769$^{***}$}} & \textbf{\textcolor{blue}{0.698$^{***}$}} & \textbf{\textcolor{blue}{0.666$^{***}$}} & \textbf{\textcolor{blue}{0.641$^{***}$}} & \textbf{\textcolor{blue}{0.645$^{***}$}} & \textbf{\textcolor{blue}{0.635$^{***}$}} & \textbf{\textcolor{blue}{0.615$^{***}$}} & \textbf{\textcolor{blue}{0.596$^{***}$}} \\
\midrule
\multicolumn{9}{c}{\cellcolor{gray!15}\textbf{Panel B: Left Tail - GW Conditional}} \\
\midrule
\multicolumn{9}{c}{\textbf{vs. $VAR^{\sigma,\kappa}_{restricted}$}} \\
\midrule
\textit{GNP} & 1.066$^{**}$ & 1.021 & \textbf{\textcolor{blue}{0.992}} & 1.058$^{*}$ & 1.034 & 1.027 & 1.011 & 1.013$^{**}$ \\
\textit{GNP def} & 1.019 & 1.022 & 1.010$^{**}$ & 1.016$^{**}$ & 1.004$^{***}$ & 1.004$^{***}$ & 1.003$^{***}$ & 1.003$^{***}$ \\
\textit{SPREAD} & \textbf{\textcolor{blue}{0.973$^{**}$}} & \textbf{\textcolor{blue}{0.948$^{**}$}} & \textbf{\textcolor{blue}{0.926$^{**}$}} & \textbf{\textcolor{blue}{0.935$^{***}$}} & \textbf{\textcolor{blue}{0.928$^{***}$}} & \textbf{\textcolor{blue}{0.941$^{***}$}} & \textbf{\textcolor{blue}{0.918$^{***}$}} & \textbf{\textcolor{blue}{0.924$^{***}$}} \\
\midrule
\multicolumn{9}{c}{\textbf{vs. $VAR^{\sigma}$}} \\
\midrule
\textit{GNP} & \textbf{\textcolor{blue}{0.888$^{***}$}} & \textbf{\textcolor{blue}{0.882$^{**}$}} & \textbf{\textcolor{blue}{0.879$^{**}$}} & \textbf{\textcolor{blue}{0.898$^{*}$}} & \textbf{\textcolor{blue}{0.899}} & \textbf{\textcolor{blue}{0.904$^{*}$}} & \textbf{\textcolor{blue}{0.887$^{*}$}} & \textbf{\textcolor{blue}{0.893$^{*}$}} \\
\textit{GNP def} & \textbf{\textcolor{blue}{0.726$^{***}$}} & \textbf{\textcolor{blue}{0.754$^{***}$}} & \textbf{\textcolor{blue}{0.766$^{***}$}} & \textbf{\textcolor{blue}{0.797$^{***}$}} & \textbf{\textcolor{blue}{0.813$^{***}$}} & \textbf{\textcolor{blue}{0.817$^{***}$}} & \textbf{\textcolor{blue}{0.820$^{***}$}} & \textbf{\textcolor{blue}{0.828$^{***}$}} \\
\textit{SPREAD} & \textbf{\textcolor{blue}{0.769$^{***}$}} & \textbf{\textcolor{blue}{0.698$^{***}$}} & \textbf{\textcolor{blue}{0.666$^{***}$}} & \textbf{\textcolor{blue}{0.641$^{***}$}} & \textbf{\textcolor{blue}{0.645$^{***}$}} & \textbf{\textcolor{blue}{0.635$^{***}$}} & \textbf{\textcolor{blue}{0.615$^{***}$}} & \textbf{\textcolor{blue}{0.596$^{***}$}} \\
\midrule
\multicolumn{9}{c}{\cellcolor{gray!15}\textbf{Panel C: Right Tail - GW Unconditional}} \\
\midrule
\multicolumn{9}{c}{\textbf{vs. $VAR^{\sigma,\kappa}_{restricted}$}} \\
\midrule
\textit{GNP} & \textbf{\textcolor{blue}{0.916}} & \textbf{\textcolor{blue}{0.968}} & 1.039 & \textbf{\textcolor{blue}{0.961}} & \textbf{\textcolor{blue}{0.979}} & \textbf{\textcolor{blue}{0.985}} & 1.018 & 1.030 \\
\textit{GNP def} & \textbf{\textcolor{blue}{0.975}} & \textbf{\textcolor{blue}{0.946}} & \textbf{\textcolor{blue}{0.942}} & \textbf{\textcolor{blue}{0.908}} & \textbf{\textcolor{blue}{0.901}} & \textbf{\textcolor{blue}{0.890}} & \textbf{\textcolor{blue}{0.875}} & \textbf{\textcolor{blue}{0.865}} \\
\textit{SPREAD} & 1.010 & 1.009 & 1.004 & 1.010 & 1.004 & \textbf{\textcolor{blue}{0.993}} & \textbf{\textcolor{blue}{0.999}} & \textbf{\textcolor{blue}{0.989}} \\
\midrule
\multicolumn{9}{c}{\textbf{vs. $VAR^{\sigma}$}} \\
\midrule
\textit{GNP} & \textbf{\textcolor{blue}{0.873$^{*}$}} & \textbf{\textcolor{blue}{0.857$^{***}$}} & \textbf{\textcolor{blue}{0.878$^{***}$}} & \textbf{\textcolor{blue}{0.880$^{**}$}} & \textbf{\textcolor{blue}{0.894$^{*}$}} & \textbf{\textcolor{blue}{0.883$^{**}$}} & \textbf{\textcolor{blue}{0.909}} & \textbf{\textcolor{blue}{0.903}} \\
\textit{GNP def} & \textbf{\textcolor{blue}{0.858$^{***}$}} & \textbf{\textcolor{blue}{0.850$^{***}$}} & \textbf{\textcolor{blue}{0.862$^{***}$}} & \textbf{\textcolor{blue}{0.852$^{***}$}} & \textbf{\textcolor{blue}{0.854$^{**}$}} & \textbf{\textcolor{blue}{0.855$^{**}$}} & \textbf{\textcolor{blue}{0.856$^{**}$}} & \textbf{\textcolor{blue}{0.863$^{***}$}} \\
\textit{SPREAD} & \textbf{\textcolor{blue}{0.898$^{***}$}} & \textbf{\textcolor{blue}{0.906$^{***}$}} & \textbf{\textcolor{blue}{0.907$^{**}$}} & \textbf{\textcolor{blue}{0.901$^{**}$}} & \textbf{\textcolor{blue}{0.908$^{*}$}} & \textbf{\textcolor{blue}{0.902$^{**}$}} & \textbf{\textcolor{blue}{0.899$^{**}$}} & \textbf{\textcolor{blue}{0.896$^{**}$}} \\
\midrule
\multicolumn{9}{c}{\cellcolor{gray!15}\textbf{Panel D: Right Tail - GW Conditional}} \\
\midrule
\multicolumn{9}{c}{\textbf{vs. $VAR^{\sigma,\kappa}_{restricted}$}} \\
\midrule
\textit{GNP} & \textbf{\textcolor{blue}{0.916}} & \textbf{\textcolor{blue}{0.968}} & 1.039 & \textbf{\textcolor{blue}{0.961}} & \textbf{\textcolor{blue}{0.979}} & \textbf{\textcolor{blue}{0.985}} & 1.018$^{**}$ & 1.030$^{**}$ \\
\textit{GNP def} & \textbf{\textcolor{blue}{0.975}} & \textbf{\textcolor{blue}{0.946}} & \textbf{\textcolor{blue}{0.942}} & \textbf{\textcolor{blue}{0.908}} & \textbf{\textcolor{blue}{0.901}} & \textbf{\textcolor{blue}{0.890}} & \textbf{\textcolor{blue}{0.875}} & \textbf{\textcolor{blue}{0.865}} \\
\textit{SPREAD} & 1.010 & 1.009 & 1.004 & 1.010 & 1.004 & \textbf{\textcolor{blue}{0.993}} & \textbf{\textcolor{blue}{0.999$^{*}$}} & \textbf{\textcolor{blue}{0.989}} \\
\midrule
\multicolumn{9}{c}{\textbf{vs. $VAR^{\sigma}$}} \\
\midrule
\textit{GNP} & \textbf{\textcolor{blue}{0.873}} & \textbf{\textcolor{blue}{0.857$^{**}$}} & \textbf{\textcolor{blue}{0.878$^{**}$}} & \textbf{\textcolor{blue}{0.880$^{*}$}} & \textbf{\textcolor{blue}{0.894$^{*}$}} & \textbf{\textcolor{blue}{0.883$^{**}$}} & \textbf{\textcolor{blue}{0.909$^{*}$}} & \textbf{\textcolor{blue}{0.903$^{*}$}} \\
\textit{GNP def} & \textbf{\textcolor{blue}{0.858$^{***}$}} & \textbf{\textcolor{blue}{0.850$^{***}$}} & \textbf{\textcolor{blue}{0.862$^{***}$}} & \textbf{\textcolor{blue}{0.852$^{***}$}} & \textbf{\textcolor{blue}{0.854$^{**}$}} & \textbf{\textcolor{blue}{0.855$^{**}$}} & \textbf{\textcolor{blue}{0.856$^{**}$}} & \textbf{\textcolor{blue}{0.863$^{***}$}} \\
\textit{SPREAD} & \textbf{\textcolor{blue}{0.898$^{***}$}} & \textbf{\textcolor{blue}{0.906$^{***}$}} & \textbf{\textcolor{blue}{0.907$^{***}$}} & \textbf{\textcolor{blue}{0.901$^{***}$}} & \textbf{\textcolor{blue}{0.908$^{***}$}} & \textbf{\textcolor{blue}{0.902$^{***}$}} & \textbf{\textcolor{blue}{0.899$^{***}$}} & \textbf{\textcolor{blue}{0.896$^{***}$}} \\
\bottomrule
\end{tabular}
\begin{tablenotes}
\tiny
\item \textbf{Notes:} Values shown are CRPS ratios ($VAR^{\sigma,\kappa}$ model/Competitor); values less than 1 (\textcolor{blue}{\textbf{bold blue}}) indicate the proposed model ($VAR^{\sigma,\kappa}$) outperforms the competitor. Panels A and B show left tail weighted CRPS for unconditional and conditional Giacomini-White tests. Panels C and D show right tail weighted CRPS for unconditional and conditional Giacomini-White tests. Significance levels: *** p$<$0.01, ** p$<$0.05, * p$<$0.1.
\end{tablenotes}
\end{threeparttable}
\end{table}

\subsection{UK Forecast Performance Over Time}
\subsubsection{Baseline results with Conditional Giacomini-White test}

\begin{table}[H]
\centering
\caption{Weighted Both Tails Performance}
\label{tab:weighted_both_tails_uk_gw_conditional}
\begin{threeparttable}
\scriptsize
\begin{tabular}{@{}l*{6}{c}@{}}
\toprule
 & \multicolumn{6}{c}{\textbf{Variables}} \\
\cmidrule(lr){2-7}
\textbf{Period} & \multicolumn{2}{c}{\textbf{GNP}} & \multicolumn{2}{c}{\textbf{GNP Deflator}} & \multicolumn{2}{c}{\textbf{SPREAD}} \\
\cmidrule(lr){2-3} \cmidrule(lr){4-5} \cmidrule(lr){6-7}
 & \textbf{$VAR^{\sigma,\kappa}_{rest}$} & \textbf{$VAR^{\sigma}$} & \textbf{$VAR^{\sigma,\kappa}_{rest}$} & \textbf{$VAR^{\sigma}$} & \textbf{$VAR^{\sigma,\kappa}_{rest}$} & \textbf{$VAR^{\sigma}$} \\
\midrule
\multicolumn{7}{c}{\cellcolor{gray!15}\textbf{Panel A: Cumulative Weighted Both Tails Log Score Differences}} \\
\midrule
Full Sample & \textbf{\textcolor{blue}{490.0}} & -4781.9 & \textbf{\textcolor{blue}{161.9}}$^{*}$ & \textbf{\textcolor{blue}{262.4}}$^{***}$ & -2092.0$^{*}$ & -6453.8 \\
1975Q2--2023Q1 & (\textbf{\textcolor{blue}{2.6}}) & (-25.5) & (\textbf{\textcolor{blue}{0.9}}) & (\textbf{\textcolor{blue}{1.4}}) & (-11.1) & (-34.5) \\
\addlinespace
Stagflation & -2.4 & \textbf{\textcolor{blue}{7.5}}$^{**}$ & \textbf{\textcolor{blue}{29.9}}$^{**}$ & \textbf{\textcolor{blue}{21.1}}$^{*}$ & -51.5$^{*}$ & -169.4$^{*}$ \\
1975Q2--1979Q4 & (-0.0) & (\textbf{\textcolor{blue}{0.5}}) & (\textbf{\textcolor{blue}{4.9}}) & (\textbf{\textcolor{blue}{2.7}}) & (-3.3) & (-10.7) \\
\addlinespace
Thatcher & \textbf{\textcolor{blue}{2.7}}$^{*}$ & \textbf{\textcolor{blue}{22.4}}$^{***}$ & -101.0 & -144.6 & -51.9 & -23.9$^{***}$ \\
1979Q1--1989Q4 & (\textbf{\textcolor{blue}{0.1}}) & (\textbf{\textcolor{blue}{0.5}}) & (-2.3) & (-3.3) & (-1.2) & (-0.5) \\
\addlinespace
ERM crisis & \textbf{\textcolor{blue}{13.3}}$^{***}$ & \textbf{\textcolor{blue}{25.2}}$^{***}$ & \textbf{\textcolor{blue}{8.3}}$^{**}$ & \textbf{\textcolor{blue}{14.3}}$^{**}$ & -7.0$^{***}$ & \textbf{\textcolor{blue}{99.5}}$^{***}$ \\
1990Q1--1992Q4 & (\textbf{\textcolor{blue}{1.1}}) & (\textbf{\textcolor{blue}{2.1}}) & (\textbf{\textcolor{blue}{0.7}}) & (\textbf{\textcolor{blue}{1.2}}) & (-0.6) & (\textbf{\textcolor{blue}{8.3}}) \\
\addlinespace
Great Moderation & -2.9$^{***}$ & \textbf{\textcolor{blue}{19.8}}$^{***}$ & -5.0$^{**}$ & \textbf{\textcolor{blue}{110.5}}$^{***}$ & -253.3$^{*}$ & \textbf{\textcolor{blue}{35.4}}$^{**}$ \\
1993Q1--2007Q4 & (-0.0) & (\textbf{\textcolor{blue}{0.3}}) & (-0.1) & (\textbf{\textcolor{blue}{1.8}}) & (-4.2) & (\textbf{\textcolor{blue}{0.6}}) \\
\addlinespace
GFC & \textbf{\textcolor{blue}{171.2}}$^{*}$ & -216.8 & \textbf{\textcolor{blue}{5.8}}$^{***}$ & \textbf{\textcolor{blue}{25.4}}$^{***}$ & -1765.2 & -6287.8 \\
2007Q1--2012Q4 & (\textbf{\textcolor{blue}{7.1}}) & (-9.0) & (\textbf{\textcolor{blue}{0.2}}) & (\textbf{\textcolor{blue}{1.1}}) & (-73.5) & (-262.0) \\
\addlinespace
Brexit & -6.1$^{***}$ & \textbf{\textcolor{blue}{11.6}}$^{***}$ & -3.5$^{***}$ & \textbf{\textcolor{blue}{75.1}}$^{***}$ & \textbf{\textcolor{blue}{25.3}}$^{***}$ & \textbf{\textcolor{blue}{249.3}}$^{***}$ \\
2013Q1--2019Q4 & (-0.2) & (\textbf{\textcolor{blue}{0.4}}) & (-0.1) & (\textbf{\textcolor{blue}{2.7}}) & (\textbf{\textcolor{blue}{0.9}}) & (\textbf{\textcolor{blue}{8.9}}) \\
\addlinespace
COVID-19 & \textbf{\textcolor{blue}{287.0}}$^{***}$ & -4710.2$^{**}$ & \textbf{\textcolor{blue}{5.1}}$^{***}$ & \textbf{\textcolor{blue}{19.6}}$^{***}$ & -69.0$^{***}$ & -208.9$^{***}$ \\
2020Q1--2021Q2 & (\textbf{\textcolor{blue}{43.8}}) & (-966.0) & (\textbf{\textcolor{blue}{1.0}}) & (\textbf{\textcolor{blue}{3.9}}) & (-13.8) & (-41.8) \\
\addlinespace
Since 2021Q3 & \textbf{\textcolor{blue}{7.3}}$^{***}$ & \textbf{\textcolor{blue}{3.1}}$^{***}$ & \textbf{\textcolor{blue}{54.4}}$^{***}$ & -9.8 & \textbf{\textcolor{blue}{72.3}}$^{***}$ & -322.8$^{**}$ \\
2021Q3--2024Q1 & (\textbf{\textcolor{blue}{0.5}}) & (\textbf{\textcolor{blue}{0.3}}) & (\textbf{\textcolor{blue}{7.1}}) & (-1.2) & (\textbf{\textcolor{blue}{8.9}}) & (-41.8) \\
\midrule
\multicolumn{7}{c}{\cellcolor{gray!15}\textbf{Panel B: Weighted CRPS Both Tails Ratios}} \\
\midrule
Full Sample & \textbf{\textcolor{blue}{0.964}} & \textbf{\textcolor{blue}{0.979}}$^{*}$ & \textbf{\textcolor{blue}{0.814}}$^{*}$ & \textbf{\textcolor{blue}{0.880}}$^{*}$ & 1.007$^{**}$ & 1.039 \\
1975Q2--2023Q1 &  & &  & &  & \\
\addlinespace
Stagflation & 1.017 & \textbf{\textcolor{blue}{0.897}}$^{***}$ & \textbf{\textcolor{blue}{0.752}} & \textbf{\textcolor{blue}{0.823}} & 1.011$^{**}$ & 1.045$^{***}$ \\
1975Q2--1979Q4 &  & &  & &  & \\
\addlinespace
Thatcher & \textbf{\textcolor{blue}{0.992}} & \textbf{\textcolor{blue}{0.959}}$^{*}$ & \textbf{\textcolor{blue}{0.863}} & \textbf{\textcolor{blue}{0.976}} & \textbf{\textcolor{blue}{1.000}} & \textbf{\textcolor{blue}{0.971}}$^{***}$ \\
1979Q1--1989Q4 &  & &  & &  & \\
\addlinespace
ERM crisis & \textbf{\textcolor{blue}{0.931}}$^{***}$ & \textbf{\textcolor{blue}{0.781}}$^{***}$ & \textbf{\textcolor{blue}{0.920}}$^{***}$ & \textbf{\textcolor{blue}{0.922}}$^{**}$ & 1.021$^{***}$ & \textbf{\textcolor{blue}{0.821}}$^{***}$ \\
1990Q1--1992Q4 &  & &  & &  & \\
\addlinespace
Great Moderation & 1.001 & \textbf{\textcolor{blue}{0.902}}$^{***}$ & 1.044$^{***}$ & \textbf{\textcolor{blue}{0.699}}$^{***}$ & 1.075$^{**}$ & \textbf{\textcolor{blue}{0.991}}$^{*}$ \\
1993Q1--2007Q4 &  & &  & &  & \\
\addlinespace
GFC & 1.066$^{**}$ & 1.091 & \textbf{\textcolor{blue}{0.902}}$^{***}$ & \textbf{\textcolor{blue}{0.783}}$^{***}$ & \textbf{\textcolor{blue}{0.997}}$^{**}$ & 1.168 \\
2007Q1--2012Q4 &  & &  & &  & \\
\addlinespace
Brexit & 1.129$^{***}$ & \textbf{\textcolor{blue}{0.856}}$^{***}$ & 1.048$^{*}$ & \textbf{\textcolor{blue}{0.732}}$^{***}$ & \textbf{\textcolor{blue}{0.971}}$^{**}$ & \textbf{\textcolor{blue}{0.819}}$^{***}$ \\
2013Q1--2019Q4 &  & &  & &  & \\
\addlinespace
COVID-19 & \textbf{\textcolor{blue}{0.962}}$^{***}$ & 1.042$^{***}$ & \textbf{\textcolor{blue}{0.968}}$^{***}$ & \textbf{\textcolor{blue}{0.716}}$^{***}$ & \textbf{\textcolor{blue}{0.988}}$^{***}$ & 1.174$^{***}$ \\
2020Q1--2021Q2 &  & &  & &  & \\
\addlinespace
Since 2021Q3 & 1.007$^{***}$ & \textbf{\textcolor{blue}{0.968}}$^{***}$ & \textbf{\textcolor{blue}{0.744}}$^{***}$ & 1.081$^{**}$ & \textbf{\textcolor{blue}{0.958}}$^{***}$ & 1.209$^{***}$ \\
2021Q3--2024Q1 &  & &  & &  & \\
\bottomrule
\end{tabular}
\begin{tablenotes}
\tiny
\item \textbf{Notes:} Panel A shows cumulative weighted both tails log score differences ($VAR^{\sigma,\kappa}$ model - Competitor) using Giacomini-White conditional test. Values in parentheses show average score differences per quarter. Positive values (\textcolor{blue}{\textbf{bold blue}}) indicate superior performance. Panel B shows weighted CRPS both tails ratios. Values less than 1 (\textcolor{blue}{\textbf{bold blue}}) indicate superior performance. All values are averaged across forecast horizons 1-8. Significance levels: *** p$<$0.01, ** p$<$0.05, * p$<$0.1.
\end{tablenotes}
\end{threeparttable}
\end{table}

\subsubsection{Additional results - standard LS, CRPS and other weighted tails}

\begin{table}[H]
\centering
\caption{RMSE Ratios}
\label{tab:rmse_uk_gw_comparison}
\begin{threeparttable}
\scriptsize
\begin{tabular}{@{}l*{6}{c}@{}}
\toprule
 & \multicolumn{6}{c}{\textbf{Variables}} \\
\cmidrule(lr){2-7}
\textbf{Period} & \multicolumn{2}{c}{\textbf{GNP}} & \multicolumn{2}{c}{\textbf{GNP Deflator}} & \multicolumn{2}{c}{\textbf{SPREAD}} \\
\cmidrule(lr){2-3} \cmidrule(lr){4-5} \cmidrule(lr){6-7}
 & \textbf{$VAR^{\sigma,\kappa}_{rest}$} & \textbf{$VAR^{\sigma}$} & \textbf{$VAR^{\sigma,\kappa}_{rest}$} & \textbf{$VAR^{\sigma}$} & \textbf{$VAR^{\sigma,\kappa}_{rest}$} & \textbf{$VAR^{\sigma}$} \\
\midrule
\multicolumn{7}{c}{\cellcolor{gray!15}\textbf{Panel A: GW Unconditional Test}} \\
\midrule
Full Sample & \textbf{\textcolor{blue}{0.948}} & \textbf{\textcolor{blue}{0.871}}$^{*}$ & \textbf{\textcolor{blue}{0.920}} & \textbf{\textcolor{blue}{0.954}} & \textbf{\textcolor{blue}{0.995}} & \textbf{\textcolor{blue}{0.985}} \\
1975Q2--2023Q1 &  & &  & &  & \\
\addlinespace
Stagflation & 1.237 & \textbf{\textcolor{blue}{0.704}}$^{***}$ & \textbf{\textcolor{blue}{0.849}} & \textbf{\textcolor{blue}{0.889}} & \textbf{\textcolor{blue}{0.990}}$^{**}$ & \textbf{\textcolor{blue}{0.995}}$^{*}$ \\
1975Q2--1979Q4 &  & &  & &  & \\
\addlinespace
Thatcher & 1.018 & \textbf{\textcolor{blue}{0.881}} & \textbf{\textcolor{blue}{0.967}} & \textbf{\textcolor{blue}{0.990}} & 1.005 & \textbf{\textcolor{blue}{0.990}} \\
1979Q1--1989Q4 &  & &  & &  & \\
\addlinespace
ERM crisis & \textbf{\textcolor{blue}{0.943}}$^{***}$ & \textbf{\textcolor{blue}{0.736}}$^{**}$ & \textbf{\textcolor{blue}{0.931}}$^{*}$ & \textbf{\textcolor{blue}{0.982}} & 1.017$^{**}$ & \textbf{\textcolor{blue}{0.944}}$^{***}$ \\
1990Q1--1992Q4 &  & &  & &  & \\
\addlinespace
Great Moderation & \textbf{\textcolor{blue}{0.947}}$^{*}$ & \textbf{\textcolor{blue}{0.953}} & 1.117$^{*}$ & \textbf{\textcolor{blue}{0.923}}$^{**}$ & 1.040$^{***}$ & 1.039$^{**}$ \\
1993Q1--2007Q4 &  & &  & &  & \\
\addlinespace
GFC & 1.113$^{*}$ & 1.048$^{**}$ & \textbf{\textcolor{blue}{0.857}}$^{**}$ & 1.090$^{*}$ & \textbf{\textcolor{blue}{0.996}} & \textbf{\textcolor{blue}{0.994}}$^{***}$ \\
2007Q1--2012Q4 &  & &  & &  & \\
\addlinespace
Brexit & 1.097$^{**}$ & 1.047 & \textbf{\textcolor{blue}{0.975}} & 1.138$^{***}$ & \textbf{\textcolor{blue}{0.963}}$^{**}$ & \textbf{\textcolor{blue}{0.928}}$^{***}$ \\
2013Q1--2019Q4 &  & &  & &  & \\
\addlinespace
COVID-19 & \textbf{\textcolor{blue}{0.967}}$^{***}$ & \textbf{\textcolor{blue}{0.980}}$^{***}$ & 1.052$^{***}$ & 1.062$^{**}$ & \textbf{\textcolor{blue}{0.967}}$^{**}$ & \textbf{\textcolor{blue}{0.955}}$^{*}$ \\
2020Q1--2021Q2 &  & &  & &  & \\
\addlinespace
Since 2021Q3 & 1.123$^{*}$ & \textbf{\textcolor{blue}{0.949}}$^{*}$ & \textbf{\textcolor{blue}{0.757}}$^{***}$ & 1.034$^{***}$ & \textbf{\textcolor{blue}{0.971}}$^{***}$ & \textbf{\textcolor{blue}{0.978}}$^{***}$ \\
2021Q3--2024Q1 &  & &  & &  & \\
\midrule
\multicolumn{7}{c}{\cellcolor{gray!15}\textbf{Panel B: GW Conditional Test}} \\
\midrule
Full Sample & \textbf{\textcolor{blue}{0.948}} & \textbf{\textcolor{blue}{0.871}} & \textbf{\textcolor{blue}{0.920}} & \textbf{\textcolor{blue}{0.954}} & \textbf{\textcolor{blue}{0.995}} & \textbf{\textcolor{blue}{0.985}} \\
1975Q2--2023Q1 &  & &  & &  & \\
\addlinespace
Stagflation & 1.237 & \textbf{\textcolor{blue}{0.704}}$^{***}$ & \textbf{\textcolor{blue}{0.849}}$^{**}$ & \textbf{\textcolor{blue}{0.889}}$^{***}$ & \textbf{\textcolor{blue}{0.990}}$^{***}$ & \textbf{\textcolor{blue}{0.995}}$^{**}$ \\
1975Q2--1979Q4 &  & &  & &  & \\
\addlinespace
Thatcher & 1.018 & \textbf{\textcolor{blue}{0.881}} & \textbf{\textcolor{blue}{0.967}}$^{*}$ & \textbf{\textcolor{blue}{0.990}} & 1.005 & \textbf{\textcolor{blue}{0.990}}$^{*}$ \\
1979Q1--1989Q4 &  & &  & &  & \\
\addlinespace
ERM crisis & \textbf{\textcolor{blue}{0.943}}$^{***}$ & \textbf{\textcolor{blue}{0.736}}$^{**}$ & \textbf{\textcolor{blue}{0.931}} & \textbf{\textcolor{blue}{0.982}} & 1.017$^{*}$ & \textbf{\textcolor{blue}{0.944}}$^{***}$ \\
1990Q1--1992Q4 &  & &  & &  & \\
\addlinespace
Great Moderation & \textbf{\textcolor{blue}{0.947}} & \textbf{\textcolor{blue}{0.953}}$^{**}$ & 1.117 & \textbf{\textcolor{blue}{0.923}} & 1.040$^{***}$ & 1.039$^{*}$ \\
1993Q1--2007Q4 &  & &  & &  & \\
\addlinespace
GFC & 1.113$^{*}$ & 1.048$^{*}$ & \textbf{\textcolor{blue}{0.857}} & 1.090$^{*}$ & \textbf{\textcolor{blue}{0.996}} & \textbf{\textcolor{blue}{0.994}}$^{***}$ \\
2007Q1--2012Q4 &  & &  & &  & \\
\addlinespace
Brexit & 1.097$^{**}$ & 1.047 & \textbf{\textcolor{blue}{0.975}} & 1.138$^{***}$ & \textbf{\textcolor{blue}{0.963}}$^{**}$ & \textbf{\textcolor{blue}{0.928}}$^{***}$ \\
2013Q1--2019Q4 &  & &  & &  & \\
\addlinespace
COVID-19 & \textbf{\textcolor{blue}{0.967}}$^{***}$ & \textbf{\textcolor{blue}{0.980}}$^{***}$ & 1.052$^{***}$ & 1.062$^{***}$ & \textbf{\textcolor{blue}{0.967}}$^{***}$ & \textbf{\textcolor{blue}{0.955}}$^{***}$ \\
2020Q1--2021Q2 &  & &  & &  & \\
\addlinespace
Since 2021Q3 & 1.123$^{*}$ & \textbf{\textcolor{blue}{0.949}}$^{*}$ & \textbf{\textcolor{blue}{0.757}}$^{***}$ & 1.034$^{***}$ & \textbf{\textcolor{blue}{0.971}}$^{***}$ & \textbf{\textcolor{blue}{0.978}}$^{***}$ \\
2021Q3--2024Q1 &  & &  & &  & \\
\bottomrule
\end{tabular}
\begin{tablenotes}
\tiny
\item \textbf{Notes:} Panel A shows RMSE ratios ($VAR^{\sigma,\kappa}$ RMSE / Competitor RMSE) using Giacomini-White unconditional test. Panel B shows RMSE ratios using Giacomini-White conditional test. Values less than 1 (\textcolor{blue}{\textbf{bold blue}}) indicate the proposed model outperforms the competitor. All ratios are averaged across forecast horizons 1-8. Significance levels: *** p$<$0.01, ** p$<$0.05, * p$<$0.1.
\end{tablenotes}
\end{threeparttable}
\end{table}

\begin{table}[H]
\centering
\caption{Cumulative Log Score Differences: UK GW Tests Comparison}
\label{tab:cumulative_logscore_uk_gw_comparison}
\begin{threeparttable}
\scriptsize
\begin{tabular}{@{}l*{6}{c}@{}}
\toprule
 & \multicolumn{6}{c}{\textbf{Variables}} \\
\cmidrule(lr){2-7}
\textbf{Period} & \multicolumn{2}{c}{\textbf{GNP}} & \multicolumn{2}{c}{\textbf{GNP Deflator}} & \multicolumn{2}{c}{\textbf{SPREAD}} \\
\cmidrule(lr){2-3} \cmidrule(lr){4-5} \cmidrule(lr){6-7}
 & \textbf{$VAR^{\sigma,\kappa}_{rest}$} & \textbf{$VAR^{\sigma}$} & \textbf{$VAR^{\sigma,\kappa}_{rest}$} & \textbf{$VAR^{\sigma}$} & \textbf{$VAR^{\sigma,\kappa}_{rest}$} & \textbf{$VAR^{\sigma}$} \\
\midrule
\multicolumn{7}{c}{\cellcolor{gray!15}\textbf{Panel A: GW Unconditional Test}} \\
\midrule
Full Sample & -815.0$^{***}$ & -1898.3$^{***}$ & \textbf{\textcolor{blue}{481.1}}$^{**}$ & \textbf{\textcolor{blue}{7135.7}}$^{***}$ & -1437.8 & \textbf{\textcolor{blue}{2140.9}}$^{***}$ \\
1975Q2--2023Q1 & (-4.3) & (-10.2) & (\textbf{\textcolor{blue}{2.6}}) & (\textbf{\textcolor{blue}{37.9}}) & (-7.7) & (\textbf{\textcolor{blue}{11.3}}) \\
\addlinespace
Stagflation & -111.8$^{*}$ & \textbf{\textcolor{blue}{326.8}}$^{***}$ & \textbf{\textcolor{blue}{33.0}} & \textbf{\textcolor{blue}{47.7}} & \textbf{\textcolor{blue}{23.4}} & \textbf{\textcolor{blue}{34.3}} \\
1975Q2--1979Q4 & (-6.8) & (\textbf{\textcolor{blue}{22.0}}) & (\textbf{\textcolor{blue}{5.8}}) & (\textbf{\textcolor{blue}{5.1}}) & (\textbf{\textcolor{blue}{1.5}}) & (\textbf{\textcolor{blue}{2.6}}) \\
\addlinespace
Thatcher & -90.2 & \textbf{\textcolor{blue}{274.3}} & -229.0 & \textbf{\textcolor{blue}{401.4}}$^{***}$ & \textbf{\textcolor{blue}{270.8}}$^{***}$ & \textbf{\textcolor{blue}{1721.9}}$^{***}$ \\
1979Q1--1989Q4 & (-2.1) & (\textbf{\textcolor{blue}{6.2}}) & (-5.2) & (\textbf{\textcolor{blue}{9.1}}) & (\textbf{\textcolor{blue}{6.2}}) & (\textbf{\textcolor{blue}{39.1}}) \\
\addlinespace
ERM crisis & \textbf{\textcolor{blue}{65.6}}$^{**}$ & \textbf{\textcolor{blue}{288.5}}$^{***}$ & \textbf{\textcolor{blue}{63.8}}$^{***}$ & \textbf{\textcolor{blue}{238.9}}$^{***}$ & \textbf{\textcolor{blue}{10.2}}$^{**}$ & \textbf{\textcolor{blue}{649.0}}$^{***}$ \\
1990Q1--1992Q4 & (\textbf{\textcolor{blue}{5.5}}) & (\textbf{\textcolor{blue}{24.0}}) & (\textbf{\textcolor{blue}{5.3}}) & (\textbf{\textcolor{blue}{19.9}}) & (\textbf{\textcolor{blue}{0.9}}) & (\textbf{\textcolor{blue}{54.1}}) \\
\addlinespace
Great Moderation & -739.7$^{***}$ & \textbf{\textcolor{blue}{1302.6}}$^{***}$ & -172.2$^{***}$ & \textbf{\textcolor{blue}{2990.8}}$^{***}$ & -265.1$^{**}$ & \textbf{\textcolor{blue}{3590.0}}$^{***}$ \\
1993Q1--2007Q4 & (-12.3) & (\textbf{\textcolor{blue}{21.7}}) & (-2.9) & (\textbf{\textcolor{blue}{49.8}}) & (-4.4) & (\textbf{\textcolor{blue}{59.8}}) \\
\addlinespace
GFC & \textbf{\textcolor{blue}{110.8}}$^{***}$ & -177.9 & \textbf{\textcolor{blue}{250.3}}$^{***}$ & \textbf{\textcolor{blue}{1119.5}}$^{***}$ & -1661.3 & -5540.9 \\
2007Q1--2012Q4 & (\textbf{\textcolor{blue}{4.6}}) & (-7.4) & (\textbf{\textcolor{blue}{10.4}}) & (\textbf{\textcolor{blue}{46.6}}) & (-69.2) & (-230.9) \\
\addlinespace
Brexit & -543.0$^{***}$ & \textbf{\textcolor{blue}{892.7}}$^{***}$ & \textbf{\textcolor{blue}{82.1}} & \textbf{\textcolor{blue}{1984.5}}$^{***}$ & \textbf{\textcolor{blue}{170.3}}$^{***}$ & \textbf{\textcolor{blue}{1843.9}}$^{***}$ \\
2013Q1--2019Q4 & (-19.4) & (\textbf{\textcolor{blue}{31.9}}) & (\textbf{\textcolor{blue}{2.9}}) & (\textbf{\textcolor{blue}{70.9}}) & (\textbf{\textcolor{blue}{6.1}}) & (\textbf{\textcolor{blue}{65.9}}) \\
\addlinespace
COVID-19 & \textbf{\textcolor{blue}{354.6}}$^{***}$ & -4792.9$^{**}$ & \textbf{\textcolor{blue}{49.4}}$^{***}$ & \textbf{\textcolor{blue}{293.0}}$^{***}$ & -51.4 & \textbf{\textcolor{blue}{36.5}} \\
2020Q1--2021Q2 & (\textbf{\textcolor{blue}{59.4}}) & (-983.3) & (\textbf{\textcolor{blue}{9.9}}) & (\textbf{\textcolor{blue}{58.6}}) & (-10.3) & (\textbf{\textcolor{blue}{7.3}}) \\
\addlinespace
Since 2021Q3 & \textbf{\textcolor{blue}{72.9}}$^{**}$ & \textbf{\textcolor{blue}{26.6}}$^{***}$ & \textbf{\textcolor{blue}{249.0}}$^{***}$ & \textbf{\textcolor{blue}{34.8}} & \textbf{\textcolor{blue}{99.3}}$^{***}$ & -126.1 \\
2021Q3--2024Q1 & (\textbf{\textcolor{blue}{4.0}}) & (\textbf{\textcolor{blue}{2.2}}) & (\textbf{\textcolor{blue}{33.7}}) & (\textbf{\textcolor{blue}{6.4}}) & (\textbf{\textcolor{blue}{12.5}}) & (-13.9) \\
\midrule
\multicolumn{7}{c}{\cellcolor{gray!15}\textbf{Panel B: GW Conditional Test}} \\
\midrule
Full Sample & -815.0$^{***}$ & -1898.3$^{***}$ & \textbf{\textcolor{blue}{481.1}}$^{**}$ & \textbf{\textcolor{blue}{7135.7}}$^{***}$ & -1437.8$^{*}$ & \textbf{\textcolor{blue}{2140.9}}$^{***}$ \\
1975Q2--2023Q1 & (-4.3) & (-10.2) & (\textbf{\textcolor{blue}{2.6}}) & (\textbf{\textcolor{blue}{37.9}}) & (-7.7) & (\textbf{\textcolor{blue}{11.3}}) \\
\addlinespace
Stagflation & -111.8$^{***}$ & \textbf{\textcolor{blue}{326.8}}$^{***}$ & \textbf{\textcolor{blue}{33.0}}$^{**}$ & \textbf{\textcolor{blue}{47.7}}$^{**}$ & \textbf{\textcolor{blue}{23.4}}$^{***}$ & \textbf{\textcolor{blue}{34.3}}$^{***}$ \\
1975Q2--1979Q4 & (-6.8) & (\textbf{\textcolor{blue}{22.0}}) & (\textbf{\textcolor{blue}{5.8}}) & (\textbf{\textcolor{blue}{5.1}}) & (\textbf{\textcolor{blue}{1.5}}) & (\textbf{\textcolor{blue}{2.6}}) \\
\addlinespace
Thatcher & -90.2$^{***}$ & \textbf{\textcolor{blue}{274.3}}$^{***}$ & -229.0$^{*}$ & \textbf{\textcolor{blue}{401.4}}$^{***}$ & \textbf{\textcolor{blue}{270.8}}$^{***}$ & \textbf{\textcolor{blue}{1721.9}}$^{***}$ \\
1979Q1--1989Q4 & (-2.1) & (\textbf{\textcolor{blue}{6.2}}) & (-5.2) & (\textbf{\textcolor{blue}{9.1}}) & (\textbf{\textcolor{blue}{6.2}}) & (\textbf{\textcolor{blue}{39.1}}) \\
\addlinespace
ERM crisis & \textbf{\textcolor{blue}{65.6}}$^{***}$ & \textbf{\textcolor{blue}{288.5}}$^{***}$ & \textbf{\textcolor{blue}{63.8}}$^{***}$ & \textbf{\textcolor{blue}{238.9}}$^{***}$ & \textbf{\textcolor{blue}{10.2}}$^{***}$ & \textbf{\textcolor{blue}{649.0}}$^{***}$ \\
1990Q1--1992Q4 & (\textbf{\textcolor{blue}{5.5}}) & (\textbf{\textcolor{blue}{24.0}}) & (\textbf{\textcolor{blue}{5.3}}) & (\textbf{\textcolor{blue}{19.9}}) & (\textbf{\textcolor{blue}{0.9}}) & (\textbf{\textcolor{blue}{54.1}}) \\
\addlinespace
Great Moderation & -739.7$^{***}$ & \textbf{\textcolor{blue}{1302.6}}$^{***}$ & -172.2$^{***}$ & \textbf{\textcolor{blue}{2990.8}}$^{***}$ & -265.1$^{*}$ & \textbf{\textcolor{blue}{3590.0}}$^{***}$ \\
1993Q1--2007Q4 & (-12.3) & (\textbf{\textcolor{blue}{21.7}}) & (-2.9) & (\textbf{\textcolor{blue}{49.8}}) & (-4.4) & (\textbf{\textcolor{blue}{59.8}}) \\
\addlinespace
GFC & \textbf{\textcolor{blue}{110.8}}$^{***}$ & -177.9$^{*}$ & \textbf{\textcolor{blue}{250.3}}$^{***}$ & \textbf{\textcolor{blue}{1119.5}}$^{***}$ & -1661.3 & -5540.9$^{***}$ \\
2007Q1--2012Q4 & (\textbf{\textcolor{blue}{4.6}}) & (-7.4) & (\textbf{\textcolor{blue}{10.4}}) & (\textbf{\textcolor{blue}{46.6}}) & (-69.2) & (-230.9) \\
\addlinespace
Brexit & -543.0$^{***}$ & \textbf{\textcolor{blue}{892.7}}$^{***}$ & \textbf{\textcolor{blue}{82.1}}$^{***}$ & \textbf{\textcolor{blue}{1984.5}}$^{***}$ & \textbf{\textcolor{blue}{170.3}}$^{***}$ & \textbf{\textcolor{blue}{1843.9}}$^{***}$ \\
2013Q1--2019Q4 & (-19.4) & (\textbf{\textcolor{blue}{31.9}}) & (\textbf{\textcolor{blue}{2.9}}) & (\textbf{\textcolor{blue}{70.9}}) & (\textbf{\textcolor{blue}{6.1}}) & (\textbf{\textcolor{blue}{65.9}}) \\
\addlinespace
COVID-19 & \textbf{\textcolor{blue}{354.6}}$^{***}$ & -4792.9$^{**}$ & \textbf{\textcolor{blue}{49.4}}$^{***}$ & \textbf{\textcolor{blue}{293.0}}$^{***}$ & -51.4$^{***}$ & \textbf{\textcolor{blue}{36.5}}$^{***}$ \\
2020Q1--2021Q2 & (\textbf{\textcolor{blue}{59.4}}) & (-983.3) & (\textbf{\textcolor{blue}{9.9}}) & (\textbf{\textcolor{blue}{58.6}}) & (-10.3) & (\textbf{\textcolor{blue}{7.3}}) \\
\addlinespace
Since 2021Q3 & \textbf{\textcolor{blue}{72.9}}$^{*}$ & \textbf{\textcolor{blue}{26.6}}$^{***}$ & \textbf{\textcolor{blue}{249.0}}$^{***}$ & \textbf{\textcolor{blue}{34.8}}$^{***}$ & \textbf{\textcolor{blue}{99.3}}$^{***}$ & -126.1$^{***}$ \\
2021Q3--2024Q1 & (\textbf{\textcolor{blue}{4.0}}) & (\textbf{\textcolor{blue}{2.2}}) & (\textbf{\textcolor{blue}{33.7}}) & (\textbf{\textcolor{blue}{6.4}}) & (\textbf{\textcolor{blue}{12.5}}) & (-13.9) \\
\bottomrule
\end{tabular}
\begin{tablenotes}
\tiny
\item \textbf{Notes:} Panel A shows Cumulative log score differences ($VAR^{\sigma,\kappa}$ - Competitor) using Giacomini-White unconditional test. Panel B shows Cumulative log score differences using Giacomini-White conditional test. Values in parentheses show average score differences per quarter within each period. Positive values (\textcolor{blue}{\textbf{bold blue}}) indicate the proposed model outperforms the competitor. All values are averaged across forecast horizons 1-8. Significance levels: *** p$<$0.01, ** p$<$0.05, * p$<$0.1.
\end{tablenotes}
\end{threeparttable}
\end{table}

\begin{table}[H]
\centering
\caption{Cumulative Weighted Left Tail Log Score Differences}
\label{tab:cumulative_weighted_left_tail_logscore_uk_gw_comparison}
\begin{threeparttable}
\scriptsize
\begin{tabular}{@{}l*{6}{c}@{}}
\toprule
 & \multicolumn{6}{c}{\textbf{Variables}} \\
\cmidrule(lr){2-7}
\textbf{Period} & \multicolumn{2}{c}{\textbf{GNP}} & \multicolumn{2}{c}{\textbf{GNP Deflator}} & \multicolumn{2}{c}{\textbf{SPREAD}} \\
\cmidrule(lr){2-3} \cmidrule(lr){4-5} \cmidrule(lr){6-7}
 & \textbf{$VAR^{\sigma,\kappa}_{rest}$} & \textbf{$VAR^{\sigma}$} & \textbf{$VAR^{\sigma,\kappa}_{rest}$} & \textbf{$VAR^{\sigma}$} & \textbf{$VAR^{\sigma,\kappa}_{rest}$} & \textbf{$VAR^{\sigma}$} \\
\midrule
\multicolumn{7}{c}{\cellcolor{gray!15}\textbf{Panel A: GW Unconditional Test}} \\
\midrule
Full Sample & -49.5$^{***}$ & -2961.7$^{***}$ & \textbf{\textcolor{blue}{93.5}}$^{**}$ & \textbf{\textcolor{blue}{4080.6}}$^{***}$ & \textbf{\textcolor{blue}{471.0}}$^{***}$ & \textbf{\textcolor{blue}{4653.2}}$^{***}$ \\
1975Q2--2023Q1 & (-0.2) & (-15.9) & (\textbf{\textcolor{blue}{0.5}}) & (\textbf{\textcolor{blue}{21.7}}) & (\textbf{\textcolor{blue}{2.5}}) & (\textbf{\textcolor{blue}{24.8}}) \\
\addlinespace
Stagflation & -66.1 & \textbf{\textcolor{blue}{201.5}}$^{***}$ & -9.3$^{*}$ & \textbf{\textcolor{blue}{1.3}} & \textbf{\textcolor{blue}{106.9}}$^{***}$ & \textbf{\textcolor{blue}{195.4}}$^{***}$ \\
1975Q2--1979Q4 & (-4.0) & (\textbf{\textcolor{blue}{13.2}}) & (-0.3) & (\textbf{\textcolor{blue}{0.4}}) & (\textbf{\textcolor{blue}{7.0}}) & (\textbf{\textcolor{blue}{13.2}}) \\
\addlinespace
Thatcher & -67.1 & \textbf{\textcolor{blue}{290.8}}$^{**}$ & -83.0 & \textbf{\textcolor{blue}{250.4}}$^{***}$ & \textbf{\textcolor{blue}{193.4}}$^{***}$ & \textbf{\textcolor{blue}{1111.4}}$^{***}$ \\
1979Q1--1989Q4 & (-1.5) & (\textbf{\textcolor{blue}{6.6}}) & (-1.9) & (\textbf{\textcolor{blue}{5.7}}) & (\textbf{\textcolor{blue}{4.4}}) & (\textbf{\textcolor{blue}{25.3}}) \\
\addlinespace
ERM crisis & \textbf{\textcolor{blue}{50.8}}$^{**}$ & \textbf{\textcolor{blue}{187.7}}$^{***}$ & \textbf{\textcolor{blue}{18.6}}$^{***}$ & \textbf{\textcolor{blue}{97.0}}$^{***}$ & \textbf{\textcolor{blue}{18.3}}$^{***}$ & \textbf{\textcolor{blue}{324.7}}$^{***}$ \\
1990Q1--1992Q4 & (\textbf{\textcolor{blue}{4.2}}) & (\textbf{\textcolor{blue}{15.6}}) & (\textbf{\textcolor{blue}{1.5}}) & (\textbf{\textcolor{blue}{8.1}}) & (\textbf{\textcolor{blue}{1.5}}) & (\textbf{\textcolor{blue}{27.1}}) \\
\addlinespace
Great Moderation & -369.5$^{***}$ & \textbf{\textcolor{blue}{623.3}}$^{***}$ & -101.2$^{***}$ & \textbf{\textcolor{blue}{1795.3}}$^{***}$ & \textbf{\textcolor{blue}{71.9}} & \textbf{\textcolor{blue}{1975.3}}$^{***}$ \\
1993Q1--2007Q4 & (-6.2) & (\textbf{\textcolor{blue}{10.4}}) & (-1.7) & (\textbf{\textcolor{blue}{29.9}}) & (\textbf{\textcolor{blue}{1.2}}) & (\textbf{\textcolor{blue}{32.9}}) \\
\addlinespace
GFC & \textbf{\textcolor{blue}{182.4}}$^{***}$ & -234.9 & \textbf{\textcolor{blue}{137.0}}$^{***}$ & \textbf{\textcolor{blue}{622.6}}$^{***}$ & \textbf{\textcolor{blue}{36.3}}$^{**}$ & \textbf{\textcolor{blue}{377.8}}$^{**}$ \\
2007Q1--2012Q4 & (\textbf{\textcolor{blue}{7.6}}) & (-9.8) & (\textbf{\textcolor{blue}{5.7}}) & (\textbf{\textcolor{blue}{25.9}}) & (\textbf{\textcolor{blue}{1.5}}) & (\textbf{\textcolor{blue}{15.7}}) \\
\addlinespace
Brexit & -272.5$^{***}$ & \textbf{\textcolor{blue}{453.0}}$^{***}$ & \textbf{\textcolor{blue}{41.7}} & \textbf{\textcolor{blue}{1191.5}}$^{***}$ & \textbf{\textcolor{blue}{57.5}}$^{***}$ & \textbf{\textcolor{blue}{679.7}}$^{***}$ \\
2013Q1--2019Q4 & (-9.7) & (\textbf{\textcolor{blue}{16.2}}) & (\textbf{\textcolor{blue}{1.5}}) & (\textbf{\textcolor{blue}{42.6}}) & (\textbf{\textcolor{blue}{2.1}}) & (\textbf{\textcolor{blue}{24.3}}) \\
\addlinespace
COVID-19 & \textbf{\textcolor{blue}{453.6}}$^{***}$ & -4440.2$^{**}$ & \textbf{\textcolor{blue}{31.8}}$^{***}$ & \textbf{\textcolor{blue}{180.9}}$^{***}$ & \textbf{\textcolor{blue}{7.3}}$^{*}$ & \textbf{\textcolor{blue}{100.2}}$^{**}$ \\
2020Q1--2021Q2 & (\textbf{\textcolor{blue}{85.4}}) & (-899.8) & (\textbf{\textcolor{blue}{6.4}}) & (\textbf{\textcolor{blue}{36.2}}) & (\textbf{\textcolor{blue}{1.5}}) & (\textbf{\textcolor{blue}{20.0}}) \\
\addlinespace
Since 2021Q3 & \textbf{\textcolor{blue}{29.5}}$^{**}$ & \textbf{\textcolor{blue}{5.5}}$^{***}$ & \textbf{\textcolor{blue}{64.0}}$^{***}$ & \textbf{\textcolor{blue}{20.9}} & \textbf{\textcolor{blue}{9.0}}$^{***}$ & \textbf{\textcolor{blue}{111.3}} \\
2021Q3--2024Q1 & (\textbf{\textcolor{blue}{1.5}}) & (\textbf{\textcolor{blue}{0.1}}) & (\textbf{\textcolor{blue}{8.9}}) & (\textbf{\textcolor{blue}{3.4}}) & (\textbf{\textcolor{blue}{1.2}}) & (\textbf{\textcolor{blue}{16.0}}) \\
\midrule
\multicolumn{7}{c}{\cellcolor{gray!15}\textbf{Panel B: GW Conditional Test}} \\
\midrule
Full Sample & -49.5$^{***}$ & -2961.7$^{***}$ & \textbf{\textcolor{blue}{93.5}}$^{***}$ & \textbf{\textcolor{blue}{4080.6}}$^{***}$ & \textbf{\textcolor{blue}{471.0}}$^{***}$ & \textbf{\textcolor{blue}{4653.2}}$^{***}$ \\
1975Q2--2023Q1 & (-0.2) & (-15.9) & (\textbf{\textcolor{blue}{0.5}}) & (\textbf{\textcolor{blue}{21.7}}) & (\textbf{\textcolor{blue}{2.5}}) & (\textbf{\textcolor{blue}{24.8}}) \\
\addlinespace
Stagflation & -66.1$^{***}$ & \textbf{\textcolor{blue}{201.5}}$^{***}$ & -9.3$^{***}$ & \textbf{\textcolor{blue}{1.3}}$^{***}$ & \textbf{\textcolor{blue}{106.9}}$^{***}$ & \textbf{\textcolor{blue}{195.4}}$^{***}$ \\
1975Q2--1979Q4 & (-4.0) & (\textbf{\textcolor{blue}{13.2}}) & (-0.3) & (\textbf{\textcolor{blue}{0.4}}) & (\textbf{\textcolor{blue}{7.0}}) & (\textbf{\textcolor{blue}{13.2}}) \\
\addlinespace
Thatcher & -67.1$^{***}$ & \textbf{\textcolor{blue}{290.8}}$^{**}$ & -83.0$^{*}$ & \textbf{\textcolor{blue}{250.4}}$^{***}$ & \textbf{\textcolor{blue}{193.4}}$^{***}$ & \textbf{\textcolor{blue}{1111.4}}$^{***}$ \\
1979Q1--1989Q4 & (-1.5) & (\textbf{\textcolor{blue}{6.6}}) & (-1.9) & (\textbf{\textcolor{blue}{5.7}}) & (\textbf{\textcolor{blue}{4.4}}) & (\textbf{\textcolor{blue}{25.3}}) \\
\addlinespace
ERM crisis & \textbf{\textcolor{blue}{50.8}}$^{***}$ & \textbf{\textcolor{blue}{187.7}}$^{***}$ & \textbf{\textcolor{blue}{18.6}}$^{***}$ & \textbf{\textcolor{blue}{97.0}}$^{***}$ & \textbf{\textcolor{blue}{18.3}}$^{***}$ & \textbf{\textcolor{blue}{324.7}}$^{***}$ \\
1990Q1--1992Q4 & (\textbf{\textcolor{blue}{4.2}}) & (\textbf{\textcolor{blue}{15.6}}) & (\textbf{\textcolor{blue}{1.5}}) & (\textbf{\textcolor{blue}{8.1}}) & (\textbf{\textcolor{blue}{1.5}}) & (\textbf{\textcolor{blue}{27.1}}) \\
\addlinespace
Great Moderation & -369.5$^{***}$ & \textbf{\textcolor{blue}{623.3}}$^{***}$ & -101.2$^{***}$ & \textbf{\textcolor{blue}{1795.3}}$^{***}$ & \textbf{\textcolor{blue}{71.9}} & \textbf{\textcolor{blue}{1975.3}}$^{***}$ \\
1993Q1--2007Q4 & (-6.2) & (\textbf{\textcolor{blue}{10.4}}) & (-1.7) & (\textbf{\textcolor{blue}{29.9}}) & (\textbf{\textcolor{blue}{1.2}}) & (\textbf{\textcolor{blue}{32.9}}) \\
\addlinespace
GFC & \textbf{\textcolor{blue}{182.4}}$^{***}$ & -234.9 & \textbf{\textcolor{blue}{137.0}}$^{***}$ & \textbf{\textcolor{blue}{622.6}}$^{***}$ & \textbf{\textcolor{blue}{36.3}}$^{**}$ & \textbf{\textcolor{blue}{377.8}}$^{**}$ \\
2007Q1--2012Q4 & (\textbf{\textcolor{blue}{7.6}}) & (-9.8) & (\textbf{\textcolor{blue}{5.7}}) & (\textbf{\textcolor{blue}{25.9}}) & (\textbf{\textcolor{blue}{1.5}}) & (\textbf{\textcolor{blue}{15.7}}) \\
\addlinespace
Brexit & -272.5$^{***}$ & \textbf{\textcolor{blue}{453.0}}$^{***}$ & \textbf{\textcolor{blue}{41.7}}$^{***}$ & \textbf{\textcolor{blue}{1191.5}}$^{***}$ & \textbf{\textcolor{blue}{57.5}}$^{***}$ & \textbf{\textcolor{blue}{679.7}}$^{***}$ \\
2013Q1--2019Q4 & (-9.7) & (\textbf{\textcolor{blue}{16.2}}) & (\textbf{\textcolor{blue}{1.5}}) & (\textbf{\textcolor{blue}{42.6}}) & (\textbf{\textcolor{blue}{2.1}}) & (\textbf{\textcolor{blue}{24.3}}) \\
\addlinespace
COVID-19 & \textbf{\textcolor{blue}{453.6}}$^{***}$ & -4440.2$^{***}$ & \textbf{\textcolor{blue}{31.8}}$^{***}$ & \textbf{\textcolor{blue}{180.9}}$^{***}$ & \textbf{\textcolor{blue}{7.3}}$^{***}$ & \textbf{\textcolor{blue}{100.2}}$^{***}$ \\
2020Q1--2021Q2 & (\textbf{\textcolor{blue}{85.4}}) & (-899.8) & (\textbf{\textcolor{blue}{6.4}}) & (\textbf{\textcolor{blue}{36.2}}) & (\textbf{\textcolor{blue}{1.5}}) & (\textbf{\textcolor{blue}{20.0}}) \\
\addlinespace
Since 2021Q3 & \textbf{\textcolor{blue}{29.5}}$^{**}$ & \textbf{\textcolor{blue}{5.5}}$^{***}$ & \textbf{\textcolor{blue}{64.0}}$^{***}$ & \textbf{\textcolor{blue}{20.9}}$^{***}$ & \textbf{\textcolor{blue}{9.0}}$^{***}$ & \textbf{\textcolor{blue}{111.3}}$^{***}$ \\
2021Q3--2024Q1 & (\textbf{\textcolor{blue}{1.5}}) & (\textbf{\textcolor{blue}{0.1}}) & (\textbf{\textcolor{blue}{8.9}}) & (\textbf{\textcolor{blue}{3.4}}) & (\textbf{\textcolor{blue}{1.2}}) & (\textbf{\textcolor{blue}{16.0}}) \\
\bottomrule
\end{tabular}
\begin{tablenotes}
\tiny
\item \textbf{Notes:} Panel A shows Cumulative weighted left tail log score differences ($VAR^{\sigma,\kappa}$ model - Competitor) using Giacomini-White unconditional test. Panel B shows Cumulative weighted left tail log score differences using Giacomini-White conditional test. Values in parentheses show average score differences per quarter within each period. Positive values (\textcolor{blue}{\textbf{bold blue}}) indicate the proposed model outperforms the competitor on left tail forecasting. All values are averaged across forecast horizons 1-8. Significance levels: *** p$<$0.01, ** p$<$0.05, * p$<$0.1.
\end{tablenotes}
\end{threeparttable}
\end{table}

\begin{table}[H]
\centering
\caption{Cumulative Weighted Right Tail Log Score Differences}
\label{tab:cumulative_weighted_right_tail_logscore_uk_gw_comparison}
\begin{threeparttable}
\scriptsize
\begin{tabular}{@{}l*{6}{c}@{}}
\toprule
 & \multicolumn{6}{c}{\textbf{Variables}} \\
\cmidrule(lr){2-7}
\textbf{Period} & \multicolumn{2}{c}{\textbf{GNP}} & \multicolumn{2}{c}{\textbf{GNP Deflator}} & \multicolumn{2}{c}{\textbf{SPREAD}} \\
\cmidrule(lr){2-3} \cmidrule(lr){4-5} \cmidrule(lr){6-7}
 & \textbf{$VAR^{\sigma,\kappa}_{rest}$} & \textbf{$VAR^{\sigma}$} & \textbf{$VAR^{\sigma,\kappa}_{rest}$} & \textbf{$VAR^{\sigma}$} & \textbf{$VAR^{\sigma,\kappa}_{rest}$} & \textbf{$VAR^{\sigma}$} \\
\midrule
\multicolumn{7}{c}{\cellcolor{gray!15}\textbf{Panel A: GW Unconditional Test}} \\
\midrule
Full Sample & -765.5$^{***}$ & \textbf{\textcolor{blue}{1063.4}}$^{***}$ & \textbf{\textcolor{blue}{387.7}}$^{**}$ & \textbf{\textcolor{blue}{3055.1}}$^{***}$ & -1908.7$^{*}$ & -2512.3$^{***}$ \\
1975Q2--2023Q1 & (-4.1) & (\textbf{\textcolor{blue}{5.6}}) & (\textbf{\textcolor{blue}{2.1}}) & (\textbf{\textcolor{blue}{16.2}}) & (-10.2) & (-13.5) \\
\addlinespace
Stagflation & -45.7$^{*}$ & \textbf{\textcolor{blue}{125.3}}$^{**}$ & \textbf{\textcolor{blue}{42.3}} & \textbf{\textcolor{blue}{46.4}} & -83.5$^{***}$ & -161.2$^{*}$ \\
1975Q2--1979Q4 & (-2.8) & (\textbf{\textcolor{blue}{8.7}}) & (\textbf{\textcolor{blue}{6.2}}) & (\textbf{\textcolor{blue}{4.7}}) & (-5.5) & (-10.6) \\
\addlinespace
Thatcher & -23.1 & -16.6 & -146.0 & \textbf{\textcolor{blue}{151.1}}$^{***}$ & \textbf{\textcolor{blue}{77.3}}$^{***}$ & \textbf{\textcolor{blue}{610.5}}$^{***}$ \\
1979Q1--1989Q4 & (-0.5) & (-0.4) & (-3.3) & (\textbf{\textcolor{blue}{3.4}}) & (\textbf{\textcolor{blue}{1.8}}) & (\textbf{\textcolor{blue}{13.9}}) \\
\addlinespace
ERM crisis & \textbf{\textcolor{blue}{14.8}}$^{*}$ & \textbf{\textcolor{blue}{100.8}}$^{***}$ & \textbf{\textcolor{blue}{45.3}}$^{***}$ & \textbf{\textcolor{blue}{141.9}}$^{***}$ & -8.1$^{***}$ & \textbf{\textcolor{blue}{324.2}}$^{***}$ \\
1990Q1--1992Q4 & (\textbf{\textcolor{blue}{1.2}}) & (\textbf{\textcolor{blue}{8.4}}) & (\textbf{\textcolor{blue}{3.8}}) & (\textbf{\textcolor{blue}{11.8}}) & (-0.7) & (\textbf{\textcolor{blue}{27.0}}) \\
\addlinespace
Great Moderation & -370.2$^{***}$ & \textbf{\textcolor{blue}{679.3}}$^{***}$ & -71.0$^{***}$ & \textbf{\textcolor{blue}{1195.5}}$^{***}$ & -337.0$^{***}$ & \textbf{\textcolor{blue}{1614.7}}$^{***}$ \\
1993Q1--2007Q4 & (-6.2) & (\textbf{\textcolor{blue}{11.3}}) & (-1.2) & (\textbf{\textcolor{blue}{19.9}}) & (-5.6) & (\textbf{\textcolor{blue}{26.9}}) \\
\addlinespace
GFC & -71.6$^{***}$ & \textbf{\textcolor{blue}{56.9}}$^{***}$ & \textbf{\textcolor{blue}{113.3}}$^{***}$ & \textbf{\textcolor{blue}{496.9}}$^{***}$ & -1697.5 & -5918.7 \\
2007Q1--2012Q4 & (-3.0) & (\textbf{\textcolor{blue}{2.4}}) & (\textbf{\textcolor{blue}{4.7}}) & (\textbf{\textcolor{blue}{20.7}}) & (-70.7) & (-246.6) \\
\addlinespace
Brexit & -270.5$^{***}$ & \textbf{\textcolor{blue}{439.7}}$^{***}$ & \textbf{\textcolor{blue}{40.5}}$^{*}$ & \textbf{\textcolor{blue}{793.0}}$^{***}$ & \textbf{\textcolor{blue}{112.8}}$^{***}$ & \textbf{\textcolor{blue}{1164.1}}$^{***}$ \\
2013Q1--2019Q4 & (-9.7) & (\textbf{\textcolor{blue}{15.7}}) & (\textbf{\textcolor{blue}{1.4}}) & (\textbf{\textcolor{blue}{28.3}}) & (\textbf{\textcolor{blue}{4.0}}) & (\textbf{\textcolor{blue}{41.6}}) \\
\addlinespace
COVID-19 & -99.0$^{***}$ & -352.6$^{***}$ & \textbf{\textcolor{blue}{17.6}}$^{***}$ & \textbf{\textcolor{blue}{112.1}}$^{***}$ & -58.7 & -63.6 \\
2020Q1--2021Q2 & (-25.9) & (-83.4) & (\textbf{\textcolor{blue}{3.5}}) & (\textbf{\textcolor{blue}{22.4}}) & (-11.7) & (-12.7) \\
\addlinespace
Since 2021Q3 & \textbf{\textcolor{blue}{43.5}}$^{**}$ & \textbf{\textcolor{blue}{21.1}}$^{**}$ & \textbf{\textcolor{blue}{185.0}}$^{***}$ & \textbf{\textcolor{blue}{13.9}} & \textbf{\textcolor{blue}{90.3}}$^{***}$ & -237.5 \\
2021Q3--2024Q1 & (\textbf{\textcolor{blue}{2.4}}) & (\textbf{\textcolor{blue}{2.1}}) & (\textbf{\textcolor{blue}{24.8}}) & (\textbf{\textcolor{blue}{3.0}}) & (\textbf{\textcolor{blue}{11.3}}) & (-29.8) \\
\midrule
\multicolumn{7}{c}{\cellcolor{gray!15}\textbf{Panel B: GW Conditional Test}} \\
\midrule
Full Sample & -765.5$^{***}$ & \textbf{\textcolor{blue}{1063.4}}$^{***}$ & \textbf{\textcolor{blue}{387.7}}$^{*}$ & \textbf{\textcolor{blue}{3055.1}}$^{***}$ & -1908.7 & -2512.3$^{***}$ \\
1975Q2--2023Q1 & (-4.1) & (\textbf{\textcolor{blue}{5.6}}) & (\textbf{\textcolor{blue}{2.1}}) & (\textbf{\textcolor{blue}{16.2}}) & (-10.2) & (-13.5) \\
\addlinespace
Stagflation & -45.7$^{***}$ & \textbf{\textcolor{blue}{125.3}}$^{**}$ & \textbf{\textcolor{blue}{42.3}}$^{**}$ & \textbf{\textcolor{blue}{46.4}}$^{**}$ & -83.5$^{***}$ & -161.2$^{***}$ \\
1975Q2--1979Q4 & (-2.8) & (\textbf{\textcolor{blue}{8.7}}) & (\textbf{\textcolor{blue}{6.2}}) & (\textbf{\textcolor{blue}{4.7}}) & (-5.5) & (-10.6) \\
\addlinespace
Thatcher & -23.1$^{***}$ & -16.6$^{***}$ & -146.0 & \textbf{\textcolor{blue}{151.1}}$^{***}$ & \textbf{\textcolor{blue}{77.3}}$^{***}$ & \textbf{\textcolor{blue}{610.5}}$^{***}$ \\
1979Q1--1989Q4 & (-0.5) & (-0.4) & (-3.3) & (\textbf{\textcolor{blue}{3.4}}) & (\textbf{\textcolor{blue}{1.8}}) & (\textbf{\textcolor{blue}{13.9}}) \\
\addlinespace
ERM crisis & \textbf{\textcolor{blue}{14.8}}$^{***}$ & \textbf{\textcolor{blue}{100.8}}$^{***}$ & \textbf{\textcolor{blue}{45.3}}$^{***}$ & \textbf{\textcolor{blue}{141.9}}$^{***}$ & -8.1$^{***}$ & \textbf{\textcolor{blue}{324.2}}$^{***}$ \\
1990Q1--1992Q4 & (\textbf{\textcolor{blue}{1.2}}) & (\textbf{\textcolor{blue}{8.4}}) & (\textbf{\textcolor{blue}{3.8}}) & (\textbf{\textcolor{blue}{11.8}}) & (-0.7) & (\textbf{\textcolor{blue}{27.0}}) \\
\addlinespace
Great Moderation & -370.2$^{***}$ & \textbf{\textcolor{blue}{679.3}}$^{***}$ & -71.0$^{***}$ & \textbf{\textcolor{blue}{1195.5}}$^{***}$ & -337.0$^{**}$ & \textbf{\textcolor{blue}{1614.7}}$^{***}$ \\
1993Q1--2007Q4 & (-6.2) & (\textbf{\textcolor{blue}{11.3}}) & (-1.2) & (\textbf{\textcolor{blue}{19.9}}) & (-5.6) & (\textbf{\textcolor{blue}{26.9}}) \\
\addlinespace
GFC & -71.6$^{***}$ & \textbf{\textcolor{blue}{56.9}}$^{**}$ & \textbf{\textcolor{blue}{113.3}}$^{***}$ & \textbf{\textcolor{blue}{496.9}}$^{***}$ & -1697.5 & -5918.7$^{**}$ \\
2007Q1--2012Q4 & (-3.0) & (\textbf{\textcolor{blue}{2.4}}) & (\textbf{\textcolor{blue}{4.7}}) & (\textbf{\textcolor{blue}{20.7}}) & (-70.7) & (-246.6) \\
\addlinespace
Brexit & -270.5$^{***}$ & \textbf{\textcolor{blue}{439.7}}$^{***}$ & \textbf{\textcolor{blue}{40.5}}$^{***}$ & \textbf{\textcolor{blue}{793.0}}$^{***}$ & \textbf{\textcolor{blue}{112.8}}$^{***}$ & \textbf{\textcolor{blue}{1164.1}}$^{***}$ \\
2013Q1--2019Q4 & (-9.7) & (\textbf{\textcolor{blue}{15.7}}) & (\textbf{\textcolor{blue}{1.4}}) & (\textbf{\textcolor{blue}{28.3}}) & (\textbf{\textcolor{blue}{4.0}}) & (\textbf{\textcolor{blue}{41.6}}) \\
\addlinespace
COVID-19 & -99.0 & -352.6$^{***}$ & \textbf{\textcolor{blue}{17.6}}$^{***}$ & \textbf{\textcolor{blue}{112.1}}$^{***}$ & -58.7$^{***}$ & -63.6$^{***}$ \\
2020Q1--2021Q2 & (-25.9) & (-83.4) & (\textbf{\textcolor{blue}{3.5}}) & (\textbf{\textcolor{blue}{22.4}}) & (-11.7) & (-12.7) \\
\addlinespace
Since 2021Q3 & \textbf{\textcolor{blue}{43.5}}$^{*}$ & \textbf{\textcolor{blue}{21.1}}$^{***}$ & \textbf{\textcolor{blue}{185.0}}$^{***}$ & \textbf{\textcolor{blue}{13.9}}$^{***}$ & \textbf{\textcolor{blue}{90.3}}$^{***}$ & -237.5$^{***}$ \\
2021Q3--2024Q1 & (\textbf{\textcolor{blue}{2.4}}) & (\textbf{\textcolor{blue}{2.1}}) & (\textbf{\textcolor{blue}{24.8}}) & (\textbf{\textcolor{blue}{3.0}}) & (\textbf{\textcolor{blue}{11.3}}) & (-29.8) \\
\bottomrule
\end{tabular}
\begin{tablenotes}
\tiny
\item \textbf{Notes:} Panel A shows Cumulative weighted right tail log score differences ($VAR^{\sigma,\kappa}$ model - Competitor) using Giacomini-White unconditional test. Panel B shows Cumulative weighted right tail log score differences using Giacomini-White conditional test. Values in parentheses show average score differences per quarter within each period. Positive values (\textcolor{blue}{\textbf{bold blue}}) indicate the proposed model outperforms the competitor on right tail forecasting. All values are averaged across forecast horizons 1-8. Significance levels: *** p$<$0.01, ** p$<$0.05, * p$<$0.1.
\end{tablenotes}
\end{threeparttable}
\end{table}

\begin{table}[H]
\centering
\caption{CRPS Ratios}
\label{tab:crps_uk_gw_comparison}
\begin{threeparttable}
\scriptsize
\begin{tabular}{@{}l*{6}{c}@{}}
\toprule
 & \multicolumn{6}{c}{\textbf{Variables}} \\
\cmidrule(lr){2-7}
\textbf{Period} & \multicolumn{2}{c}{\textbf{GNP}} & \multicolumn{2}{c}{\textbf{GNP Deflator}} & \multicolumn{2}{c}{\textbf{SPREAD}} \\
\cmidrule(lr){2-3} \cmidrule(lr){4-5} \cmidrule(lr){6-7}
 & \textbf{$VAR^{\sigma,\kappa}_{rest}$} & \textbf{$VAR^{\sigma}$} & \textbf{$VAR^{\sigma,\kappa}_{rest}$} & \textbf{$VAR^{\sigma}$} & \textbf{$VAR^{\sigma,\kappa}_{rest}$} & \textbf{$VAR^{\sigma}$} \\
\midrule
\multicolumn{7}{c}{\cellcolor{gray!15}\textbf{Panel A: GW Unconditional Test}} \\
\midrule
Full Sample & 1.008 & \textbf{\textcolor{blue}{0.887}}$^{***}$ & \textbf{\textcolor{blue}{0.950}} & \textbf{\textcolor{blue}{0.828}}$^{***}$ & \textbf{\textcolor{blue}{0.985}}$^{*}$ & \textbf{\textcolor{blue}{0.818}}$^{***}$ \\
1975Q2--2023Q1 & & & & & & \\
\addlinespace
Stagflation & 1.131 & \textbf{\textcolor{blue}{0.722}}$^{***}$ & \textbf{\textcolor{blue}{0.861}} & \textbf{\textcolor{blue}{0.892}} & \textbf{\textcolor{blue}{0.985}}$^{**}$ & \textbf{\textcolor{blue}{0.976}} \\
1975Q2--1979Q4 & & & & & & \\
\addlinespace
Thatcher & 1.017 & \textbf{\textcolor{blue}{0.917}} & \textbf{\textcolor{blue}{0.988}} & \textbf{\textcolor{blue}{0.946}}$^{***}$ & \textbf{\textcolor{blue}{0.960}}$^{***}$ & \textbf{\textcolor{blue}{0.798}}$^{***}$ \\
1979Q1--1989Q4 & & & & & & \\
\addlinespace
ERM crisis & \textbf{\textcolor{blue}{0.945}}$^{***}$ & \textbf{\textcolor{blue}{0.738}}$^{***}$ & \textbf{\textcolor{blue}{0.944}}$^{***}$ & \textbf{\textcolor{blue}{0.896}}$^{***}$ & 1.003$^{**}$ & \textbf{\textcolor{blue}{0.697}}$^{***}$ \\
1990Q1--1992Q4 & & & & & & \\
\addlinespace
Great Moderation & 1.046$^{***}$ & \textbf{\textcolor{blue}{0.860}}$^{***}$ & 1.056$^{*}$ & \textbf{\textcolor{blue}{0.700}}$^{***}$ & 1.042$^{**}$ & \textbf{\textcolor{blue}{0.719}}$^{***}$ \\
1993Q1--2007Q4 & & & & & & \\
\addlinespace
GFC & 1.113$^{**}$ & 1.042 & \textbf{\textcolor{blue}{0.883}}$^{***}$ & \textbf{\textcolor{blue}{0.762}}$^{***}$ & \textbf{\textcolor{blue}{0.982}}$^{*}$ & 1.009 \\
2007Q1--2012Q4 & & & & & & \\
\addlinespace
Brexit & 1.160$^{***}$ & \textbf{\textcolor{blue}{0.808}}$^{***}$ & \textbf{\textcolor{blue}{0.978}} & \textbf{\textcolor{blue}{0.647}}$^{***}$ & \textbf{\textcolor{blue}{0.953}}$^{***}$ & \textbf{\textcolor{blue}{0.653}}$^{***}$ \\
2013Q1--2019Q4 & & & & & & \\
\addlinespace
COVID-19 & \textbf{\textcolor{blue}{0.929}}$^{**}$ & 1.024$^{***}$ & \textbf{\textcolor{blue}{0.967}}$^{***}$ & \textbf{\textcolor{blue}{0.691}}$^{***}$ & \textbf{\textcolor{blue}{0.969}}$^{**}$ & \textbf{\textcolor{blue}{0.875}} \\
2020Q1--2021Q2 & & & & & & \\
\addlinespace
Since 2021Q3 & \textbf{\textcolor{blue}{0.984}}$^{**}$ & \textbf{\textcolor{blue}{0.956}}$^{*}$ & \textbf{\textcolor{blue}{0.737}}$^{***}$ & 1.005$^{*}$ & \textbf{\textcolor{blue}{0.958}}$^{***}$ & 1.009 \\
2021Q3--2024Q1 & & & & & & \\
\midrule
\multicolumn{7}{c}{\cellcolor{gray!15}\textbf{Panel B: GW Conditional Test}} \\
\midrule
Full Sample & 1.008$^{**}$ & \textbf{\textcolor{blue}{0.887}}$^{***}$ & \textbf{\textcolor{blue}{0.950}}$^{*}$ & \textbf{\textcolor{blue}{0.828}}$^{***}$ & \textbf{\textcolor{blue}{0.985}}$^{**}$ & \textbf{\textcolor{blue}{0.818}}$^{***}$ \\
1975Q2--2023Q1 & & & & & & \\
\addlinespace
Stagflation & 1.131 & \textbf{\textcolor{blue}{0.722}}$^{***}$ & \textbf{\textcolor{blue}{0.861}}$^{**}$ & \textbf{\textcolor{blue}{0.892}}$^{***}$ & \textbf{\textcolor{blue}{0.985}}$^{***}$ & \textbf{\textcolor{blue}{0.976}}$^{***}$ \\
1975Q2--1979Q4 & & & & & & \\
\addlinespace
Thatcher & 1.017$^{*}$ & \textbf{\textcolor{blue}{0.917}}$^{**}$ & \textbf{\textcolor{blue}{0.988}}$^{**}$ & \textbf{\textcolor{blue}{0.946}}$^{**}$ & \textbf{\textcolor{blue}{0.960}}$^{***}$ & \textbf{\textcolor{blue}{0.798}}$^{***}$ \\
1979Q1--1989Q4 & & & & & & \\
\addlinespace
ERM crisis & \textbf{\textcolor{blue}{0.945}}$^{***}$ & \textbf{\textcolor{blue}{0.738}}$^{***}$ & \textbf{\textcolor{blue}{0.944}}$^{**}$ & \textbf{\textcolor{blue}{0.896}}$^{***}$ & 1.003$^{***}$ & \textbf{\textcolor{blue}{0.697}}$^{***}$ \\
1990Q1--1992Q4 & & & & & & \\
\addlinespace
Great Moderation & 1.046$^{***}$ & \textbf{\textcolor{blue}{0.860}}$^{***}$ & 1.056$^{*}$ & \textbf{\textcolor{blue}{0.700}}$^{***}$ & 1.042$^{**}$ & \textbf{\textcolor{blue}{0.719}}$^{***}$ \\
1993Q1--2007Q4 & & & & & & \\
\addlinespace
GFC & 1.113$^{**}$ & 1.042 & \textbf{\textcolor{blue}{0.883}}$^{***}$ & \textbf{\textcolor{blue}{0.762}}$^{***}$ & \textbf{\textcolor{blue}{0.982}}$^{**}$ & 1.009$^{***}$ \\
2007Q1--2012Q4 & & & & & & \\
\addlinespace
Brexit & 1.160$^{***}$ & \textbf{\textcolor{blue}{0.808}}$^{***}$ & \textbf{\textcolor{blue}{0.978}}$^{**}$ & \textbf{\textcolor{blue}{0.647}}$^{***}$ & \textbf{\textcolor{blue}{0.953}}$^{***}$ & \textbf{\textcolor{blue}{0.653}}$^{***}$ \\
2013Q1--2019Q4 & & & & & & \\
\addlinespace
COVID-19 & \textbf{\textcolor{blue}{0.929}}$^{***}$ & 1.024$^{***}$ & \textbf{\textcolor{blue}{0.967}}$^{***}$ & \textbf{\textcolor{blue}{0.691}}$^{***}$ & \textbf{\textcolor{blue}{0.969}}$^{***}$ & \textbf{\textcolor{blue}{0.875}}$^{***}$ \\
2020Q1--2021Q2 & & & & & & \\
\addlinespace
Since 2021Q3 & \textbf{\textcolor{blue}{0.984}}$^{**}$ & \textbf{\textcolor{blue}{0.956}}$^{***}$ & \textbf{\textcolor{blue}{0.737}}$^{***}$ & 1.005$^{***}$ & \textbf{\textcolor{blue}{0.958}}$^{***}$ & 1.009$^{***}$ \\
2021Q3--2024Q1 & & & & & & \\
\bottomrule
\end{tabular}
\begin{tablenotes}
\tiny
\item \textbf{Notes:} Panel A shows CRPS ratios ($VAR^{\sigma,\kappa}$ CRPS / Competitor CRPS) using Giacomini-White unconditional test. Panel B shows CRPS ratios using Giacomini-White conditional test. Values less than 1 (\textcolor{blue}{\textbf{bold blue}}) indicate the proposed model outperforms the competitor. All ratios are averaged across forecast horizons 1-8. Significance levels: *** p$<$0.01, ** p$<$0.05, * p$<$0.1.
\end{tablenotes}
\end{threeparttable}
\end{table}

\begin{table}[H]
\centering
\caption{Weighted CRPS Left Tail Ratios}
\label{tab:wcrps_left_tail_uk_gw_comparison}
\begin{threeparttable}
\scriptsize
\begin{tabular}{@{}l*{6}{c}@{}}
\toprule
 & \multicolumn{6}{c}{\textbf{Variables}} \\
\cmidrule(lr){2-7}
\textbf{Period} & \multicolumn{2}{c}{\textbf{GNP}} & \multicolumn{2}{c}{\textbf{GNP Deflator}} & \multicolumn{2}{c}{\textbf{SPREAD}} \\
\cmidrule(lr){2-3} \cmidrule(lr){4-5} \cmidrule(lr){6-7}
 & \textbf{$VAR^{\sigma,\kappa}_{rest}$} & \textbf{$VAR^{\sigma}$} & \textbf{$VAR^{\sigma,\kappa}_{rest}$} & \textbf{$VAR^{\sigma}$} & \textbf{$VAR^{\sigma,\kappa}_{rest}$} & \textbf{$VAR^{\sigma}$} \\
\midrule
\multicolumn{7}{c}{\cellcolor{gray!15}\textbf{Panel A: GW Unconditional Test}} \\
\midrule
Full Sample & 1.027$^{***}$ & \textbf{\textcolor{blue}{0.889}}$^{***}$ & 1.012 & \textbf{\textcolor{blue}{0.788}}$^{***}$ & \textbf{\textcolor{blue}{0.936}}$^{***}$ & \textbf{\textcolor{blue}{0.658}}$^{***}$ \\
1975Q2--2023Q1 & & & & & & \\
\addlinespace
Stagflation & 1.166 & \textbf{\textcolor{blue}{0.683}}$^{***}$ & 1.130$^{*}$ & 1.085 & \textbf{\textcolor{blue}{0.878}}$^{***}$ & \textbf{\textcolor{blue}{0.805}}$^{***}$ \\
1975Q2--1979Q4 & & & & & & \\
\addlinespace
Thatcher & 1.027 & \textbf{\textcolor{blue}{0.852}}$^{*}$ & 1.081$^{*}$ & \textbf{\textcolor{blue}{0.930}}$^{***}$ & \textbf{\textcolor{blue}{0.934}}$^{***}$ & \textbf{\textcolor{blue}{0.732}}$^{***}$ \\
1979Q1--1989Q4 & & & & & & \\
\addlinespace
ERM crisis & \textbf{\textcolor{blue}{0.941}}$^{***}$ & \textbf{\textcolor{blue}{0.740}}$^{***}$ & \textbf{\textcolor{blue}{0.961}}$^{**}$ & \textbf{\textcolor{blue}{0.892}}$^{***}$ & \textbf{\textcolor{blue}{0.983}}$^{***}$ & \textbf{\textcolor{blue}{0.613}}$^{***}$ \\
1990Q1--1992Q4 & & & & & & \\
\addlinespace
Great Moderation & 1.054$^{***}$ & \textbf{\textcolor{blue}{0.857}}$^{***}$ & 1.055$^{*}$ & \textbf{\textcolor{blue}{0.700}}$^{***}$ & \textbf{\textcolor{blue}{0.967}} & \textbf{\textcolor{blue}{0.587}}$^{***}$ \\
1993Q1--2007Q4 & & & & & & \\
\addlinespace
GFC & 1.104$^{**}$ & 1.051$^{*}$ & \textbf{\textcolor{blue}{0.884}}$^{***}$ & \textbf{\textcolor{blue}{0.758}}$^{***}$ & \textbf{\textcolor{blue}{0.912}}$^{**}$ & \textbf{\textcolor{blue}{0.511}}$^{***}$ \\
2007Q1--2012Q4 & & & & & & \\
\addlinespace
Brexit & 1.159$^{***}$ & \textbf{\textcolor{blue}{0.812}}$^{***}$ & \textbf{\textcolor{blue}{0.985}} & \textbf{\textcolor{blue}{0.653}}$^{***}$ & \textbf{\textcolor{blue}{0.952}}$^{***}$ & \textbf{\textcolor{blue}{0.588}}$^{***}$ \\
2013Q1--2019Q4 & & & & & & \\
\addlinespace
COVID-19 & \textbf{\textcolor{blue}{0.962}}$^{*}$ & 1.030$^{***}$ & \textbf{\textcolor{blue}{0.966}}$^{***}$ & \textbf{\textcolor{blue}{0.695}}$^{***}$ & \textbf{\textcolor{blue}{0.924}}$^{***}$ & \textbf{\textcolor{blue}{0.473}}$^{**}$ \\
2020Q1--2021Q2 & & & & & & \\
\addlinespace
Since 2021Q3 & \textbf{\textcolor{blue}{0.999}}$^{***}$ & \textbf{\textcolor{blue}{0.972}}$^{**}$ & \textbf{\textcolor{blue}{0.736}}$^{***}$ & \textbf{\textcolor{blue}{0.970}}$^{*}$ & \textbf{\textcolor{blue}{0.958}}$^{***}$ & \textbf{\textcolor{blue}{0.581}} \\
2021Q3--2024Q1 & & & & & & \\
\midrule
\multicolumn{7}{c}{\cellcolor{gray!15}\textbf{Panel B: GW Conditional Test}} \\
\midrule
Full Sample & 1.027$^{**}$ & \textbf{\textcolor{blue}{0.889}}$^{***}$ & 1.012$^{***}$ & \textbf{\textcolor{blue}{0.788}}$^{***}$ & \textbf{\textcolor{blue}{0.936}}$^{***}$ & \textbf{\textcolor{blue}{0.658}}$^{***}$ \\
1975Q2--2023Q1 & & & & & & \\
\addlinespace
Stagflation & 1.166 & \textbf{\textcolor{blue}{0.683}}$^{***}$ & 1.130$^{***}$ & 1.085$^{***}$ & \textbf{\textcolor{blue}{0.878}}$^{***}$ & \textbf{\textcolor{blue}{0.805}}$^{***}$ \\
1975Q2--1979Q4 & & & & & & \\
\addlinespace
Thatcher & 1.027 & \textbf{\textcolor{blue}{0.852}}$^{*}$ & 1.081$^{**}$ & \textbf{\textcolor{blue}{0.930}}$^{**}$ & \textbf{\textcolor{blue}{0.934}}$^{***}$ & \textbf{\textcolor{blue}{0.732}}$^{***}$ \\
1979Q1--1989Q4 & & & & & & \\
\addlinespace
ERM crisis & \textbf{\textcolor{blue}{0.941}}$^{***}$ & \textbf{\textcolor{blue}{0.740}}$^{***}$ & \textbf{\textcolor{blue}{0.961}}$^{**}$ & \textbf{\textcolor{blue}{0.892}}$^{***}$ & \textbf{\textcolor{blue}{0.983}}$^{**}$ & \textbf{\textcolor{blue}{0.613}}$^{***}$ \\
1990Q1--1992Q4 & & & & & & \\
\addlinespace
Great Moderation & 1.054$^{***}$ & \textbf{\textcolor{blue}{0.857}}$^{***}$ & 1.055$^{*}$ & \textbf{\textcolor{blue}{0.700}}$^{***}$ & \textbf{\textcolor{blue}{0.967}}$^{*}$ & \textbf{\textcolor{blue}{0.587}}$^{***}$ \\
1993Q1--2007Q4 & & & & & & \\
\addlinespace
GFC & 1.104$^{**}$ & 1.051$^{*}$ & \textbf{\textcolor{blue}{0.884}}$^{***}$ & \textbf{\textcolor{blue}{0.758}}$^{***}$ & \textbf{\textcolor{blue}{0.912}}$^{**}$ & \textbf{\textcolor{blue}{0.511}}$^{**}$ \\
2007Q1--2012Q4 & & & & & & \\
\addlinespace
Brexit & 1.159$^{***}$ & \textbf{\textcolor{blue}{0.812}}$^{***}$ & \textbf{\textcolor{blue}{0.985}}$^{**}$ & \textbf{\textcolor{blue}{0.653}}$^{***}$ & \textbf{\textcolor{blue}{0.952}}$^{***}$ & \textbf{\textcolor{blue}{0.588}}$^{***}$ \\
2013Q1--2019Q4 & & & & & & \\
\addlinespace
COVID-19 & \textbf{\textcolor{blue}{0.962}}$^{***}$ & 1.030$^{***}$ & \textbf{\textcolor{blue}{0.966}}$^{***}$ & \textbf{\textcolor{blue}{0.695}}$^{***}$ & \textbf{\textcolor{blue}{0.924}}$^{***}$ & \textbf{\textcolor{blue}{0.473}}$^{***}$ \\
2020Q1--2021Q2 & & & & & & \\
\addlinespace
Since 2021Q3 & \textbf{\textcolor{blue}{0.999}}$^{**}$ & \textbf{\textcolor{blue}{0.972}}$^{***}$ & \textbf{\textcolor{blue}{0.736}}$^{***}$ & \textbf{\textcolor{blue}{0.970}}$^{***}$ & \textbf{\textcolor{blue}{0.958}}$^{***}$ & \textbf{\textcolor{blue}{0.581}}$^{***}$ \\
2021Q3--2024Q1 & & & & & & \\
\bottomrule
\end{tabular}
\begin{tablenotes}
\tiny
\item \textbf{Notes:} Panel A shows Weighted CRPS left tail ratios ($VAR^{\sigma,\kappa}$ WCRPS-L / Competitor WCRPS-L) using Giacomini-White unconditional test. Panel B shows Weighted CRPS left tail ratios using Giacomini-White conditional test. Values less than 1 (\textcolor{blue}{\textbf{bold blue}}) indicate the proposed model outperforms the competitor on left tail forecasting. All ratios are averaged across forecast horizons 1-8. Significance levels: *** p$<$0.01, ** p$<$0.05, * p$<$0.1.
\end{tablenotes}
\end{threeparttable}
\end{table}

\begin{table}[H]
\centering
\caption{Weighted CRPS Right Tail Ratios}
\label{tab:wcrps_right_tail_uk_gw_comparison}
\begin{threeparttable}
\scriptsize
\begin{tabular}{@{}l*{6}{c}@{}}
\toprule
 & \multicolumn{6}{c}{\textbf{Variables}} \\
\cmidrule(lr){2-7}
\textbf{Period} & \multicolumn{2}{c}{\textbf{GNP}} & \multicolumn{2}{c}{\textbf{GNP Deflator}} & \multicolumn{2}{c}{\textbf{SPREAD}} \\
\cmidrule(lr){2-3} \cmidrule(lr){4-5} \cmidrule(lr){6-7}
 & \textbf{$VAR^{\sigma,\kappa}_{rest}$} & \textbf{$VAR^{\sigma}$} & \textbf{$VAR^{\sigma,\kappa}_{rest}$} & \textbf{$VAR^{\sigma}$} & \textbf{$VAR^{\sigma,\kappa}_{rest}$} & \textbf{$VAR^{\sigma}$} \\
\midrule
\multicolumn{7}{c}{\cellcolor{gray!15}\textbf{Panel A: GW Unconditional Test}} \\
\midrule
Full Sample & \textbf{\textcolor{blue}{0.987}} & \textbf{\textcolor{blue}{0.883}}$^{***}$ & \textbf{\textcolor{blue}{0.914}} & \textbf{\textcolor{blue}{0.854}}$^{***}$ & 1.004 & \textbf{\textcolor{blue}{0.897}}$^{***}$ \\
1975Q2--2023Q1 & & & & & & \\
\addlinespace
Stagflation & 1.101 & \textbf{\textcolor{blue}{0.767}}$^{***}$ & \textbf{\textcolor{blue}{0.833}} & \textbf{\textcolor{blue}{0.874}} & 1.054$^{***}$ & 1.100$^{***}$ \\
1975Q2--1979Q4 & & & & & & \\
\addlinespace
Thatcher & 1.006 & \textbf{\textcolor{blue}{0.998}} & \textbf{\textcolor{blue}{0.953}} & \textbf{\textcolor{blue}{0.952}}$^{***}$ & \textbf{\textcolor{blue}{0.981}}$^{*}$ & \textbf{\textcolor{blue}{0.859}}$^{***}$ \\
1979Q1--1989Q4 & & & & & & \\
\addlinespace
ERM crisis & \textbf{\textcolor{blue}{0.952}}$^{**}$ & \textbf{\textcolor{blue}{0.735}}$^{***}$ & \textbf{\textcolor{blue}{0.935}}$^{***}$ & \textbf{\textcolor{blue}{0.898}}$^{***}$ & 1.012$^{***}$ & \textbf{\textcolor{blue}{0.744}}$^{***}$ \\
1990Q1--1992Q4 & & & & & & \\
\addlinespace
Great Moderation & 1.040$^{***}$ & \textbf{\textcolor{blue}{0.861}}$^{***}$ & 1.058 & \textbf{\textcolor{blue}{0.701}}$^{***}$ & 1.078$^{***}$ & \textbf{\textcolor{blue}{0.799}}$^{***}$ \\
1993Q1--2007Q4 & & & & & & \\
\addlinespace
GFC & 1.130$^{**}$ & 1.018 & \textbf{\textcolor{blue}{0.883}}$^{***}$ & \textbf{\textcolor{blue}{0.766}}$^{***}$ & \textbf{\textcolor{blue}{0.987}} & 1.069 \\
2007Q1--2012Q4 & & & & & & \\
\addlinespace
Brexit & 1.161$^{***}$ & \textbf{\textcolor{blue}{0.803}}$^{***}$ & \textbf{\textcolor{blue}{0.968}} & \textbf{\textcolor{blue}{0.637}}$^{***}$ & \textbf{\textcolor{blue}{0.953}}$^{***}$ & \textbf{\textcolor{blue}{0.681}}$^{***}$ \\
2013Q1--2019Q4 & & & & & & \\
\addlinespace
COVID-19 & \textbf{\textcolor{blue}{0.879}}$^{*}$ & 1.031$^{***}$ & \textbf{\textcolor{blue}{0.968}}$^{***}$ & \textbf{\textcolor{blue}{0.686}}$^{***}$ & \textbf{\textcolor{blue}{0.975}}$^{**}$ & \textbf{\textcolor{blue}{0.957}} \\
2020Q1--2021Q2 & & & & & & \\
\addlinespace
Since 2021Q3 & \textbf{\textcolor{blue}{0.976}}$^{*}$ & \textbf{\textcolor{blue}{0.948}} & \textbf{\textcolor{blue}{0.736}}$^{***}$ & 1.019$^{*}$ & \textbf{\textcolor{blue}{0.957}}$^{***}$ & 1.087 \\
2021Q3--2024Q1 & & & & & & \\
\midrule
\multicolumn{7}{c}{\cellcolor{gray!15}\textbf{Panel B: GW Conditional Test}} \\
\midrule
Full Sample & \textbf{\textcolor{blue}{0.987}}$^{**}$ & \textbf{\textcolor{blue}{0.883}}$^{**}$ & \textbf{\textcolor{blue}{0.914}} & \textbf{\textcolor{blue}{0.854}}$^{***}$ & 1.004 & \textbf{\textcolor{blue}{0.897}}$^{***}$ \\
1975Q2--2023Q1 & & & & & & \\
\addlinespace
Stagflation & 1.101 & \textbf{\textcolor{blue}{0.767}}$^{***}$ & \textbf{\textcolor{blue}{0.833}}$^{*}$ & \textbf{\textcolor{blue}{0.874}}$^{**}$ & 1.054$^{***}$ & 1.100$^{***}$ \\
1975Q2--1979Q4 & & & & & & \\
\addlinespace
Thatcher & 1.006$^{***}$ & \textbf{\textcolor{blue}{0.998}}$^{***}$ & \textbf{\textcolor{blue}{0.953}} & \textbf{\textcolor{blue}{0.952}}$^{**}$ & \textbf{\textcolor{blue}{0.981}}$^{***}$ & \textbf{\textcolor{blue}{0.859}}$^{***}$ \\
1979Q1--1989Q4 & & & & & & \\
\addlinespace
ERM crisis & \textbf{\textcolor{blue}{0.952}}$^{***}$ & \textbf{\textcolor{blue}{0.735}}$^{***}$ & \textbf{\textcolor{blue}{0.935}}$^{**}$ & \textbf{\textcolor{blue}{0.898}}$^{***}$ & 1.012$^{***}$ & \textbf{\textcolor{blue}{0.744}}$^{***}$ \\
1990Q1--1992Q4 & & & & & & \\
\addlinespace
Great Moderation & 1.040$^{***}$ & \textbf{\textcolor{blue}{0.861}}$^{***}$ & 1.058$^{*}$ & \textbf{\textcolor{blue}{0.701}}$^{***}$ & 1.078$^{**}$ & \textbf{\textcolor{blue}{0.799}}$^{***}$ \\
1993Q1--2007Q4 & & & & & & \\
\addlinespace
GFC & 1.130$^{***}$ & 1.018$^{**}$ & \textbf{\textcolor{blue}{0.883}}$^{***}$ & \textbf{\textcolor{blue}{0.766}}$^{***}$ & \textbf{\textcolor{blue}{0.987}}$^{**}$ & 1.069$^{*}$ \\
2007Q1--2012Q4 & & & & & & \\
\addlinespace
Brexit & 1.161$^{***}$ & \textbf{\textcolor{blue}{0.803}}$^{***}$ & \textbf{\textcolor{blue}{0.968}}$^{***}$ & \textbf{\textcolor{blue}{0.637}}$^{***}$ & \textbf{\textcolor{blue}{0.953}}$^{***}$ & \textbf{\textcolor{blue}{0.681}}$^{***}$ \\
2013Q1--2019Q4 & & & & & & \\
\addlinespace
COVID-19 & \textbf{\textcolor{blue}{0.879}}$^{***}$ & 1.031$^{***}$ & \textbf{\textcolor{blue}{0.968}}$^{***}$ & \textbf{\textcolor{blue}{0.686}}$^{***}$ & \textbf{\textcolor{blue}{0.975}}$^{***}$ & \textbf{\textcolor{blue}{0.957}}$^{***}$ \\
2020Q1--2021Q2 & & & & & & \\
\addlinespace
Since 2021Q3 & \textbf{\textcolor{blue}{0.976}}$^{**}$ & \textbf{\textcolor{blue}{0.948}}$^{***}$ & \textbf{\textcolor{blue}{0.736}}$^{***}$ & 1.019$^{**}$ & \textbf{\textcolor{blue}{0.957}}$^{***}$ & 1.087$^{***}$ \\
2021Q3--2024Q1 & & & & & & \\
\bottomrule
\end{tabular}
\begin{tablenotes}
\tiny
\item \textbf{Notes:} Panel A shows Weighted CRPS right tail ratios ($VAR^{\sigma,\kappa}$ WCRPS-R / Competitor WCRPS-R) using Giacomini-White unconditional test. Panel B shows Weighted CRPS right tail ratios using Giacomini-White conditional test. Values less than 1 (\textcolor{blue}{\textbf{bold blue}}) indicate the proposed model outperforms the competitor on right tail forecasting. All ratios are averaged across forecast horizons 1-8. Significance levels: *** p$<$0.01, ** p$<$0.05, * p$<$0.1.
\end{tablenotes}
\end{threeparttable}
\end{table}

\clearpage

\clearpage

\end{document}